\documentclass[longauth]{aa}  
\usepackage{graphicx}
\usepackage{natbib}
\usepackage{scalerel}

\usepackage[table]{xcolor}

\usepackage{txfonts}

\makeatletter
\renewcommand*\aa@pageof{, page \thepage{} of \pageref*{LastPage}}
\makeatother

\usepackage[utf8]{inputenc}

\usepackage[switch, modulo]{lineno}

\usepackage{euclid}
\usepackage{orcidlink}

\usepackage{amsmath,amssymb,amsfonts}
\usepackage{enumerate} 
\usepackage{bm}
\usepackage{float}
\usepackage{subfigure}
\usepackage{braket}
\usepackage{placeins}
\usepackage{tabularx}
\usepackage{ulem}
\usepackage{fontawesome}
\usepackage{multirow}	
\usepackage{booktabs}
\usepackage{csquotes}
\usepackage{soul}

\let\oldequation\equation
\let\oldendequation\endequation

\renewenvironment{equation}
  {\linenomathNonumbers\oldequation}
  {\oldendequation\endlinenomath}
  
\let\oldalign\align
\let\oldendalign\endalign

\renewenvironment{align}
  {\linenomathNonumbers\oldalign}
  {\oldendalign\endlinenomath}

\definecolor{cyan}{rgb}{0.88,1,1}
\definecolor{gray}{gray}{0.9}
\definecolor{crisp}{HTML}{c7ddf2}
\definecolor{crispier}{HTML}{e8f1fa}

\definecolor{amaranth}{rgb}{0.9, 0.17, 0.31}
\definecolor{forestgreen(web)}{rgb}{0.13, 0.55, 0.13}
\definecolor{lavender(web)}{rgb}{0.9, 0.9, 0.98}
\definecolor{cosmiclatte}{rgb}{1.0, 0.97, 0.91}
\definecolor{jonquil}{rgb}{0.98, 0.85, 0.37}
\definecolor{khaki(x11)(lightkhaki)}{rgb}{0.94, 0.9, 0.55}
\definecolor{thistle}{rgb}{0.85, 0.75, 0.85}
\definecolor{fuchia}{rgb}{1.0,0.0,1.0}
 
\newcommand{\threextwo}{\text{3$\times$2pt}}
\newcommand{\GCsp}{\text{GC}\ensuremath{_\mathrm{sp}}}
\newcommand{\GCph}{\text{GC}\ensuremath{_\mathrm{ph}}}
\newcommand{\XCph}{\text{XC}\ensuremath{_\mathrm{ph}}}
\newcommand{\Omegam}
{\ensuremath{\Omega_{\mathrm{m}}}}
\newcommand{\Omegab}{\ensuremath{\Omega_{\mathrm{b}}}}
\newcommand{\Omegac}
{\ensuremath{\Omega_{\mathrm{c}}}}
\newcommand{\Omeganu}{\ensuremath{\Omega_{\nu}}}
\newcommand{\OmegaDE}{\ensuremath{\Omega_{\mathrm{DE}}}}

\newcommand{\lcdm}{\ensuremath{\Lambda\mathrm{CDM}}}
\newcommand{\logfr}{\ensuremath{\logten|f_{R0}|}}
\newcommand{\fr}{\ensuremath{|f_{R0}|}}

\newcommand{\de}{\mathrm{d}}
\DeclareMathOperator{\Tr}{Tr}

\newcommand{\arepo}{\texttt{Arepo}}
\newcommand{\mgarepo}{\texttt{MG-AREPO}}
\newcommand{\ecosmog}{\texttt{ECOSMOG}}
\newcommand{\forge}{\texttt{FORGE}}
\newcommand{\react}{\texttt{ReACT}}
\newcommand{\emantis}{\texttt{e-Mantis}}
\newcommand{\ramses}{\texttt{RAMSES}}
\newcommand{\halofit}{\texttt{halofit}}
\newcommand{\montepython}{\texttt{MontePython}}
\newcommand{\bcemu}{\texttt{BCemu}}
\newcommand{\eetwo}{\texttt{EuclidEmulator2}}
\newcommand{\bacco}{\texttt{Bacco}}
\newcommand{\hmcode}{\texttt{HMCode2020}}
\newcommand{\fofrfitting}{\texttt{fitting}}
\newcommand{\dustgrain}{\texttt{DUSTGRAIN}}
\newcommand{\elephant}{\texttt{ELEPHANT}}
\newcommand{\cosmopower}{\texttt{Cosmopower}}
\newcommand{\mggadget}{\texttt{MG-Gadget}}

\defcitealias{Blanchard:2019oqi}{EC20}
\defcitealias{paper1}{paper 1}
\defcitealias{paper2}{paper 2}

\usepackage{xparse}
\ExplSyntaxOn
\NewDocumentCommand{\citelist}{m}
 {
  \bool_set_true:N \l_tmpa_bool 
  \clist_map_inline:nn { #1 }
   {
    \bool_if:NF \l_tmpa_bool {,~}
    \cite{##1}
   \bool_set_false:N \l_tmpa_bool

   }
 }
\ExplSyntaxOff

\hypersetup{
    colorlinks=true,
    linkcolor=blue,
    filecolor=magenta,      
    urlcolor=blue,
    citecolor=blue
}

\usepackage[nameinlink, capitalise]{cleveref}
\usepackage[switch, modulo]{lineno}
\Crefname{chapter}{Chap.}{Chaps.}
\Crefname{section}{Sect.}{Sects.}
\Crefname{figure}{Fig.}{Figs.}
\Crefname{chapter}{Chapter}{Chapters}
\Crefname{section}{Section}{Sections}
\Crefname{figure}{Figure}{Figures}

\begin{document}
\title{\Euclid preparation}
\subtitle{LXXI. Simulations and nonlinearities beyond $\mathsf{\Lambda}$CDM. 3. Constraints on $f(R)$ models from the photometric primary probes}    
\newcommand{\orcid}[1]{} 


\author{Euclid Collaboration: K.~Koyama\orcid{0000-0001-6727-6915}\thanks{\email{kazuya.koyama@port.ac.uk}}\inst{\ref{aff1}}
	\and S.~Pamuk\orcid{0009-0004-0852-8624}\inst{\ref{aff2}}
	\and S.~Casas\orcid{0000-0002-4751-5138}\inst{\ref{aff2}}
	\and B.~Bose\orcid{0000-0003-1965-8614}\inst{\ref{aff3}}
	\and P.~Carrilho\orcid{0000-0003-1339-0194}\inst{\ref{aff3}}
	\and I.~S\'aez-Casares\orcid{0000-0003-0013-5266}\inst{\ref{aff4}}
	\and L.~Atayde\orcid{0000-0001-6373-9193}\inst{\ref{aff5},\ref{aff6}}
	\and M.~Cataneo\orcid{0000-0002-7992-0656}\inst{\ref{aff7},\ref{aff8}}
	\and B.~Fiorini\orcid{0000-0002-0092-4321}\inst{\ref{aff1}}
	\and C.~Giocoli\orcid{0000-0002-9590-7961}\inst{\ref{aff9},\ref{aff10}}
	\and A.~M.~C.~Le~Brun\orcid{0000-0002-0936-4594}\inst{\ref{aff4}}
	\and F.~Pace\orcid{0000-0001-8039-0480}\inst{\ref{aff11},\ref{aff12},\ref{aff13}}
	\and A.~Pourtsidou\orcid{0000-0001-9110-5550}\inst{\ref{aff3},\ref{aff14}}
	\and Y.~Rasera\orcid{0000-0003-3424-6941}\inst{\ref{aff4},\ref{aff15}}
	\and Z.~Sakr\orcid{0000-0002-4823-3757}\inst{\ref{aff16},\ref{aff17},\ref{aff18}}
	\and H.-A.~Winther\orcid{0000-0002-6325-2710}\inst{\ref{aff19}}
	\and E.~Altamura\orcid{0000-0001-6973-1897}\inst{\ref{aff20}}
	\and J.~Adamek\orcid{0000-0002-0723-6740}\inst{\ref{aff21}}
	\and M.~Baldi\orcid{0000-0003-4145-1943}\inst{\ref{aff22},\ref{aff9},\ref{aff23}}
	\and M.-A.~Breton\inst{\ref{aff24},\ref{aff25},\ref{aff4}}
	\and G.~R\'acz\orcid{0000-0003-3906-5699}\inst{\ref{aff26}}
	\and F.~Vernizzi\orcid{0000-0003-3426-2802}\inst{\ref{aff27}}
	\and A.~Amara\inst{\ref{aff28}}
	\and S.~Andreon\orcid{0000-0002-2041-8784}\inst{\ref{aff29}}
	\and N.~Auricchio\orcid{0000-0003-4444-8651}\inst{\ref{aff9}}
	\and C.~Baccigalupi\orcid{0000-0002-8211-1630}\inst{\ref{aff30},\ref{aff31},\ref{aff32},\ref{aff33}}
	\and S.~Bardelli\orcid{0000-0002-8900-0298}\inst{\ref{aff9}}
	\and F.~Bernardeau\inst{\ref{aff27},\ref{aff34}}
	\and A.~Biviano\orcid{0000-0002-0857-0732}\inst{\ref{aff31},\ref{aff30}}
	\and C.~Bodendorf\inst{\ref{aff35}}
	\and D.~Bonino\orcid{0000-0002-3336-9977}\inst{\ref{aff13}}
	\and E.~Branchini\orcid{0000-0002-0808-6908}\inst{\ref{aff36},\ref{aff37},\ref{aff29}}
	\and M.~Brescia\orcid{0000-0001-9506-5680}\inst{\ref{aff38},\ref{aff39},\ref{aff40}}
	\and J.~Brinchmann\orcid{0000-0003-4359-8797}\inst{\ref{aff41},\ref{aff42}}
	\and A.~Caillat\inst{\ref{aff43}}
	\and S.~Camera\orcid{0000-0003-3399-3574}\inst{\ref{aff11},\ref{aff12},\ref{aff13}}
	\and G.~Ca\~nas-Herrera\orcid{0000-0003-2796-2149}\inst{\ref{aff44},\ref{aff45}}
	\and V.~Capobianco\orcid{0000-0002-3309-7692}\inst{\ref{aff13}}
	\and C.~Carbone\orcid{0000-0003-0125-3563}\inst{\ref{aff46}}
	\and J.~Carretero\orcid{0000-0002-3130-0204}\inst{\ref{aff47},\ref{aff48}}
	\and M.~Castellano\orcid{0000-0001-9875-8263}\inst{\ref{aff49}}
	\and G.~Castignani\orcid{0000-0001-6831-0687}\inst{\ref{aff9}}
	\and S.~Cavuoti\orcid{0000-0002-3787-4196}\inst{\ref{aff39},\ref{aff40}}
	\and K.~C.~Chambers\orcid{0000-0001-6965-7789}\inst{\ref{aff50}}
	\and A.~Cimatti\inst{\ref{aff51}}
	\and C.~Colodro-Conde\inst{\ref{aff52}}
	\and G.~Congedo\orcid{0000-0003-2508-0046}\inst{\ref{aff3}}
	\and C.~J.~Conselice\orcid{0000-0003-1949-7638}\inst{\ref{aff20}}
	\and L.~Conversi\orcid{0000-0002-6710-8476}\inst{\ref{aff53},\ref{aff54}}
	\and Y.~Copin\orcid{0000-0002-5317-7518}\inst{\ref{aff55}}
	\and F.~Courbin\orcid{0000-0003-0758-6510}\inst{\ref{aff56},\ref{aff57},\ref{aff58}}
	\and H.~M.~Courtois\orcid{0000-0003-0509-1776}\inst{\ref{aff59}}
	\and A.~Da~Silva\orcid{0000-0002-6385-1609}\inst{\ref{aff6},\ref{aff5}}
	\and H.~Degaudenzi\orcid{0000-0002-5887-6799}\inst{\ref{aff60}}
	\and G.~De~Lucia\orcid{0000-0002-6220-9104}\inst{\ref{aff31}}
	\and H.~Dole\orcid{0000-0002-9767-3839}\inst{\ref{aff61}}
	\and M.~Douspis\orcid{0000-0003-4203-3954}\inst{\ref{aff61}}
	\and F.~Dubath\orcid{0000-0002-6533-2810}\inst{\ref{aff60}}
	\and C.~A.~J.~Duncan\inst{\ref{aff20}}
	\and X.~Dupac\inst{\ref{aff54}}
	\and S.~Dusini\orcid{0000-0002-1128-0664}\inst{\ref{aff62}}
	\and S.~Escoffier\orcid{0000-0002-2847-7498}\inst{\ref{aff63}}
	\and M.~Farina\orcid{0000-0002-3089-7846}\inst{\ref{aff64}}
	\and R.~Farinelli\inst{\ref{aff9}}
	\and S.~Farrens\orcid{0000-0002-9594-9387}\inst{\ref{aff65}}
	\and S.~Ferriol\inst{\ref{aff55}}
	\and F.~Finelli\orcid{0000-0002-6694-3269}\inst{\ref{aff9},\ref{aff66}}
	\and P.~Fosalba\orcid{0000-0002-1510-5214}\inst{\ref{aff67},\ref{aff24}}
	\and M.~Frailis\orcid{0000-0002-7400-2135}\inst{\ref{aff31}}
	\and E.~Franceschi\orcid{0000-0002-0585-6591}\inst{\ref{aff9}}
	\and S.~Galeotta\orcid{0000-0002-3748-5115}\inst{\ref{aff31}}
	\and B.~Gillis\orcid{0000-0002-4478-1270}\inst{\ref{aff3}}
	\and P.~G\'omez-Alvarez\orcid{0000-0002-8594-5358}\inst{\ref{aff68},\ref{aff54}}
	\and J.~Gracia-Carpio\inst{\ref{aff35}}
	\and A.~Grazian\orcid{0000-0002-5688-0663}\inst{\ref{aff69}}
	\and F.~Grupp\inst{\ref{aff35},\ref{aff70}}
	\and L.~Guzzo\orcid{0000-0001-8264-5192}\inst{\ref{aff71},\ref{aff29}}
	\and M.~Hailey\inst{\ref{aff72}}
	\and S.~V.~H.~Haugan\orcid{0000-0001-9648-7260}\inst{\ref{aff19}}
	\and W.~Holmes\inst{\ref{aff26}}
	\and F.~Hormuth\inst{\ref{aff73}}
	\and A.~Hornstrup\orcid{0000-0002-3363-0936}\inst{\ref{aff74},\ref{aff75}}
	\and P.~Hudelot\inst{\ref{aff34}}
	\and S.~Ili\'c\orcid{0000-0003-4285-9086}\inst{\ref{aff76},\ref{aff17}}
	\and K.~Jahnke\orcid{0000-0003-3804-2137}\inst{\ref{aff77}}
	\and M.~Jhabvala\inst{\ref{aff78}}
	\and B.~Joachimi\orcid{0000-0001-7494-1303}\inst{\ref{aff79}}
	\and E.~Keih\"anen\orcid{0000-0003-1804-7715}\inst{\ref{aff80}}
	\and S.~Kermiche\orcid{0000-0002-0302-5735}\inst{\ref{aff63}}
	\and A.~Kiessling\orcid{0000-0002-2590-1273}\inst{\ref{aff26}}
	\and M.~Kilbinger\orcid{0000-0001-9513-7138}\inst{\ref{aff65}}
	\and B.~Kubik\orcid{0009-0006-5823-4880}\inst{\ref{aff55}}
	\and M.~Kunz\orcid{0000-0002-3052-7394}\inst{\ref{aff81}}
	\and H.~Kurki-Suonio\orcid{0000-0002-4618-3063}\inst{\ref{aff82},\ref{aff83}}
	\and P.~B.~Lilje\orcid{0000-0003-4324-7794}\inst{\ref{aff19}}
	\and V.~Lindholm\orcid{0000-0003-2317-5471}\inst{\ref{aff82},\ref{aff83}}
	\and I.~Lloro\inst{\ref{aff84}}
	\and G.~Mainetti\orcid{0000-0003-2384-2377}\inst{\ref{aff85}}
	\and D.~Maino\inst{\ref{aff71},\ref{aff46},\ref{aff86}}
	\and E.~Maiorano\orcid{0000-0003-2593-4355}\inst{\ref{aff9}}
	\and O.~Mansutti\orcid{0000-0001-5758-4658}\inst{\ref{aff31}}
	\and O.~Marggraf\orcid{0000-0001-7242-3852}\inst{\ref{aff8}}
	\and K.~Markovic\orcid{0000-0001-6764-073X}\inst{\ref{aff26}}
	\and M.~Martinelli\orcid{0000-0002-6943-7732}\inst{\ref{aff49},\ref{aff87}}
	\and N.~Martinet\orcid{0000-0003-2786-7790}\inst{\ref{aff43}}
	\and F.~Marulli\orcid{0000-0002-8850-0303}\inst{\ref{aff88},\ref{aff9},\ref{aff23}}
	\and R.~Massey\orcid{0000-0002-6085-3780}\inst{\ref{aff89}}
	\and E.~Medinaceli\orcid{0000-0002-4040-7783}\inst{\ref{aff9}}
	\and S.~Mei\orcid{0000-0002-2849-559X}\inst{\ref{aff90}}
	\and M.~Melchior\inst{\ref{aff91}}
	\and Y.~Mellier\inst{\ref{aff92},\ref{aff34}}
	\and M.~Meneghetti\orcid{0000-0003-1225-7084}\inst{\ref{aff9},\ref{aff23}}
	\and E.~Merlin\orcid{0000-0001-6870-8900}\inst{\ref{aff49}}
	\and G.~Meylan\inst{\ref{aff56}}
	\and A.~Mora\orcid{0000-0002-1922-8529}\inst{\ref{aff93}}
	\and M.~Moresco\orcid{0000-0002-7616-7136}\inst{\ref{aff88},\ref{aff9}}
	\and L.~Moscardini\orcid{0000-0002-3473-6716}\inst{\ref{aff88},\ref{aff9},\ref{aff23}}
	\and E.~Munari\orcid{0000-0002-1751-5946}\inst{\ref{aff31},\ref{aff30}}
	\and C.~Neissner\orcid{0000-0001-8524-4968}\inst{\ref{aff94},\ref{aff48}}
	\and S.-M.~Niemi\inst{\ref{aff44}}
	\and C.~Padilla\orcid{0000-0001-7951-0166}\inst{\ref{aff94}}
	\and S.~Paltani\orcid{0000-0002-8108-9179}\inst{\ref{aff60}}
	\and F.~Pasian\orcid{0000-0002-4869-3227}\inst{\ref{aff31}}
	\and K.~Pedersen\inst{\ref{aff95}}
	\and W.~J.~Percival\orcid{0000-0002-0644-5727}\inst{\ref{aff96},\ref{aff97},\ref{aff98}}
	\and V.~Pettorino\inst{\ref{aff44}}
	\and S.~Pires\orcid{0000-0002-0249-2104}\inst{\ref{aff65}}
	\and G.~Polenta\orcid{0000-0003-4067-9196}\inst{\ref{aff99}}
	\and M.~Poncet\inst{\ref{aff100}}
	\and L.~A.~Popa\inst{\ref{aff101}}
	\and L.~Pozzetti\orcid{0000-0001-7085-0412}\inst{\ref{aff9}}
	\and F.~Raison\orcid{0000-0002-7819-6918}\inst{\ref{aff35}}
	\and A.~Renzi\orcid{0000-0001-9856-1970}\inst{\ref{aff102},\ref{aff62}}
	\and J.~Rhodes\orcid{0000-0002-4485-8549}\inst{\ref{aff26}}
	\and G.~Riccio\inst{\ref{aff39}}
	\and E.~Romelli\orcid{0000-0003-3069-9222}\inst{\ref{aff31}}
	\and M.~Roncarelli\orcid{0000-0001-9587-7822}\inst{\ref{aff9}}
	\and R.~Saglia\orcid{0000-0003-0378-7032}\inst{\ref{aff70},\ref{aff35}}
	\and J.-C.~Salvignol\inst{\ref{aff44}}
	\and A.~G.~S\'anchez\orcid{0000-0003-1198-831X}\inst{\ref{aff35}}
	\and D.~Sapone\orcid{0000-0001-7089-4503}\inst{\ref{aff103}}
	\and B.~Sartoris\orcid{0000-0003-1337-5269}\inst{\ref{aff70},\ref{aff31}}
	\and M.~Schirmer\orcid{0000-0003-2568-9994}\inst{\ref{aff77}}
	\and T.~Schrabback\orcid{0000-0002-6987-7834}\inst{\ref{aff104}}
	\and A.~Secroun\orcid{0000-0003-0505-3710}\inst{\ref{aff63}}
	\and G.~Seidel\orcid{0000-0003-2907-353X}\inst{\ref{aff77}}
	\and S.~Serrano\orcid{0000-0002-0211-2861}\inst{\ref{aff67},\ref{aff105},\ref{aff24}}
	\and C.~Sirignano\orcid{0000-0002-0995-7146}\inst{\ref{aff102},\ref{aff62}}
	\and G.~Sirri\orcid{0000-0003-2626-2853}\inst{\ref{aff23}}
	\and A.~Spurio~Mancini\orcid{0000-0001-5698-0990}\inst{\ref{aff106},\ref{aff72}}
	\and L.~Stanco\orcid{0000-0002-9706-5104}\inst{\ref{aff62}}
	\and J.~Steinwagner\orcid{0000-0001-7443-1047}\inst{\ref{aff35}}
	\and P.~Tallada-Cresp\'{i}\orcid{0000-0002-1336-8328}\inst{\ref{aff47},\ref{aff48}}
	\and A.~N.~Taylor\inst{\ref{aff3}}
	\and I.~Tereno\inst{\ref{aff6},\ref{aff107}}
	\and N.~Tessore\orcid{0000-0002-9696-7931}\inst{\ref{aff79}}
	\and S.~Toft\orcid{0000-0003-3631-7176}\inst{\ref{aff108},\ref{aff109}}
	\and R.~Toledo-Moreo\orcid{0000-0002-2997-4859}\inst{\ref{aff110}}
	\and F.~Torradeflot\orcid{0000-0003-1160-1517}\inst{\ref{aff48},\ref{aff47}}
	\and I.~Tutusaus\orcid{0000-0002-3199-0399}\inst{\ref{aff17}}
	\and L.~Valenziano\orcid{0000-0002-1170-0104}\inst{\ref{aff9},\ref{aff66}}
	\and J.~Valiviita\orcid{0000-0001-6225-3693}\inst{\ref{aff82},\ref{aff83}}
	\and T.~Vassallo\orcid{0000-0001-6512-6358}\inst{\ref{aff70},\ref{aff31}}
	\and G.~Verdoes~Kleijn\orcid{0000-0001-5803-2580}\inst{\ref{aff111}}
	\and A.~Veropalumbo\orcid{0000-0003-2387-1194}\inst{\ref{aff29},\ref{aff37},\ref{aff112}}
	\and Y.~Wang\orcid{0000-0002-4749-2984}\inst{\ref{aff113}}
	\and J.~Weller\orcid{0000-0002-8282-2010}\inst{\ref{aff70},\ref{aff35}}
	\and G.~Zamorani\orcid{0000-0002-2318-301X}\inst{\ref{aff9}}
	\and E.~Zucca\orcid{0000-0002-5845-8132}\inst{\ref{aff9}}
	\and E.~Bozzo\orcid{0000-0002-8201-1525}\inst{\ref{aff60}}
	\and C.~Burigana\orcid{0000-0002-3005-5796}\inst{\ref{aff114},\ref{aff66}}
	\and M.~Calabrese\orcid{0000-0002-2637-2422}\inst{\ref{aff115},\ref{aff46}}
	\and D.~Di~Ferdinando\inst{\ref{aff23}}
	\and J.~A.~Escartin~Vigo\inst{\ref{aff35}}
	\and G.~Fabbian\orcid{0000-0002-3255-4695}\inst{\ref{aff116},\ref{aff117}}
	\and S.~Matthew\orcid{0000-0001-8448-1697}\inst{\ref{aff3}}
	\and N.~Mauri\orcid{0000-0001-8196-1548}\inst{\ref{aff51},\ref{aff23}}
	\and A.~Pezzotta\orcid{0000-0003-0726-2268}\inst{\ref{aff35}}
	\and M.~P\"ontinen\orcid{0000-0001-5442-2530}\inst{\ref{aff82}}
	\and V.~Scottez\inst{\ref{aff92},\ref{aff118}}
	\and M.~Tenti\orcid{0000-0002-4254-5901}\inst{\ref{aff23}}
	\and M.~Viel\orcid{0000-0002-2642-5707}\inst{\ref{aff30},\ref{aff31},\ref{aff33},\ref{aff32},\ref{aff119}}
	\and M.~Wiesmann\orcid{0009-0000-8199-5860}\inst{\ref{aff19}}
	\and Y.~Akrami\orcid{0000-0002-2407-7956}\inst{\ref{aff120},\ref{aff121}}
	\and S.~Anselmi\orcid{0000-0002-3579-9583}\inst{\ref{aff62},\ref{aff102},\ref{aff4}}
	\and M.~Archidiacono\orcid{0000-0003-4952-9012}\inst{\ref{aff71},\ref{aff86}}
	\and F.~Atrio-Barandela\orcid{0000-0002-2130-2513}\inst{\ref{aff122}}
	\and M.~Ballardini\orcid{0000-0003-4481-3559}\inst{\ref{aff123},\ref{aff9},\ref{aff124}}
	\and D.~Bertacca\orcid{0000-0002-2490-7139}\inst{\ref{aff102},\ref{aff69},\ref{aff62}}
	\and A.~Blanchard\orcid{0000-0001-8555-9003}\inst{\ref{aff17}}
	\and L.~Blot\orcid{0000-0002-9622-7167}\inst{\ref{aff125},\ref{aff4}}
	\and H.~B\"ohringer\orcid{0000-0001-8241-4204}\inst{\ref{aff35},\ref{aff126},\ref{aff127}}
	\and S.~Bruton\orcid{0000-0002-6503-5218}\inst{\ref{aff128}}
	\and R.~Cabanac\orcid{0000-0001-6679-2600}\inst{\ref{aff17}}
	\and A.~Calabro\orcid{0000-0003-2536-1614}\inst{\ref{aff49}}
	\and B.~Camacho~Quevedo\orcid{0000-0002-8789-4232}\inst{\ref{aff67},\ref{aff24}}
	\and A.~Cappi\inst{\ref{aff9},\ref{aff129}}
	\and F.~Caro\inst{\ref{aff49}}
	\and C.~S.~Carvalho\inst{\ref{aff107}}
	\and T.~Castro\orcid{0000-0002-6292-3228}\inst{\ref{aff31},\ref{aff32},\ref{aff30},\ref{aff119}}
	\and S.~Contarini\orcid{0000-0002-9843-723X}\inst{\ref{aff35}}
	\and A.~R.~Cooray\orcid{0000-0002-3892-0190}\inst{\ref{aff130}}
	\and G.~Desprez\orcid{0000-0001-8325-1742}\inst{\ref{aff131}}
	\and A.~D\'iaz-S\'anchez\orcid{0000-0003-0748-4768}\inst{\ref{aff132}}
	\and J.~J.~Diaz\inst{\ref{aff133}}
	\and S.~Di~Domizio\orcid{0000-0003-2863-5895}\inst{\ref{aff36},\ref{aff37}}
	\and M.~Ezziati\orcid{0009-0003-6065-1585}\inst{\ref{aff43}}
	\and A.~G.~Ferrari\orcid{0009-0005-5266-4110}\inst{\ref{aff51},\ref{aff23}}
	\and P.~G.~Ferreira\orcid{0000-0002-3021-2851}\inst{\ref{aff134}}
	\and I.~Ferrero\orcid{0000-0002-1295-1132}\inst{\ref{aff19}}
	\and A.~Finoguenov\orcid{0000-0002-4606-5403}\inst{\ref{aff82}}
	\and A.~Fontana\orcid{0000-0003-3820-2823}\inst{\ref{aff49}}
	\and F.~Fornari\orcid{0000-0003-2979-6738}\inst{\ref{aff66}}
	\and L.~Gabarra\orcid{0000-0002-8486-8856}\inst{\ref{aff134}}
	\and K.~Ganga\orcid{0000-0001-8159-8208}\inst{\ref{aff90}}
	\and J.~Garc\'ia-Bellido\orcid{0000-0002-9370-8360}\inst{\ref{aff120}}
	\and T.~Gasparetto\orcid{0000-0002-7913-4866}\inst{\ref{aff31}}
	\and V.~Gautard\inst{\ref{aff135}}
	\and E.~Gaztanaga\orcid{0000-0001-9632-0815}\inst{\ref{aff24},\ref{aff67},\ref{aff1}}
	\and F.~Giacomini\orcid{0000-0002-3129-2814}\inst{\ref{aff23}}
	\and F.~Gianotti\orcid{0000-0003-4666-119X}\inst{\ref{aff9}}
	\and G.~Gozaliasl\orcid{0000-0002-0236-919X}\inst{\ref{aff136},\ref{aff82}}
	\and C.~M.~Gutierrez\orcid{0000-0001-7854-783X}\inst{\ref{aff137}}
	\and A.~Hall\orcid{0000-0002-3139-8651}\inst{\ref{aff3}}
	\and H.~Hildebrandt\orcid{0000-0002-9814-3338}\inst{\ref{aff7}}
	\and J.~Hjorth\orcid{0000-0002-4571-2306}\inst{\ref{aff95}}
	\and A.~Jimenez~Mu\~noz\orcid{0009-0004-5252-185X}\inst{\ref{aff138}}
	\and S.~Joudaki\orcid{0000-0001-8820-673X}\inst{\ref{aff1}}
	\and J.~J.~E.~Kajava\orcid{0000-0002-3010-8333}\inst{\ref{aff139},\ref{aff140}}
	\and V.~Kansal\orcid{0000-0002-4008-6078}\inst{\ref{aff141},\ref{aff142}}
	\and D.~Karagiannis\orcid{0000-0002-4927-0816}\inst{\ref{aff143},\ref{aff144}}
	\and C.~C.~Kirkpatrick\inst{\ref{aff80}}
	\and J.~Le~Graet\orcid{0000-0001-6523-7971}\inst{\ref{aff63}}
	\and L.~Legrand\orcid{0000-0003-0610-5252}\inst{\ref{aff145}}
	\and J.~Lesgourgues\orcid{0000-0001-7627-353X}\inst{\ref{aff2}}
	\and T.~I.~Liaudat\orcid{0000-0002-9104-314X}\inst{\ref{aff146}}
	\and S.~J.~Liu\orcid{0000-0001-7680-2139}\inst{\ref{aff64}}
	\and A.~Loureiro\orcid{0000-0002-4371-0876}\inst{\ref{aff147},\ref{aff148}}
	\and G.~Maggio\orcid{0000-0003-4020-4836}\inst{\ref{aff31}}
	\and M.~Magliocchetti\orcid{0000-0001-9158-4838}\inst{\ref{aff64}}
	\and F.~Mannucci\orcid{0000-0002-4803-2381}\inst{\ref{aff149}}
	\and R.~Maoli\orcid{0000-0002-6065-3025}\inst{\ref{aff150},\ref{aff49}}
	\and J.~Mart\'{i}n-Fleitas\orcid{0000-0002-8594-569X}\inst{\ref{aff93}}
	\and C.~J.~A.~P.~Martins\orcid{0000-0002-4886-9261}\inst{\ref{aff151},\ref{aff41}}
	\and L.~Maurin\orcid{0000-0002-8406-0857}\inst{\ref{aff61}}
	\and R.~B.~Metcalf\orcid{0000-0003-3167-2574}\inst{\ref{aff88},\ref{aff9}}
	\and M.~Miluzio\inst{\ref{aff54},\ref{aff152}}
	\and P.~Monaco\orcid{0000-0003-2083-7564}\inst{\ref{aff153},\ref{aff31},\ref{aff32},\ref{aff30}}
	\and A.~Montoro\orcid{0000-0003-4730-8590}\inst{\ref{aff24},\ref{aff67}}
	\and C.~Moretti\orcid{0000-0003-3314-8936}\inst{\ref{aff33},\ref{aff119},\ref{aff31},\ref{aff30},\ref{aff32}}
	\and G.~Morgante\inst{\ref{aff9}}
	\and C.~Murray\inst{\ref{aff90}}
	\and S.~Nadathur\orcid{0000-0001-9070-3102}\inst{\ref{aff1}}
	\and L.~Pagano\orcid{0000-0003-1820-5998}\inst{\ref{aff123},\ref{aff124}}
	\and L.~Patrizii\inst{\ref{aff23}}
	\and V.~Popa\orcid{0000-0002-9118-8330}\inst{\ref{aff101}}
	\and D.~Potter\orcid{0000-0002-0757-5195}\inst{\ref{aff21}}
	\and P.~Reimberg\orcid{0000-0003-3410-0280}\inst{\ref{aff92}}
	\and I.~Risso\orcid{0000-0003-2525-7761}\inst{\ref{aff112}}
	\and P.-F.~Rocci\inst{\ref{aff61}}
	\and M.~Sahl\'en\orcid{0000-0003-0973-4804}\inst{\ref{aff154}}
	\and E.~Sarpa\orcid{0000-0002-1256-655X}\inst{\ref{aff33},\ref{aff119},\ref{aff32}}
	\and A.~Schneider\orcid{0000-0001-7055-8104}\inst{\ref{aff21}}
	\and M.~Sereno\orcid{0000-0003-0302-0325}\inst{\ref{aff9},\ref{aff23}}
	\and A.~Silvestri\orcid{0000-0001-6904-5061}\inst{\ref{aff45}}
	\and J.~Stadel\orcid{0000-0001-7565-8622}\inst{\ref{aff21}}
	\and K.~Tanidis\inst{\ref{aff134}}
	\and C.~Tao\orcid{0000-0001-7961-8177}\inst{\ref{aff63}}
	\and G.~Testera\inst{\ref{aff37}}
	\and R.~Teyssier\orcid{0000-0001-7689-0933}\inst{\ref{aff155}}
	\and S.~Tosi\orcid{0000-0002-7275-9193}\inst{\ref{aff36},\ref{aff37}}
	\and A.~Troja\orcid{0000-0003-0239-4595}\inst{\ref{aff102},\ref{aff62}}
	\and M.~Tucci\inst{\ref{aff60}}
	\and D.~Vergani\orcid{0000-0003-0898-2216}\inst{\ref{aff9}}
	\and G.~Verza\orcid{0000-0002-1886-8348}\inst{\ref{aff156},\ref{aff157}}
	\and P.~Vielzeuf\orcid{0000-0003-2035-9339}\inst{\ref{aff63}}
	\and N.~A.~Walton\orcid{0000-0003-3983-8778}\inst{\ref{aff116}}}

\institute{Institute of Cosmology and Gravitation, University of Portsmouth, Portsmouth PO1 3FX, UK\label{aff1}
	\and
	Institute for Theoretical Particle Physics and Cosmology (TTK), RWTH Aachen University, 52056 Aachen, Germany\label{aff2}
	\and
	Institute for Astronomy, University of Edinburgh, Royal Observatory, Blackford Hill, Edinburgh EH9 3HJ, UK\label{aff3}
	\and
	Laboratoire Univers et Th\'eorie, Observatoire de Paris, Universit\'e PSL, Universit\'e Paris Cit\'e, CNRS, 92190 Meudon, France\label{aff4}
	\and
	Instituto de Astrof\'isica e Ci\^encias do Espa\c{c}o, Faculdade de Ci\^encias, Universidade de Lisboa, Campo Grande, 1749-016 Lisboa, Portugal\label{aff5}
	\and
	Departamento de F\'isica, Faculdade de Ci\^encias, Universidade de Lisboa, Edif\'icio C8, Campo Grande, PT1749-016 Lisboa, Portugal\label{aff6}
	\and
	Ruhr University Bochum, Faculty of Physics and Astronomy, Astronomical Institute (AIRUB), German Centre for Cosmological Lensing (GCCL), 44780 Bochum, Germany\label{aff7}
	\and
	Universit\"at Bonn, Argelander-Institut f\"ur Astronomie, Auf dem H\"ugel 71, 53121 Bonn, Germany\label{aff8}
	\and
	INAF-Osservatorio di Astrofisica e Scienza dello Spazio di Bologna, Via Piero Gobetti 93/3, 40129 Bologna, Italy\label{aff9}
	\and
	Istituto Nazionale di Fisica Nucleare, Sezione di Bologna, Via Irnerio 46, 40126 Bologna, Italy\label{aff10}
	\and
	Dipartimento di Fisica, Universit\`a degli Studi di Torino, Via P. Giuria 1, 10125 Torino, Italy\label{aff11}
	\and
	INFN-Sezione di Torino, Via P. Giuria 1, 10125 Torino, Italy\label{aff12}
	\and
	INAF-Osservatorio Astrofisico di Torino, Via Osservatorio 20, 10025 Pino Torinese (TO), Italy\label{aff13}
	\and
	Higgs Centre for Theoretical Physics, School of Physics and Astronomy, The University of Edinburgh, Edinburgh EH9 3FD, UK\label{aff14}
	\and
	Institut universitaire de France (IUF), 1 rue Descartes, 75231 PARIS CEDEX 05, France\label{aff15}
	\and
	Institut f\"ur Theoretische Physik, University of Heidelberg, Philosophenweg 16, 69120 Heidelberg, Germany\label{aff16}
	\and
	Institut de Recherche en Astrophysique et Plan\'etologie (IRAP), Universit\'e de Toulouse, CNRS, UPS, CNES, 14 Av. Edouard Belin, 31400 Toulouse, France\label{aff17}
	\and
	Universit\'e St Joseph; Faculty of Sciences, Beirut, Lebanon\label{aff18}
	\and
	Institute of Theoretical Astrophysics, University of Oslo, P.O. Box 1029 Blindern, 0315 Oslo, Norway\label{aff19}
	\and
	Jodrell Bank Centre for Astrophysics, Department of Physics and Astronomy, University of Manchester, Oxford Road, Manchester M13 9PL, UK\label{aff20}
	\and
	Department of Astrophysics, University of Zurich, Winterthurerstrasse 190, 8057 Zurich, Switzerland\label{aff21}
	\and
	Dipartimento di Fisica e Astronomia, Universit\`a di Bologna, Via Gobetti 93/2, 40129 Bologna, Italy\label{aff22}
	\and
	INFN-Sezione di Bologna, Viale Berti Pichat 6/2, 40127 Bologna, Italy\label{aff23}
	\and
	Institute of Space Sciences (ICE, CSIC), Campus UAB, Carrer de Can Magrans, s/n, 08193 Barcelona, Spain\label{aff24}
	\and
	Institut de Ciencies de l'Espai (IEEC-CSIC), Campus UAB, Carrer de Can Magrans, s/n Cerdanyola del Vall\'es, 08193 Barcelona, Spain\label{aff25}
	\and
	Jet Propulsion Laboratory, California Institute of Technology, 4800 Oak Grove Drive, Pasadena, CA, 91109, USA\label{aff26}
	\and
	Institut de Physique Th\'eorique, CEA, CNRS, Universit\'e Paris-Saclay 91191 Gif-sur-Yvette Cedex, France\label{aff27}
	\and
	School of Mathematics and Physics, University of Surrey, Guildford, Surrey, GU2 7XH, UK\label{aff28}
	\and
	INAF-Osservatorio Astronomico di Brera, Via Brera 28, 20122 Milano, Italy\label{aff29}
	\and
	IFPU, Institute for Fundamental Physics of the Universe, via Beirut 2, 34151 Trieste, Italy\label{aff30}
	\and
	INAF-Osservatorio Astronomico di Trieste, Via G. B. Tiepolo 11, 34143 Trieste, Italy\label{aff31}
	\and
	INFN, Sezione di Trieste, Via Valerio 2, 34127 Trieste TS, Italy\label{aff32}
	\and
	SISSA, International School for Advanced Studies, Via Bonomea 265, 34136 Trieste TS, Italy\label{aff33}
	\and
	Institut d'Astrophysique de Paris, UMR 7095, CNRS, and Sorbonne Universit\'e, 98 bis boulevard Arago, 75014 Paris, France\label{aff34}
	\and
	Max Planck Institute for Extraterrestrial Physics, Giessenbachstr. 1, 85748 Garching, Germany\label{aff35}
	\and
	Dipartimento di Fisica, Universit\`a di Genova, Via Dodecaneso 33, 16146, Genova, Italy\label{aff36}
	\and
	INFN-Sezione di Genova, Via Dodecaneso 33, 16146, Genova, Italy\label{aff37}
	\and
	Department of Physics "E. Pancini", University Federico II, Via Cinthia 6, 80126, Napoli, Italy\label{aff38}
	\and
	INAF-Osservatorio Astronomico di Capodimonte, Via Moiariello 16, 80131 Napoli, Italy\label{aff39}
	\and
	INFN section of Naples, Via Cinthia 6, 80126, Napoli, Italy\label{aff40}
	\and
	Instituto de Astrof\'isica e Ci\^encias do Espa\c{c}o, Universidade do Porto, CAUP, Rua das Estrelas, PT4150-762 Porto, Portugal\label{aff41}
	\and
	Faculdade de Ci\^encias da Universidade do Porto, Rua do Campo de Alegre, 4150-007 Porto, Portugal\label{aff42}
	\and
	Aix-Marseille Universit\'e, CNRS, CNES, LAM, Marseille, France\label{aff43}
	\and
	European Space Agency/ESTEC, Keplerlaan 1, 2201 AZ Noordwijk, The Netherlands\label{aff44}
	\and
	Institute Lorentz, Leiden University, Niels Bohrweg 2, 2333 CA Leiden, The Netherlands\label{aff45}
	\and
	INAF-IASF Milano, Via Alfonso Corti 12, 20133 Milano, Italy\label{aff46}
	\and
	Centro de Investigaciones Energ\'eticas, Medioambientales y Tecnol\'ogicas (CIEMAT), Avenida Complutense 40, 28040 Madrid, Spain\label{aff47}
	\and
	Port d'Informaci\'{o} Cient\'{i}fica, Campus UAB, C. Albareda s/n, 08193 Bellaterra (Barcelona), Spain\label{aff48}
	\and
	INAF-Osservatorio Astronomico di Roma, Via Frascati 33, 00078 Monteporzio Catone, Italy\label{aff49}
	\and
	Institute for Astronomy, University of Hawaii, 2680 Woodlawn Drive, Honolulu, HI 96822, USA\label{aff50}
	\and
	Dipartimento di Fisica e Astronomia "Augusto Righi" - Alma Mater Studiorum Universit\`a di Bologna, Viale Berti Pichat 6/2, 40127 Bologna, Italy\label{aff51}
	\and
	Instituto de Astrof\'{\i}sica de Canarias, V\'{\i}a L\'actea, 38205 La Laguna, Tenerife, Spain\label{aff52}
	\and
	European Space Agency/ESRIN, Largo Galileo Galilei 1, 00044 Frascati, Roma, Italy\label{aff53}
	\and
	ESAC/ESA, Camino Bajo del Castillo, s/n., Urb. Villafranca del Castillo, 28692 Villanueva de la Ca\~nada, Madrid, Spain\label{aff54}
	\and
	Universit\'e Claude Bernard Lyon 1, CNRS/IN2P3, IP2I Lyon, UMR 5822, Villeurbanne, F-69100, France\label{aff55}
	\and
	Institute of Physics, Laboratory of Astrophysics, Ecole Polytechnique F\'ed\'erale de Lausanne (EPFL), Observatoire de Sauverny, 1290 Versoix, Switzerland\label{aff56}
	\and
	Institut de Ci\`{e}ncies del Cosmos (ICCUB), Universitat de Barcelona (IEEC-UB), Mart\'{i} i Franqu\`{e}s 1, 08028 Barcelona, Spain\label{aff57}
	\and
	Instituci\'o Catalana de Recerca i Estudis Avan\c{c}ats (ICREA), Passeig de Llu\'{\i}s Companys 23, 08010 Barcelona, Spain\label{aff58}
	\and
	UCB Lyon 1, CNRS/IN2P3, IUF, IP2I Lyon, 4 rue Enrico Fermi, 69622 Villeurbanne, France\label{aff59}
	\and
	Department of Astronomy, University of Geneva, ch. d'Ecogia 16, 1290 Versoix, Switzerland\label{aff60}
	\and
	Universit\'e Paris-Saclay, CNRS, Institut d'astrophysique spatiale, 91405, Orsay, France\label{aff61}
	\and
	INFN-Padova, Via Marzolo 8, 35131 Padova, Italy\label{aff62}
	\and
	Aix-Marseille Universit\'e, CNRS/IN2P3, CPPM, Marseille, France\label{aff63}
	\and
	INAF-Istituto di Astrofisica e Planetologia Spaziali, via del Fosso del Cavaliere, 100, 00100 Roma, Italy\label{aff64}
	\and
	Universit\'e Paris-Saclay, Universit\'e Paris Cit\'e, CEA, CNRS, AIM, 91191, Gif-sur-Yvette, France\label{aff65}
	\and
	INFN-Bologna, Via Irnerio 46, 40126 Bologna, Italy\label{aff66}
	\and
	Institut d'Estudis Espacials de Catalunya (IEEC),  Edifici RDIT, Campus UPC, 08860 Castelldefels, Barcelona, Spain\label{aff67}
	\and
	FRACTAL S.L.N.E., calle Tulip\'an 2, Portal 13 1A, 28231, Las Rozas de Madrid, Spain\label{aff68}
	\and
	INAF-Osservatorio Astronomico di Padova, Via dell'Osservatorio 5, 35122 Padova, Italy\label{aff69}
	\and
	Universit\"ats-Sternwarte M\"unchen, Fakult\"at f\"ur Physik, Ludwig-Maximilians-Universit\"at M\"unchen, Scheinerstrasse 1, 81679 M\"unchen, Germany\label{aff70}
	\and
	Dipartimento di Fisica "Aldo Pontremoli", Universit\`a degli Studi di Milano, Via Celoria 16, 20133 Milano, Italy\label{aff71}
	\and
	Mullard Space Science Laboratory, University College London, Holmbury St Mary, Dorking, Surrey RH5 6NT, UK\label{aff72}
	\and
	Felix Hormuth Engineering, Goethestr. 17, 69181 Leimen, Germany\label{aff73}
	\and
	Technical University of Denmark, Elektrovej 327, 2800 Kgs. Lyngby, Denmark\label{aff74}
	\and
	Cosmic Dawn Center (DAWN), Denmark\label{aff75}
	\and
	Universit\'e Paris-Saclay, CNRS/IN2P3, IJCLab, 91405 Orsay, France\label{aff76}
	\and
	Max-Planck-Institut f\"ur Astronomie, K\"onigstuhl 17, 69117 Heidelberg, Germany\label{aff77}
	\and
	NASA Goddard Space Flight Center, Greenbelt, MD 20771, USA\label{aff78}
	\and
	Department of Physics and Astronomy, University College London, Gower Street, London WC1E 6BT, UK\label{aff79}
	\and
	Department of Physics and Helsinki Institute of Physics, Gustaf H\"allstr\"omin katu 2, 00014 University of Helsinki, Finland\label{aff80}
	\and
	Universit\'e de Gen\`eve, D\'epartement de Physique Th\'eorique and Centre for Astroparticle Physics, 24 quai Ernest-Ansermet, CH-1211 Gen\`eve 4, Switzerland\label{aff81}
	\and
	Department of Physics, P.O. Box 64, 00014 University of Helsinki, Finland\label{aff82}
	\and
	Helsinki Institute of Physics, Gustaf H{\"a}llstr{\"o}min katu 2, University of Helsinki, Helsinki, Finland\label{aff83}
	\and
	NOVA optical infrared instrumentation group at ASTRON, Oude Hoogeveensedijk 4, 7991PD, Dwingeloo, The Netherlands\label{aff84}
	\and
	Centre de Calcul de l'IN2P3/CNRS, 21 avenue Pierre de Coubertin 69627 Villeurbanne Cedex, France\label{aff85}
	\and
	INFN-Sezione di Milano, Via Celoria 16, 20133 Milano, Italy\label{aff86}
	\and
	INFN-Sezione di Roma, Piazzale Aldo Moro, 2 - c/o Dipartimento di Fisica, Edificio G. Marconi, 00185 Roma, Italy\label{aff87}
	\and
	Dipartimento di Fisica e Astronomia "Augusto Righi" - Alma Mater Studiorum Universit\`a di Bologna, via Piero Gobetti 93/2, 40129 Bologna, Italy\label{aff88}
	\and
	Department of Physics, Institute for Computational Cosmology, Durham University, South Road, Durham, DH1 3LE, UK\label{aff89}
	\and
	Universit\'e Paris Cit\'e, CNRS, Astroparticule et Cosmologie, 75013 Paris, France\label{aff90}
	\and
	University of Applied Sciences and Arts of Northwestern Switzerland, School of Engineering, 5210 Windisch, Switzerland\label{aff91}
	\and
	Institut d'Astrophysique de Paris, 98bis Boulevard Arago, 75014, Paris, France\label{aff92}
	\and
	Aurora Technology for European Space Agency (ESA), Camino bajo del Castillo, s/n, Urbanizacion Villafranca del Castillo, Villanueva de la Ca\~nada, 28692 Madrid, Spain\label{aff93}
	\and
	Institut de F\'{i}sica d'Altes Energies (IFAE), The Barcelona Institute of Science and Technology, Campus UAB, 08193 Bellaterra (Barcelona), Spain\label{aff94}
	\and
	DARK, Niels Bohr Institute, University of Copenhagen, Jagtvej 155, 2200 Copenhagen, Denmark\label{aff95}
	\and
	Waterloo Centre for Astrophysics, University of Waterloo, Waterloo, Ontario N2L 3G1, Canada\label{aff96}
	\and
	Department of Physics and Astronomy, University of Waterloo, Waterloo, Ontario N2L 3G1, Canada\label{aff97}
	\and
	Perimeter Institute for Theoretical Physics, Waterloo, Ontario N2L 2Y5, Canada\label{aff98}
	\and
	Space Science Data Center, Italian Space Agency, via del Politecnico snc, 00133 Roma, Italy\label{aff99}
	\and
	Centre National d'Etudes Spatiales -- Centre spatial de Toulouse, 18 avenue Edouard Belin, 31401 Toulouse Cedex 9, France\label{aff100}
	\and
	Institute of Space Science, Str. Atomistilor, nr. 409 M\u{a}gurele, Ilfov, 077125, Romania\label{aff101}
	\and
	Dipartimento di Fisica e Astronomia "G. Galilei", Universit\`a di Padova, Via Marzolo 8, 35131 Padova, Italy\label{aff102}
	\and
	Departamento de F\'isica, FCFM, Universidad de Chile, Blanco Encalada 2008, Santiago, Chile\label{aff103}
	\and
	Universit\"at Innsbruck, Institut f\"ur Astro- und Teilchenphysik, Technikerstr. 25/8, 6020 Innsbruck, Austria\label{aff104}
	\and
	Satlantis, University Science Park, Sede Bld 48940, Leioa-Bilbao, Spain\label{aff105}
	\and
	Department of Physics, Royal Holloway, University of London, TW20 0EX, UK\label{aff106}
	\and
	Instituto de Astrof\'isica e Ci\^encias do Espa\c{c}o, Faculdade de Ci\^encias, Universidade de Lisboa, Tapada da Ajuda, 1349-018 Lisboa, Portugal\label{aff107}
	\and
	Cosmic Dawn Center (DAWN)\label{aff108}
	\and
	Niels Bohr Institute, University of Copenhagen, Jagtvej 128, 2200 Copenhagen, Denmark\label{aff109}
	\and
	Universidad Polit\'ecnica de Cartagena, Departamento de Electr\'onica y Tecnolog\'ia de Computadoras,  Plaza del Hospital 1, 30202 Cartagena, Spain\label{aff110}
	\and
	Kapteyn Astronomical Institute, University of Groningen, PO Box 800, 9700 AV Groningen, The Netherlands\label{aff111}
	\and
	Dipartimento di Fisica, Universit\`a degli studi di Genova, and INFN-Sezione di Genova, via Dodecaneso 33, 16146, Genova, Italy\label{aff112}
	\and
	Infrared Processing and Analysis Center, California Institute of Technology, Pasadena, CA 91125, USA\label{aff113}
	\and
	INAF, Istituto di Radioastronomia, Via Piero Gobetti 101, 40129 Bologna, Italy\label{aff114}
	\and
	Astronomical Observatory of the Autonomous Region of the Aosta Valley (OAVdA), Loc. Lignan 39, I-11020, Nus (Aosta Valley), Italy\label{aff115}
	\and
	Institute of Astronomy, University of Cambridge, Madingley Road, Cambridge CB3 0HA, UK\label{aff116}
	\and
	School of Physics and Astronomy, Cardiff University, The Parade, Cardiff, CF24 3AA, UK\label{aff117}
	\and
	ICL, Junia, Universit\'e Catholique de Lille, LITL, 59000 Lille, France\label{aff118}
	\and
	ICSC - Centro Nazionale di Ricerca in High Performance Computing, Big Data e Quantum Computing, Via Magnanelli 2, Bologna, Italy\label{aff119}
	\and
	Instituto de F\'isica Te\'orica UAM-CSIC, Campus de Cantoblanco, 28049 Madrid, Spain\label{aff120}
	\and
	CERCA/ISO, Department of Physics, Case Western Reserve University, 10900 Euclid Avenue, Cleveland, OH 44106, USA\label{aff121}
	\and
	Departamento de F{\'\i}sica Fundamental. Universidad de Salamanca. Plaza de la Merced s/n. 37008 Salamanca, Spain\label{aff122}
	\and
	Dipartimento di Fisica e Scienze della Terra, Universit\`a degli Studi di Ferrara, Via Giuseppe Saragat 1, 44122 Ferrara, Italy\label{aff123}
	\and
	Istituto Nazionale di Fisica Nucleare, Sezione di Ferrara, Via Giuseppe Saragat 1, 44122 Ferrara, Italy\label{aff124}
	\and
	Center for Data-Driven Discovery, Kavli IPMU (WPI), UTIAS, The University of Tokyo, Kashiwa, Chiba 277-8583, Japan\label{aff125}
	\and
	Ludwig-Maximilians-University, Schellingstrasse 4, 80799 Munich, Germany\label{aff126}
	\and
	Max-Planck-Institut f\"ur Physik, Boltzmannstr. 8, 85748 Garching, Germany\label{aff127}
	\and
	Minnesota Institute for Astrophysics, University of Minnesota, 116 Church St SE, Minneapolis, MN 55455, USA\label{aff128}
	\and
	Universit\'e C\^{o}te d'Azur, Observatoire de la C\^{o}te d'Azur, CNRS, Laboratoire Lagrange, Bd de l'Observatoire, CS 34229, 06304 Nice cedex 4, France\label{aff129}
	\and
	Department of Physics \& Astronomy, University of California Irvine, Irvine CA 92697, USA\label{aff130}
	\and
	Department of Astronomy \& Physics and Institute for Computational Astrophysics, Saint Mary's University, 923 Robie Street, Halifax, Nova Scotia, B3H 3C3, Canada\label{aff131}
	\and
	Departamento F\'isica Aplicada, Universidad Polit\'ecnica de Cartagena, Campus Muralla del Mar, 30202 Cartagena, Murcia, Spain\label{aff132}
	\and
	Instituto de Astrof\'isica de Canarias (IAC); Departamento de Astrof\'isica, Universidad de La Laguna (ULL), 38200, La Laguna, Tenerife, Spain\label{aff133}
	\and
	Department of Physics, Oxford University, Keble Road, Oxford OX1 3RH, UK\label{aff134}
	\and
	CEA Saclay, DFR/IRFU, Service d'Astrophysique, Bat. 709, 91191 Gif-sur-Yvette, France\label{aff135}
	\and
	Department of Computer Science, Aalto University, PO Box 15400, Espoo, FI-00 076, Finland\label{aff136}
	\and
	Instituto de Astrof\'\i sica de Canarias, c/ Via Lactea s/n, La Laguna 38200, Spain. Departamento de Astrof\'\i sica de la Universidad de La Laguna, Avda. Francisco Sanchez, La Laguna, 38200, Spain\label{aff137}
	\and
	Univ. Grenoble Alpes, CNRS, Grenoble INP, LPSC-IN2P3, 53, Avenue des Martyrs, 38000, Grenoble, France\label{aff138}
	\and
	Department of Physics and Astronomy, Vesilinnantie 5, 20014 University of Turku, Finland\label{aff139}
	\and
	Serco for European Space Agency (ESA), Camino bajo del Castillo, s/n, Urbanizacion Villafranca del Castillo, Villanueva de la Ca\~nada, 28692 Madrid, Spain\label{aff140}
	\and
	ARC Centre of Excellence for Dark Matter Particle Physics, Melbourne, Australia\label{aff141}
	\and
	Centre for Astrophysics \& Supercomputing, Swinburne University of Technology,  Hawthorn, Victoria 3122, Australia\label{aff142}
	\and
	School of Physics and Astronomy, Queen Mary University of London, Mile End Road, London E1 4NS, UK\label{aff143}
	\and
	Department of Physics and Astronomy, University of the Western Cape, Bellville, Cape Town, 7535, South Africa\label{aff144}
	\and
	ICTP South American Institute for Fundamental Research, Instituto de F\'{\i}sica Te\'orica, Universidade Estadual Paulista, S\~ao Paulo, Brazil\label{aff145}
	\and
	IRFU, CEA, Universit\'e Paris-Saclay 91191 Gif-sur-Yvette Cedex, France\label{aff146}
	\and
	Oskar Klein Centre for Cosmoparticle Physics, Department of Physics, Stockholm University, Stockholm, SE-106 91, Sweden\label{aff147}
	\and
	Astrophysics Group, Blackett Laboratory, Imperial College London, London SW7 2AZ, UK\label{aff148}
	\and
	INAF-Osservatorio Astrofisico di Arcetri, Largo E. Fermi 5, 50125, Firenze, Italy\label{aff149}
	\and
	Dipartimento di Fisica, Sapienza Universit\`a di Roma, Piazzale Aldo Moro 2, 00185 Roma, Italy\label{aff150}
	\and
	Centro de Astrof\'{\i}sica da Universidade do Porto, Rua das Estrelas, 4150-762 Porto, Portugal\label{aff151}
	\and
	HE Space for European Space Agency (ESA), Camino bajo del Castillo, s/n, Urbanizacion Villafranca del Castillo, Villanueva de la Ca\~nada, 28692 Madrid, Spain\label{aff152}
	\and
	Dipartimento di Fisica - Sezione di Astronomia, Universit\`a di Trieste, Via Tiepolo 11, 34131 Trieste, Italy\label{aff153}
	\and
	Theoretical astrophysics, Department of Physics and Astronomy, Uppsala University, Box 515, 751 20 Uppsala, Sweden\label{aff154}
	\and
	Department of Astrophysical Sciences, Peyton Hall, Princeton University, Princeton, NJ 08544, USA\label{aff155}
	\and
	Center for Cosmology and Particle Physics, Department of Physics, New York University, New York, NY 10003, USA\label{aff156}
	\and
	Center for Computational Astrophysics, Flatiron Institute, 162 5th Avenue, 10010, New York, NY, USA\label{aff157}}    

\date{\today}

\authorrunning{Euclid Collaboration: K. Koyama et al.}

\titlerunning{Constraints on $f(R)$ models from the photometric primary probes}


\abstract{
We study the constraint on $f(R)$ gravity that can be obtained by photometric primary probes of the \Euclid mission. Our focus is the dependence of the constraint on the theoretical modelling of the nonlinear matter power spectrum. In the Hu--Sawicki $f(R)$ gravity model, we consider four different predictions for the ratio between the power spectrum in $f(R)$ and that in $\Lambda$ cold dark matter ($\Lambda$CDM): a fitting formula, the halo model reaction approach, \react{}, and two emulators based on dark matter only {\it N}-body simulations, \forge{} and \emantis{}. These predictions are added to the \montepython{} implementation to predict the angular power spectra for weak lensing (WL), photometric galaxy clustering, and their cross-correlation. By running Markov chain Monte Carlo, we compare constraints on parameters and investigate the bias of the recovered $f(R)$ parameter if the data are created by a different model. For the pessimistic setting of WL, one-dimensional bias for the $f(R)$ parameter, $\logfr$, is found to be $0.5 \sigma$ when \forge{} is used to create the synthetic data with $\logfr =-5.301$ and fitted by \emantis{}. The impact of baryonic physics on WL is studied by using a baryonification emulator, \bcemu{}. For the optimistic setting, the $f(R)$ parameter and two main baryonic parameters are well constrained despite the degeneracies among these parameters. However, the difference in the nonlinear dark matter prediction can be compensated for the adjustment of baryonic parameters, and the one-dimensional marginalised constraint on $\logfr$ is biased. This bias can be avoided in the pessimistic setting at the expense of weaker constraints. For the pessimistic setting, using the \lcdm\ synthetic data for WL, we obtain the prior-independent upper limit of $\logfr < -5.6$. Finally, we implement a method to include theoretical errors to avoid the bias due to inaccuracies in the nonlinear matter power spectrum prediction. }


\maketitle

\section{Introduction}\label{sec:intro}
In 1998, astronomers made the surprising discovery that the expansion of the Universe is accelerating, not slowing down \citep{Riess98,Perlmutter99}.
This late-time acceleration of the Universe has become the most challenging problem in theoretical physics. The main aim of ongoing and future cosmological surveys is to address the key questions about the origin of the late-time acceleration. The acceleration can be driven by a cosmological constant or dark energy that evolves with the expansion of the Universe, Alternatively, there could be no dark energy if general relativity (GR) itself is in error on cosmological scales. There has been significant progress in developing modified theories of gravity and these have been developed into tests of GR itself via cosmological observations  \citep{Koyama:2015vza,Ishak:2018his, Ferreira:2019xrr}. Testing gravity is one of the main objectives of stage-IV dark energy surveys \citep{Albrecht:2006um}.  

Of particular importance in these surveys is the \Euclid mission \citep{EuclidSkyOverview}. The \Euclid satellite undertakes a spectroscopic survey of galaxies and an imaging survey (targeting weak lensing (WL) which can also be used to reconstruct galaxy clustering using photometric redshifts). The combination of these two probes is essential for cosmological tests of gravity \citep[][EC20 hereafter]{Blanchard:2019oqi}.

Modified gravity models typically include an additional scalar degree of freedom that gives rise to a fifth force. To satisfy the stringent constraints on deviations from GR in the Solar System, many modified gravity models utilise screening mechanisms to hide modifications of gravity on small scales \citep{Joyce:2014kja, Brax:2021wcv}. This is accomplished by nonlinearity in the equation that governs the dynamics of the scalar degree of freedom. This significantly complicates the nonlinear modelling of matter clustering in these models as the nonlinear equation for the scalar mode coupled to nonlinear density fields needs to be solved. A systematic comparison of {\it N}-body simulations in $f(R)$ gravity and normal-branch Dvali--Gabadadze--Porratti (nDGP) models was done in \citet{Winther:2015wla}. Since then, new simulations have been developed, including those using approximate methods to accelerate simulations \citep{Valogiannis:2016ane, Winther:2017jof}. A fitting formula \citep{Winther:2019mus} and emulators for the nonlinear matter power spectrum have been developed \citep{Arnold:2019vpg, Ramachandra:2020lue, saez-casares_2023, Fiorini:2023fjl}. At the same time, a semi-analytic method based on the halo model to predict the nonlinear matter power spectrum for general dark energy and modified gravity models has been developed \citep{Cataneo:2018cic, Bose:2021mkz, Bose:2022vwi, Carrilho:2021rqo}. These nonlinear predictions were used to study modifications of the WL observables \citep{Schneider:2019xpf, Harnois-Deraps:2022bie, SpurioMancini:2023mpt, Carrion:2024itc, Tsedrik:2024cdi}, cross-correlation of galaxies, and cosmic microwave background \citep{Kou:2023gyc}, for example in $f(R)$ gravity. 

In \citelist{Euclid:2023tqw}, Fisher Matrix forecasts were performed to predict \Euclid's ability to constrain $f(R)$ gravity models. In the Hu--Sawicki $f(R)$ model, it was shown that in the optimistic setting defined in \citetalias{Blanchard:2019oqi}, and for a fiducial value of $\fr = 5 \times 10^{-6}$, \Euclid alone will be able to constrain the additional parameter $\logfr$ at the $3\%$ level, using spectroscopic galaxy clustering alone; at the $1.4\%$ level, using the combination of photometric probes on their own; and at the $1\%$ level, using the combination of spectroscopic and photometric observations. The forecast for photometric probes used the fitting formula for the nonlinear matter power spectrum obtained in \citelist{Winther:2019mus}. Further Fisher Matrix forecasts have been done for other modified gravity models with  scale-independent linear growth \citep{Euclid:2023rjj}. 

For real data analysis, it is imperative to check the effect of the accuracy of the theoretical modelling. This article is part of a series that collectively explores simulations and nonlinearities beyond the $\Lambda$ cold dark matter (\lcdm) model:
\ben
    \item Numerical methods and validation \citep[][paper 1 hereafter]{paper1}
    \item Results from non-standard simulations \citep[][paper 2 hereafter]{paper2}
    \item {Constraints on $f(R)$ models from the photometric primary probes (this work)}
    \item {Cosmological constraints on non-standard cosmologies from simulated \Euclid probes (D'Amico et al. in prep.)}
\een
In \citetalias{paper1}, the comparison of {\it N}-body simulations performed in \citelist{Winther:2015wla} in the Hu--Sawicki $f(R)$ and nDGP models was extended to add more simulations. The comparison was done for the matter power spectrum and the halo mass function. The measurements of these quantities were done using the dedicated pipeline developed by \citetalias{paper2}. In \citetalias{paper2}, additional simulations have been analysed in addition to those used in \citetalias{paper1}. In this paper, we utilise these developments and compare several predictions for the nonlinear matter power spectrum in $f(R)$ gravity. We perform Markov chain Monte Carlo (MCMC) simulations using synthetic data for \Euclid photometric probes, and assess the impact of using different nonlinear models for the matter power spectrum at the level of parameter constraints. In addition, we add baryonic effects using a baryonification method that was not included in the Fisher Matrix forecast. Although this paper focuses on $f(R)$ gravity, for which multiple public codes are available to predict the nonlinear matter power spectrum, the methodology and code developed in this paper are readily applicable to other modified gravity models. The validation of the nonlinear models for spectroscopic probes has been done in \citelist{Euclid:2023bgs} and D'Amico et al. (paper 4) will perform a similar analysis to this paper's for the spectroscopic probes. 

This paper is organised as follows. In \cref{sec:thpred}, we summarise theoretical predictions for \Euclid observables based on \citetalias{Blanchard:2019oqi} and  \citelist{Euclid:2023tqw}. In \cref{sec:MP}, we introduce the Hu--Sawicki $f(R)$ gravity model and summarise the four nonlinear models for the matter power spectrum. We then discuss the implementation of these models in the \montepython{} code introduced in \citelist{Euclid:2023pxu}. In \cref{sec:comparison}, we compare predictions for the nonlinear matter power spectrum with {\it N}-body simulations and study their impact on the angular power spectra for \Euclid photometric probes. Forecasts for errors from the combination of photometric probes considered in \citetalias{Blanchard:2019oqi} are  presented based on the synthetic data created by four different nonlinear models. We also compare the result with the Fisher Matrix forecast. We then study the bias in the recovered $f(R)$ gravity parameter when the synthetic data are created by a different nonlinear model. \Cref{sec:baryonic_effects} is devoted to the study of baryonic effects using the \bcemu{} baryonification method. We show how the bias is affected by baryonic effects and obtain the upper bound on $|f_{R0}|$ using the \lcdm\ synthetic data. In \cref{sec:theoretical_error}, we implement theoretical errors to take into account the uncertainties of theoretical predictions. We conclude in \cref{sec:conclusion}. 

\section{Theoretical predictions for Euclid observables}
\label{sec:thpred}
In this section, we discuss how moving away from the standard GR assumption impacts the predictions for the angular power spectra $C(\ell)$ that will be compared with the photometric data survey. The observables that need to be computed and compared with the data are the angular power spectra for weak lensing (WL), photometric galaxy clustering (\GCph{}) and their cross-correlation (\XCph{}).
In \citetalias{Blanchard:2019oqi}, these were calculated using the Limber and flat-sky approximations in a flat \lcdm\ Universe, as 
\begin{equation}
 C^{XY}_{ij}(\ell) = c \int_{z_{\rm min}}^{z_{\rm max}}\de z\,{\frac{W_i^X(z)\,W_j^Y(z)}{H(z)\, r^2(z)}P_{\delta\delta}(k_\ell,z)}\,,
 \label{eq:ISTrecipe}
\end{equation}
where $W_i^X(z)$ is the window function in each tomographic redshift bin, $i$, with $X=\{{\rm L,G}\}$ corresponding to WL and \GCph{}, respectively. Here, $k_\ell=(\ell+1/2)/r(z)$, $r(z)$ is the comoving distance to redshift, $z$, and $H(z)$ is  the Hubble function. The nonlinear power spectrum of matter density fluctuations, $P_{\delta\delta}(k_\ell,z)$, is evaluated at wave number $k_{\ell}$ and redshift $z$, in the redshift range of the integral from $z_{\rm min}=0.001$ to $z_{\rm max}=4$. 

However, when abandoning the assumption of GR, one has to account for changes in the evolution of both the homogeneous background and cosmological perturbations. For WL, it is the Weyl potential, $\phi_W$, that determines the angular power spectrum. The power spectrum of the Weyl potential is related to $P_{\delta\delta}$ as \citep{Euclid:2023tqw} 
\begin{equation}\label{eq:sigma}
 P_{\phi_W}(k,z) = \left[-3\,\Omegam \,\left(\frac{H_0}{c}\right)^2\,(1+z)\,\Sigma(k,z)\right]^2P_{\delta\delta}(k,z)\,,
\end{equation}
where $\Sigma(k,z)$ is a phenomenological parameterisation to account for deviations from the standard \lcdm \ lensing prediction. $\Omegam$ is the density parameter of matter and $H_0$ is the Hubble parameter -- both at the present time. Here, we assumed a standard background evolution of the matter component, that is, $\rho_{\rm m}(z)=\rho_{{\rm m},0}(1+z)^3$. 

We can therefore use the recipe of \cref{eq:ISTrecipe} with $H$, $r$ and $P_{\delta\delta}$ provided by a Boltzmann solver, but with the new window functions \citep{SpurioMancini:2019rxy}
\begin{align}
 W_i^{\rm G}(k,z) = &\; \frac{1}{c}\,b_i(k,z)\,n_i(z)\,H(z)\,, \label{eq:wg_mg}\\  
 W_i^{\rm L}(k,z) = &\; \frac{3}{2}\,\Omegam\, \left(\frac{H_0}{c}\right)^2(1+z)\,r(z)\,\Sigma(k,z) \, \nonumber\\
& \times \int_z^{z_{\rm max}}{\de z'\, n_i(z)\,\frac{r(z')-r(z)}{r(z')}} + W^{\rm IA}_i(k,z)\, , \label{eq:wl_mg}
\end{align}
where $n_i(z)$ is the normalised galaxy number-density distribution in a tomographic redshift bin, $i$, such that $ \int_{z_{\rm min}}^{z_{\rm max}} n_i(z) dz = 1$, and $W^{\rm IA}_i(k,z)$ encodes the contribution of intrinsic alignments (IA) to the WL power spectrum. We follow \citetalias{Blanchard:2019oqi} in assuming an effective scale-independent galaxy bias, constant within each redshift bin, and its values, $b_i$, are introduced as nuisance parameters in our analysis, with their fiducial values determined by $b_i = \sqrt{1+\bar{z}_i}$, where $\bar{z}_i$ is the mean redshift of each redshift bin.

The IA contribution is computed following the nonlinear alignment model with a redshift dependent amplitude \citepalias{Blanchard:2019oqi},
in which\begin{equation}\label{eq:IA}
 W^{\rm IA}_i(k,z)=-\frac{\mathcal{A}_{\rm IA}\,\mathcal{C}_{\rm IA}\,\Omegam\,\mathcal{F}_{\rm IA}(z)}{\delta(k,z)\,/\,\delta(k,0)}\,n_i(z)\,\frac{H(z)}{c}\,,
\end{equation}
where 
\begin{equation}
 \mathcal{F}_{\rm IA}(z)=(1+z)^{\eta_{\rm IA}}.
\end{equation}
The parameters $\mathcal{A}_{\rm IA}$ and $\eta_{\rm IA}$ are the nuisance parameters of the model, and $\mathcal{C}_{\rm IA}$ is a constant accounting for dimensional units. The galaxy distribution is binned into $10$ equipopulated redshift bins with an overall distribution following
\begin{equation}
    n(z)\propto\left(\frac{z}{z_0}\right)^2\,\exp\left[-\left(\frac{z}{z_0}\right)^{3/2}\right]\,,
\end{equation}
with $z_0=0.9/\sqrt{2}$ and the normalisation set by the requirement that the surface density of galaxies is $\bar{n}_{\rm g}=30\,\mathrm{arcmin}^{-2}$ \citepalias{Blanchard:2019oqi}.

Changes in the theory of gravity impact the IA contribution, introducing a scale dependence through the modified perturbations' growth. This is explicitly taken into account in \cref{eq:IA} through the matter perturbation $\delta(k,z)$, while the modifications on the clustering of matter in the $\GCph$ case are accounted for in the new $P_{\delta\delta}(k_\ell,z)$.

Finally, we present the likelihood that we used. We followed the approach presented in \citelist{Euclid:2023pxu}. We first constructed an $(N^\mathrm{G}+N^\mathrm{L})\times (N^\mathrm{G}+N^\mathrm{L})$ angular power spectrum matrix for each multipole, where the different $N$ correspond to the number of redshift bins for each probe (WL and \GCph{}):\\
\begin{equation}
    \tens{C}_\ell = \left[ \begin{array}{cc}
      C^\mathrm{LL}_{ij}(\ell)   & C^\mathrm{GL}_{ij}(\ell) \\
      C^\mathrm{LG}_{ij}(\ell)   & C^\mathrm{GG}_{ij}(\ell)
    \end{array}\right],
\end{equation}
where lower-case Latin indexes $i,\ldots$ run over all tomographic bins. Similarly, the noise contributions can also be condensed into one noise matrix, $\tens{N}$:
\begin{equation}
    \tens{N}_\ell = \left[ \begin{array}{cc}
      N^\mathrm{LL}_{ij}(\ell)   & N^\mathrm{GL}_{ij}(\ell) \\
      N^\mathrm{LG}_{ij}(\ell)   & N^\mathrm{GG}_{ij}(\ell)
    \end{array}\right],
\end{equation}
where the noise terms, $N_{ij}^{AB}(\ell)$, are given by
\begin{align}
    N_{ij}^{\rm LL}(\ell) &= \frac{\delta_{ij}^{\rm K}}{\bar{n}_i}\sigma_\epsilon^2\,,\\
    N_{ij}^{\rm GG}(\ell) &= \frac{\delta_{ij}^{\rm K}}{\bar{n}_i}\,,\\
    N_{ij}^{\rm GL}(\ell) &= 0\,,
\end{align}
where $\sigma_\epsilon^2=0.3^2$ is the variance of observed ellipticities. We can further define $\hat{\tens{C}}_\ell \coloneqq \tens{C}_\ell + \tens{N}_\ell$. This is the covariance of the spherical multipole moments, $a_{lm}$.

In the Gaussian approximation, it can be shown that the description in \citetalias{Blanchard:2019oqi}, using the covariance of the angular power spectra, is equivalent to the description found in \citelist{Euclid:2023pxu}, using the covariance of the $a_{\ell m}$. The likelihood then can be expressed in terms of the observed covariance 
\begin{equation}
    \hat{\mathrm{C}}^\mathrm{obs}_{ij}(\ell) = \frac{1}{2\,\ell+1}\,\sum_{m=-\ell}^\ell \left(a_{\ell m}\right)_i\,\left(a_{\ell m}\right)^*_j\;, 
\end{equation} and the theoretically predicted one, $\hat{ \mathrm{C}}^\mathrm{th}_{ij}(\ell)$, as 
\begin{equation}
    \chi^2 = f^\mathrm{sky} \, \sum_{\ell = \ell_\mathrm{min}}^{\ell_\mathrm{max}} (2\ell+1) \,\left[\frac{d^\mathrm{mix}}{d^\mathrm{th}} + \ln\left(\frac{d^\mathrm{th}}{d^\mathrm{obs}}\right) - N\right]\;,
\end{equation}
where the determinants, $d$, are defined as 
\begin{align}
    d^\mathrm{th}(\ell) &= \det \left[\hat{\mathrm{C}}^\mathrm{th}_{ij}(\ell)\right],
    \label{eq:ds1}
    \\
    d^\mathrm{obs}(\ell) &= \det \left[\hat{\mathrm{C}}^\mathrm{obs}_{ij}(\ell)\right], 
     \label{eq:ds2}
    \\
    d^\mathrm{mix}(\ell) &= \sum_{k}^N\det \left[ \left\{\begin{array}{cc}
        \hat{\mathrm{C}}^\mathrm{obs}_{ij}(\ell)& \text{for } k = j\\
        \hat{\mathrm{C}}^\mathrm{th}_{ij}(\ell)& \text{for } k \neq j
    \end{array}\right. \right] \;.
       \label{eq:ds3}
 \end{align}
Here, $N$ is the number of bins, and thus either $(N^\mathrm{G}+N^\mathrm{L})$ for multipoles for which we include the cross correlation, or the respective $N$ for multipoles for which we treat the two probes separately. The additional factor, $f^{\rm sky}$, comes from an approximation to account for having fewer available independent $\ell$ modes due to partial sky coverage.
In this paper, we set the observed covariance $\hat{\mathrm{C}}^{\rm obs}$ to the theoretically predicted one computed at the fiducial cosmology.
Following \citetalias{Blanchard:2019oqi} and \citelist{Euclid:2023pxu}, we do not include super-sample covariance (SSC) in this work. The SSC impact was shown to be non-negligible and will need to be included in future analyses as shown in \citelist{Euclid:2023ove}. 

We consider two different scenarios: an optimistic and a pessimistic case. In the optimistic case, we consider $\ell_{\rm max}=5000$ for WL, and $\ell_{\rm max}=3000$ for $\GCph$ and \XCph{}. Instead, in the pessimistic scenario, we consider $\ell_{\rm max}=1500$ for WL, and $\ell_{\rm max}=750$ for \GCph{} and \XCph{}. 

In the smallest photometric redshift bin, the galaxy number density distribution, $n(z)$, peaks at around $z=0.25$. Under the Limber approximation for our fiducial cosmology, the corresponding maximum values of $k$ evaluated in the power spectrum corresponding to the pessimistic and optimistic scenario for \GCph\ are $k_{\rm max}=[0.7, 2.9]\,h\,\mathrm{Mpc}^{-1}$, respectively, while for the WL, maximum wavenumbers probed are $k_{\rm max}=[1.4, 4.8]\,h\,\mathrm{Mpc}^{-1}$ at the peak redshift $z=0.25$ \citep{Euclid:2023tqw}. Here, $h$ denotes the dimensionless Hubble parameter $h \coloneqq H_0 / (100 \kmsMpc)$. For smaller values of $z$, the values of $k$ at a given $\ell$ increase, but the window functions in \cref{eq:wg_mg,eq:wl_mg} suppress the power spectrum and we set it to zero after a fixed $k_{\rm max}=30\,h\,\mathrm{Mpc}^{-1}$.
We list the specific choices of scales and settings used for each observable in \Cref{tab:specifications-ec-survey}.
Although the currently foreseen specification of the Euclid Wide Survey differs from
that assumed in \citetalias{Blanchard:2019oqi}, for example in terms of the survey area, we use their results to allow for
comparison with earlier forecasts.
 
 \begin{table}
	\centering
	\caption{\Euclid survey specifications for WL, $\GCph$ and $\GCsp$. }
	\label{tab:specifications-ec-survey}
	\begin{tabularx}{\columnwidth}{Xll}
		\hline 
		\noalign{\vskip 1pt}Survey sky coverage & $f^{\rm sky}$  & $36.36\%$  \\
  		\hline
        \rowcolor{gray}\multicolumn{3}{c}{WL}\\
\noalign{\vskip 1pt}
		Number of photo-$z$ bins & $N^\mathrm{L}$ & 10 \\
		Galaxy number density & $\bar n_{\rm gal}$  & $30\,\mathrm{arcmin}^{-2}$ \\
		Intrinsic ellipticity $\sigma$ & $\sigma_\epsilon$  & 0.30 \\
		Minimum multipole & $\ell_{\rm min}$ & 10\\
		Maximum multipole & $\ell_{\rm max}$ & \\
		-- Pessimistic & & $1500$\\
		-- Optimistic & & $5000$\\
		\hline
        \rowcolor{gray}\multicolumn{3}{c}{$\GCph$}\\
		Number of photo-$z$ bins & $N^\mathrm{G}$ & 10 \\
		Minimum multipole & $\ell_{\rm min}$ & 10 \\
		Maximum multipole & $\ell_{\rm max}$ & \\
		-- Pessimistic & & $750$\\
		-- Optimistic & & $3000$\\
	\end{tabularx}
\end{table}

\section{\montepython{} implementation}
\label{sec:MP}
In this section, we describe the implementation of the likelihood discussed in \cref{sec:thpred} in the \montepython{} code developed in \citelist{Euclid:2023pxu}, in the Hu--Sawicki $f(R)$ gravity model.

\subsection{Hu--Sawicki \texorpdfstring{$f(R)$}{fR} gravity}\label{sec:fR}
Modified gravity $f(R)$ models \citep{1970MNRAS.150....1B} are models constructed by promoting the Ricci scalar, $R$, in the Einstein--Hilbert action to a generic function of $R\to R + f(R)$, that is, 
\begin{align}
 S = \frac{c^4}{16\pi G} \int{\de^4 x\, \sqrt{-g}\, \left[R+f(R)\right]} + S_{\rm m}[g_{\mu\nu}, \Psi_{\rm m}] \,, \label{eq:EHaction}
\end{align}
where $g_{\mu\nu}$ is the metric tensor, $g$ its determinant, and $S_{\rm m}$ represents the matter sector with its matter fields, $\Psi_{\rm m}$. For further discussions, we refer to \citelist{Sotiriou_2010,Clifton_2012}, and \cite{Koyama:2015vza}.

How this modification changes gravity is more easily seen by formulating the theory in the so-called Einstein frame by performing a conformal transformation $g_{\mu\nu} = \tilde{g}_{\mu\nu}A^2(\phi)$ where $A(\phi) = \sqrt{1 + f_{R}} = e^{\phi/\sqrt{6}}$, to obtain
\begin{align}
S = &\, \frac{c^4}{16\pi G} \int{\de^4 x \sqrt{-\tilde{g}} \left[\tilde{R} + \frac{1}{2}(\partial\phi)^2 - V(\phi)\right]}  \\
&\, + S_{\rm m}[A^2(\phi)\,\tilde{g}_{\mu\nu}, \Psi_{\rm m}]\,, \nonumber
\end{align}
with the potential $V = [f_{R} R - f(R)]/2(1+f_{R})^2$ and $f_{R} \coloneqq \diff f(R)/\diff R$. This demonstrates that the theory reduces to standard GR together with an extra scalar field that is coupled to the matter sector giving rise to a fifth force. This fifth force has a finite range, $\lambda_\sfont{C}$, and within this range, it mediates a force that has a strength that, in the linear regime, corresponds to $1/3$ of the usual gravitational force:
\begin{equation}
{\bf F}_{\rm fifth} = \frac{1}{3}\, \frac{G Mm}{r^2}\, (1 + r/\lambda_\sfont{C})\, \mathrm{e}^{-r/\lambda_\sfont{C}}\,.
\end{equation}
This effectively changes $G\to \tfrac{4}{3}G$ on small scales ($r\ll \lambda_\sfont{C}$) while keeping the usual $G$ on large scales. Such a large modification would be ruled out by observations if it was not for the fact that the theory possesses a screening mechanism \citep{Khoury_2004, Brax:2008hh} which hides the modifications in high-density regions. This screening mechanism effectively suppresses the fifth force by a factor $\propto f_{R} / \Phi_{\rm N}$, where $\Phi_{\rm N}$ is the Newtonian gravitational potential. 

Not all $f(R)$ models one can write down possess such a screening mechanism, which places some restrictions on their functional form \citep[see e.g.,][]{Sotiriou_2010}. One concrete example of a model that has all the right ingredients is the model proposed by \citelist{Hu:2007nk}, which in the large-curvature limit is defined by
\begin{equation}
f(R) = - 6 \OmegaDE \frac{H_0^2}{c^2} + |f_{R0}| \frac{\bar{R}_0^2}{R}\,,
\label{eq:fR}
\end{equation}
where 
\begin{equation}
\bar{R}_0 = 3 \Omegam \frac{H_0^2}{c^2} \left(1+ 4 \frac{\OmegaDE}{\Omegam} \right)\label{eq:R}
\end{equation}
is the Ricci scalar in the cosmological background and $\OmegaDE$ is the density parameter of dark energy at the present time. The first term in \cref{eq:fR} corresponds to a cosmological constant and the only free parameter is $|f_{R0}|$. In the limit $|f_{R0}| \to 0$, we recover GR and the \lcdm \ model. This parameter controls the range of the fifth force and, in the cosmological background, we have $\lambda_\sfont{C} \simeq 32\sqrt{|f_{R0}|/10^{-4}}\,{\rm Mpc}$ at the present time. Solar System constraints require $|f_{R0}| \lesssim 10^{-6}$, cosmological constraints currently lie around $10^{-6}$--$10^{-4}$ depending on the probe in question \citep[see e.g., Fig. 28 in][for a summary]{Koyama:2015vza} while astrophysical constraints at the galaxy scale can be as tight as $\lesssim 10^{-8}$ \citep{Desmond_2020}, but see \cite{Burrage:2023eol} for a recent note of caution on such galactic-scale constraints.

The energy density of the scalar field contributes in general to the expansion of the Universe; however, for viable models, like the one we consider here, this contribution is tiny (of the order $|f_{R0}|\, \Omega_{\rm DE, 0}$) apart from the constant part of the potential, which is indistinguishable from a cosmological constant. The background evolution of such models is therefore very close to \lcdm{}. Since $f(R)$ models have a conformal coupling, light deflection is weakly affected as follows
\begin{equation}\label{eq:SigmaHS}
 \Sigma(z)=\frac{1}{1+f_{R}(z)}\,.
\end{equation}
Since the maximum value of $|f_{R}(z)|$ is given by $|f_{R0}|$, for the values of $|f_{R0}|$ we consider in this paper, we can ignore this effect. Thus gravitational lensing is also not modified in the sense that the lensing potential is sourced by matter in the same way as in GR (though the underlying density field will of course be different). The main cosmological signatures of the model therefore come from having a fifth force, acting only on small scales $r \lesssim \lambda_\sfont{C}$, in the process of structure formation. The main effect of the screening mechanism is that the prediction for the amount of clustering will generally be much smaller than what naive linear perturbation theory predicts.

\subsection{Nonlinear modelling}\label{sec:nonlinear}
We implement $\Xi(k,z)$, defined as 
\begin{equation}\label{eq:frboost}
\Xi(k,z) \coloneqq \frac{P_{f(R)}(k,z)}{P_{\lcdm}(k,z)} \,,
\end{equation}
to obtain the nonlinear $f(R)$ matter power spectrum. For the \lcdm\ power spectrum $P_{\lcdm}(k,z)$, we use the \halofit{} `Takahashi' prescription \citep{Takahashi:2012em} following \citelist{Euclid:2023tqw}.  
It includes the minimum mass for massive neutrinos in $P_{\lcdm}(k,z)$, but ignores the effect of massive neutrinos on $\Xi(k,z)$. This approximation was shown to be well justified for the minimum mass of neutrinos in \citet{Winther:2019mus} using data from \citet{Baldi_2014}. We describe below four models for  $\Xi(k,z)$ used in this paper (see Table 2). We note that we can use any \lcdm\ power spectrum prediction in our approach such as \eetwo{} \citep{Euclid:2020rfv} and \bacco{} \citep{Angulo:2020vky}. 
The exercise we perform in this paper is a comparison of the different prescriptions for $\Xi(k, z)$. Given this,  the \lcdm{} nonlinear spectrum prescription does not matter and it is common to all nonlinear models.

When we apply our pipeline to observational data, it is important to control the accuracy of the \lcdm{} nonlinear spectrum prescription. We need to implement scale cuts to account for the inaccuracy of the \lcdm{} prediction. The validation of the \lcdm{} nonlinear prescription is still underway in the Euclid consortium, and we leave the investigation of the effect of using different prescriptions in \lcdm{} for future work.

\subsubsection{Fitting formula}
A fitting formula for $\Xi(k,z)$ was developed in \citet{Winther:2019mus} and describes the enhancement in the power spectrum compared to a \lcdm\ nonlinear power spectrum as a function of the parameter $|f_{R0}|$. This fitting function has been calibrated using the \dustgrain{} \citep{Giocoli:2018gqh} {\it N}-body simulations run by \mggadget{} \citep{Puchwein:2013lza} and the \elephant{} {\it N}-body simulations \citep{Cautun:2017tkc} run by \ecosmog{} \citep{Li:2011vk, Bose_2017}. The main approximation is that the cosmological parameter dependence of $\Xi(k,z)$ is ignored. In \citet{Winther:2019mus}, this assumption was checked using simulations with different $\sigma_8$, $\Omegam$, as well as the mass of massive neutrinos. \citet{Winther:2019mus} also corrected the fitting formula to account for additional dependence on these parameters. In this paper, we do not include these corrections as the previous forecast paper \citep{Euclid:2023tqw} used the fitting formula without these corrections. 

The fitting function has in total 54 parameters for the full scale, redshift and $|f_{R0}|$ dependence. 
This fitting formula is not defined outside the range $10^{-7} < \fr < 10^{-4}$. The code is publicly available (\href{https://github.com/HAWinther/FofrFittingFunction}{\faicon{github}}). 


\subsubsection{Halo model reaction}
We further consider the halo model reaction approach of \citelist{Cataneo:2018cic} which combines the halo model and perturbation theory frameworks to model corrections coming from non-standard physics. The nonlinear power spectrum is given by
\begin{equation}
 P_{\rm NL}(k,z) = \mathcal{R}(k,z) \, P^{\rm pseudo}_{\rm NL}(k,z)\,,
 \label{eq:nlpkreact}
\end{equation}
where $P^{\rm pseudo}_{\rm NL}(k,z)$ is called the pseudo-power spectrum and is defined as a nonlinear \lcdm \ spectrum with initial conditions tuned such that the linear clustering at the target redshift $z$ matches the beyond-\lcdm\ case. This choice ensures the halo mass functions in the beyond-\lcdm \ and pseudo-universes are similar, giving a smoother transition of the power spectrum over inter- and intra-halo scales. We model the pseudo-cosmology nonlinear power spectrum using \hmcode{} \citep{Mead:2015yca,Mead:2016zqy,Mead:2020vgs} by supplying the code with the linear $f(R)$ power spectrum. 

The halo model reaction, $\mathcal{R}(k,z)$, models all the corrections to the pseudo spectrum coming from nonlinear beyond-\lcdm\ physics. We refer the reader to \citelist{Cataneo:2018cic,Bose:2021mkz} and \cite{Euclid:2023rjj} for the exact expressions for this term. The halo model reaction can be computed efficiently using the publicly available \react{} \citep[\href{https://github.com/nebblu/ACTio-ReACTio}{\faicon{github}}]{Bose:2020wch,Bose:2021mkz,Bose:2022vwi} code.

Despite \react{} being highly efficient, having been used in previous real cosmic shear analyses \citep{KiDS:2020ghu}, it is still too computationally expensive for the number of tests we wish to perform. To accelerate our inference pipeline, we created a neural network emulator using the \cosmopower{} package \citep[\href{https://github.com/alessiospuriomancini/cosmopower}{\faicon{github}}]{SpurioMancini:2021ppk} for the halo model reaction-based boost
\begin{equation}\label{eq:frboostreact}
\Xi(k,z) = \frac{\mathcal{R}(k,z) \,P_{\rm HMCode2020}^{\rm pseudo}(k,z)} {P_{\lcdm}(k,z)} \,,
\end{equation}
where $P_{\lcdm}(k,z)$ and $P_{\rm HMCode2020}^{\rm pseudo}(k,z)$ were calculated using \hmcode{} \citep[\href{https://github.com/alexander-mead/HMcode}{\faicon{github}}]{Mead:2020vgs} while $\mathcal{R}(k,z)$ was calculated using \react{}.
We chose \hmcode{} to model the pseudo-power spectrum as it has been shown to have improved accuracy and does not show suppression of power with respect to \lcdm{} ($\Xi <1$) at high redshifts, which is not expected.

To train the emulator, we followed the procedure of \citelist{SpurioMancini:2023mpt} but widened the parameter priors. We produce $\sim10^5$ boost predictions in the range  $k \in [0.01, 3]\,h\,{\rm Mpc}^{-1}$ and $z \in [0,2.5]$. We take cosmologies from the Latin hypercube given by the ranges in \autoref{tab:emu_ranges} and \autoref{tab:emu_ranges2}, with the massive neutrino  density parameter today's range being $\Omeganu\in [0.0, 0.00317]$.  Emulation of the boost speeds up the computation by 4 orders of magnitude and we find similar accuracy of our emulator as found in \citelist{SpurioMancini:2023mpt}. The emulator is publicly available (\href{https://github.com/nebblu/ReACT-emus}{\faicon{github}}). Finally, we note another small difference between our emulator and that of \citelist{SpurioMancini:2023mpt}. In this work, we assume $P_{\lcdm}$ in \cref{eq:frboostreact} with $\Omegam$ equal to the total of the true cosmology, whereas in \citelist{SpurioMancini:2023mpt} they assume that  $\Omegam = \Omegab + \Omegac$, $\Omegab$ and $\Omegac$ being, respectively the baryon and cold dark matter density parameters today. The emulators give the same output for $\Omeganu = 0$, which is what we assume in this work for $\Xi$. 


\subsubsection{\forge{}}
The \forge{} emulator was introduced in \citelist{Arnold:2021xtm}. It is based on simulations for 50 combinations of $|f_{R0}|$, $\Omegam$, $\sigma_8^{\lcdm}$ and $h$ with all other parameters fixed. We note that $\sigma_8^{\lcdm}$ is the $\sigma_8$ we would obtain in a \lcdm\ model with the same cosmological parameters and initial amplitude $A_{\rm s}$ and not $\sigma_8$ in an $f(R)$ gravity model. The emulator accuracy is better than $2.5 \%$ around the centre of the explored parameter space, up to scales of $k = 10\, h\,{\rm Mpc}^{-1}$. $f(R)$ simulations are performed by a modified version of the \arepo{} code, \mgarepo{} \citep{Springel:2009aa,Arnold:2019vpg,Weinberger:2019tbd} that solves the nonlinear scalar field equation using a relaxation method. (see \citetalias{paper1} for more details.)

The emulation was made for the ratio between the power spectrum in $f(R)$ and the \halofit{} prediction in \lcdm. We note that this is different from $\Xi(k,z)$ as the power spectrum in \lcdm\ models in these simulations can have slight deviations from the \halofit{} prediction. To account for this effect, the authors provided the ratio of the power spectrum in a reference \lcdm \ model to the \halofit{} prediction. This \lcdm\ simulation uses $\Omegam=0.31315$, $h=0.6737$, $\sigma_8^{\lcdm}=0.82172$.
This can be used to obtain $\Xi(k,z)$. However, this correction is provided only in this \lcdm\ model. Thus, the assumption here is that this correction is independent of cosmological parameters. The latest version of the \forge{} simulations has \lcdm\ counterparts to the $f(R)$ simulations using the same seed. These simulations were analysed in \citetalias{paper2}. Hence, it is in principle possible to emulate directly $\Xi(k,z)$ from these simulations. However, as this is not publicly available, we opted for using the original \forge{} emulator (\href{https://bitbucket.org/arnoldcn/forge_emulator/src/master/}{\faicon{bitbucket}}) for this paper. We use one of these simulations for validation in \cref{sec:comparison}. 


\subsubsection{\emantis{}}

We also consider the predictions given by the \emantis{} emulator presented in \citet{saez-casares_2023}, which can predict the ratio $\Xi(k,z)$ between the nonlinear matter power spectrum in $f(R)$ gravity and in \lcdm. 
The predictions are calibrated from a large suite of $N$-body simulations run with the code \ecosmog{} \citep{Li:2011vk, Bose_2017}, a modified gravity version of the Adaptive Mesh Refinement (AMR) $N$-body code \ramses{} \citep{Teyssier2002, Guillet2011}.
The \emantis{} simulation suite covers the $\left\{|f_{R_0}|, \Omegam, \sigma_{8}^{\lcdm} \right\}$ parameter space with $110$ cosmological models sampled from a Latin hypercube \citep{McKay_1979}. 
The remaining \lcdm\ parameters, $h$, $n_{\rm s}$ and, $\Omegab$, remain fixed, which means that their impact on $\Xi(k,z)$ is ignored.
In \citelist{saez-casares_2023}, it was shown that the error made by this approximation is at the sub-percent level.
The quantity $\Xi(k,z)$ is measured from pairs of $f(R)$ and \lcdm\ simulations, both run with the same initial conditions, which leads to a large cancellation of cosmic variance and numerical resolution errors.
The emulation is done through a Gaussian Process regression \citep[see e.g.][]{rasmussen_williams_2005}.
The emulator can give predictions for redshifts $z \in \left[0, 2\right]$, wavenumbers $k \in \left[0.03, 10\right]\,h\,\mathrm{Mpc}^{-1}$ and cosmological parameters in the following range: $\left|f_{R_0}\right| \in \left[10^{-7}, 10^{-4}\right]$, $\Omegam \in \left[0.2365, 0.3941\right]$ and $\sigma_8^{\lcdm{}} \in \left[0.6083, 1.0140\right]$.
The estimated accuracy of \emantis{}, including emulation errors and systematic errors in the training data, is better than $3 \%$ for scales $k\lesssim 7\,h\,\mathrm{Mpc}^{-1}$ and across the whole parameter space although in most cases, the accuracy is of order $1 \%$. The code is publicly available (\href{https://gitlab.obspm.fr/e-mantis/e-mantis}{\faicon{gitlab}}).

\begin{table*}[t]
    \centering
    \caption{A summary of four nonlinear models.} 
    \begin{tabular}{ccc}
    \hline
    method & simulations (code) & parameters\\
    \hline
    \rowcolor{gray} \multicolumn{3}{l}
    {\fofrfitting}\\
    fitting formula  & \elephant{}  (\ecosmog{}) & $|f_{R_0}|$\\
      &  \dustgrain{} (\mggadget{}) &  \\
    \hline
     \rowcolor{gray} \multicolumn{3}{l}
    {\react{}}\\
    halo model & N.A. & $|f_{R_0}|, \Omegam, \Omegab$, \\
     (emulator) &  &    
    $h, n_s, A_{\rm s}  $\\
    \hline
    \rowcolor{gray} \multicolumn{3}{l}{\forge{}}\\
    emulator & \forge{} (\mgarepo{}) & $|f_{R_0}|, \Omegam, h,  \sigma_{8}^{\lcdm}$ \\
    \hline
    \rowcolor{gray} \multicolumn{3}{l}{\emantis{}}\\
    emulator & (\ecosmog{}) & $|f_{R_0}|, \Omegam,  \sigma_{8}^{\lcdm}$ \\
    \end{tabular}
    \tablefoot{For the details of N-body simulations codes and simulation suites, see \citetalias{paper1} and \citetalias{paper2}. The prediction for $\Xi$ remains the same if we vary cosmological parameters which are not included in the prediction.}
    \label{tab:nonlinearmodels}
\end{table*}

\subsection{\montepython{} implementation}
Our implementation of the likelihood is based on the \montepython{} implementation described in \citelist{Euclid:2023pxu}. For a more detailed explanation of the numerical implementation, we refer to that work. For our purposes, we have modified this code to include our prediction for $\Xi(k,z)$ to obtain the nonlinear power spectrum in the $f(R)$ gravity model. As the different implementations have different ranges in cosmological parameters, wavenumbers, and redshifts, we have chosen an extrapolation scheme to unify the ranges. We note that the exact implementation does not affect the final results strongly. We checked that the contributions from high redshifts and high wavenumbers which used extrapolations were sub-dominant.

For the modified gravity parameter $|f_{R0}|$, we used flat priors in terms of $\logfr$ to stay within the validity range of each model. For the largest scales beyond the range of the emulators, we set $\Xi$ = 1. This is because the effect of the fifth force has to vanish on these scales. For small scales, we did a power-law extrapolation. To obtain the spectral index for the extrapolation, we proceeded as follows: we calculated $\Xi$ on a grid close to the edges of the emulators; we then fitted, for fixed $z$, a linear function in the log-log space onto this grid. This was done to average out any numerical noise at the edge. We did the same for high redshifts by fitting the power law for fixed wavenumbers. For regions of both high $k$ and $z$, we did a constant extrapolation of the spectral index. 
We checked the dependence of the extrapolation methods (the power law and constant extrapolation as well as the \lcdm-extrapolation to $\Xi =1$) at the level of the angular power spectra and confirmed that the effect was smaller than the errors on the synthetic data. To keep the extrapolated function $\Xi(k, z)$ from going to non-physical values, we set a hard lower bound of $\Xi$ = 1 and an upper bound of $\Xi$ = 2. The ranges of the different emulators are found in \Cref{tab:emu_ranges}. We did a constant extrapolation in the \lcdm \ parameters. This can be done because during a typical MCMC the majority of the suggested points are within the ranges of our emulators.
In addition, $\Xi$ is insensitive to cosmological parameters as shown by \cite{Winther:2019mus}.
The validity range of \lcdm{} cosmological parameters is summarised in \cref{tab:emu_ranges2}.

\begin{table}[]
    \centering
    \caption{Ranges of wavenumbers, redshifts, and $\logfr$ in different emulators.}
    \begin{tabular}{ccc}
    \hline
    $k$ range $[h\,\mathrm{Mpc}^{-1}]$& $z$ range & $\logfr$ range\\
    \hline
    \rowcolor{gray} \multicolumn{3}{l}
    {\fofrfitting}\\
    $[\phantom{3\times} 10^{-2}, 10]$ & $\leq 2\phantom{.5}$ & $[\phantom{0}-7 \phantom{.0}\;, -4\phantom{.0}]$\\
    \hline
     \rowcolor{gray} \multicolumn{3}{l}
    {\react{}}\\
    $[\phantom{3\times} 10^{-2}, \phantom{0}3]$ & $\leq 2.5$ & $[-10\phantom{.0}\;, -4\phantom{.0}]$ \\
    \hline
    \rowcolor{gray} \multicolumn{3}{l}{\forge{}}\\
    $[\phantom{3\times} 10^{-3}, 10]$ & $\leq 2\phantom{.5}$ & $[\phantom{0}-6.4\;, -4.5]$\\
    \hline
    \rowcolor{gray} \multicolumn{3}{l}{\emantis{}}\\
    $[3\times10^{-2}, 10]$ & $\leq 2\phantom{.5}$ & $[\phantom{0}-7\phantom{.0}\;, -4\phantom{.0}]$\\
    \end{tabular}
    \label{tab:emu_ranges}
\end{table}

\begin{table*}
\centering
\caption{Ranges for cosmological parameters implemented in the \texttt{MontePython} code. }
\begin{tabular}{  c  c  c  c  c  c }
\hline  
 $\Omegam$  & $\Omegab$  & $100h$ & $n_{\rm s}$ & $\ln(10^{10}A_{\rm s})$ & $\sigma_8^{\lcdm}$ \\
\hline
\rowcolor{gray} \multicolumn{6}{l}{\react{}}
\\
$[0.24, 0.35]$ & [0.04, 0.06] & [63,75] & [0.9, 1.01] &  [2.83, 3.22]  & N.A.   
\\
\hline
\rowcolor{gray} \multicolumn{6}{l}{\forge{}}
\\
$[0.18, 0.55]$ & N.A. & [60,80] & N.A. & N.A. & [0.6, 1.0]    
\\
\hline
\rowcolor{gray} \multicolumn{6}{l}{\emantis{}}
\\
$[0.24, 0.39]$ & N.A. & N.A. & N.A. & N.A. & [0.6, 1.0]
\\
\end{tabular}
\label{tab:emu_ranges2}
\end{table*}

We also added the effect of baryonic feedback in the form of an additional correction, $\Xi^\mathrm{BFM}$, to our power spectrum prediction. The physics of the baryonic feedback effects is discussed in \cref{sec:baryonic_effects}. We refer the reader to \citelist{Schneider:2019xpf} and \citelist{Mead:2020vgs} for further details. We obtained the correction from \bcemu{} by \citelist{Giri_2021}. This emulator is only trained for redshifts below $z=2$ and wavenumbers $k<12.5\,h\,\mathrm{Mpc}^{-1}$. In this case, we use constant extrapolation for both $k$ and $z$. We checked that this had little effect on our results. This is because the \Euclid main probes are mostly sensitive to redshifts around $z\sim 1$. For these redshifts, the extrapolation only affects very high multipoles, $\ell\gtrsim 4500$. Thus, the extrapolation has negligible effect on the angular power spectrum. We estimate the overall effect this has on our result to be at most at the percent level. In the absence of modified gravity, to obtain the nonlinear baryonic feedback power spectrum, we multiply the \lcdm\ nonlinear power spectrum by  $\Xi^\mathrm{BFM}$. When adding the effect of modified gravity, we combine both boosts as
\begin{equation*}
    P_{f(R)}^{\mathrm{BFM}}(k,z) = \Xi^\mathrm{BFM}(k, z)\:\Xi(k,z)\:P_{\lcdm{}}(k,z)\,.
\end{equation*}
We can make this approximation if both effects are independent. This was shown to be the case for small deviations from \lcdm\ in $f(R)$ gravity models considered in this paper using hydrodynamical simulations by \citelist{Arnold:2019vpg} and \cite{Arnold:2019zup}. 

Our final addition to the code is the inclusion of theoretical errors. For this, we have adjusted the prescription of \citelist{Audren_2013}. The theory and numerical implementation is further discussed in \cref{sec:theoretical_error}.

\section{Comparison of different nonlinear models}\label{sec:comparison}
In this section, we compare four different predictions for the nonlinear dark matter power spectrum in terms of $\Xi(k, z)$ defined in \cref{eq:frboost} introduced in \cref{sec:MP}. We start from a comparison of the matter power spectrum with {\it N}-body simulations. We then compare the angular power spectra in the \Euclid\ reference cosmology. We perform MCMC analysis to compare errors and investigate bias due to the difference in the prediction of the nonlinear matter power spectrum using the settings defined in \citetalias{Blanchard:2019oqi}.

\subsection{Comparison of predictions}
\subsubsection{Comparison with {\it N}-body simulations}
\begin{figure*}[ht]
\centering
\includegraphics[width=1\linewidth]{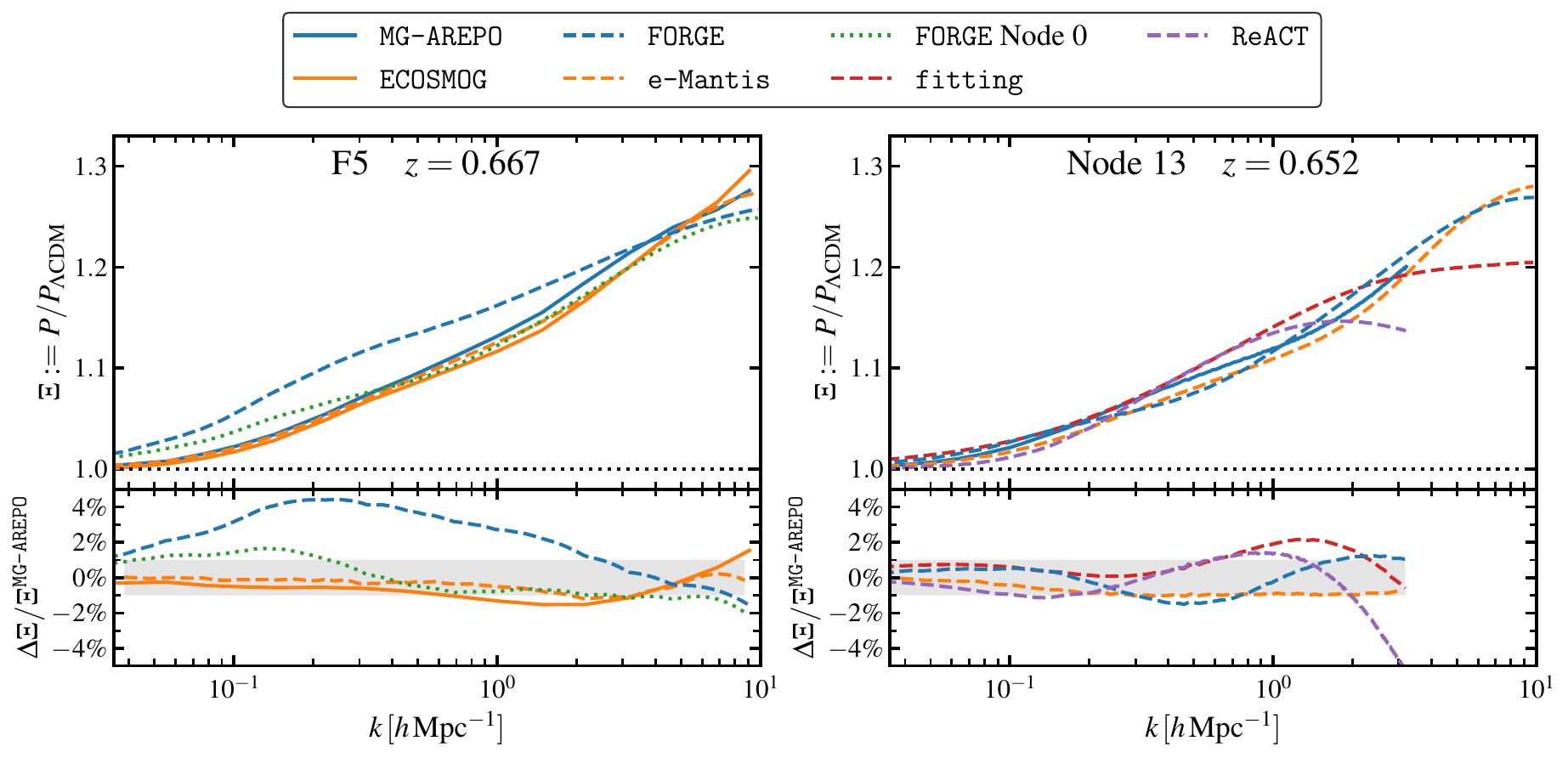}
  \caption{\textit{Left panel}:  Comparisons between $N$-body simulations, \emantis{} and \forge{} for $|f_{R0}|= 10^{-5}$. $N$-body data from simulations run by \mgarepo{} and \ecosmog{} codes are taken from the latest comparison project presented in \citetalias{paper1}. We also include the prediction of \forge{} where the cosmological parameters in the fiducial \lcdm \ simulation,  $\Omegam=0.31315$ and $h=0.6737$, are used to make the prediction (\forge{} Node 0).  \textit{Right panel}:  Comparison of four prescriptions with an {\it N}-body simulation run by \mgarepo{}. This simulation is one of the simulations used for training to construct the \forge{} emulator (Node 13).  The measurement of the power spectrum was done in \citetalias{paper2}. }
\label{fig:comparison}
\end{figure*}

\Cref{fig:comparison} shows a comparison of the ratio of the power spectra between $f(R)$ gravity and \lcdm \ measured from $N$-body simulations with the theoretical predictions for $\Xi$. 
These simulations use the same initial conditions and the ratio removes the cosmic variance and the effect of mass resolution. The left-hand side plot shows a comparison using the measurements from \citetalias{paper1}. This is based on the comparison project in \citelist{Winther:2015wla}. These simulations were run in a \lcdm{} cosmology with $\Omegam = 0.269$, $h = 0.704$, $n_\mathrm{s} = 0.966$ and $\sigma_8 = 0.801$. The simulations have $N_\mathrm{p} = 512^3$ particles of mass $M_{\rm p} \simeq 8.756 \times 10^9\,h^{-1}\,M_{\odot}$ in a box of size $B = 250\,h^{-1}\,\mathrm{Mpc}$ and start at redshift $z = 49$. We picked a model called F5 with $|f_{R0}| =  10^{-5}$ and showed the result at $z=0.667$ that is presented in \citetalias{paper1}. In the comparison, we included the measurements from \mgarepo{} and \ecosmog{}, as \forge{} is based on \mgarepo{}, while \emantis{} is based on \ecosmog{}. The prediction of \emantis{} agrees with \ecosmog{} very well. On the other hand, the \forge{} prediction deviates from \mgarepo{} as well as \ecosmog{}. We note that the \forge{} prediction is corrected using the \lcdm \ simulation (Node 0) in the \forge{} simulation suite with $\Omegam=0.31315$, $h=0.6737$, $\sigma_8^{\lcdm}=0.82172$ to obtain $\Xi$. We find that if we use $\Omegam=0.31315$ and $h=0.6737$ in the \forge{} prediction, the agreement with \mgarepo{} is much better. 

To further investigate this issue with \forge{}, we used the measurement of the power spectrum in one of the nodes in the \forge{} simulation suite (Node 13) run by \mgarepo{} that is closest to the \Euclid\ reference cosmology that we will use in this paper with non-zero $|f_{R0}|$. This simulation has the following cosmological parameters: $\Omegam = 0.34671$, $h = 0.70056$ and $\sigma_8 = 0.78191$ and we show a comparison at $z=0.652$. Both \lcdm \ and $f(R)$ simulations with $|f_{R0}| = 10^{-4.90056}$ are available so that we can measure $\Xi$ directly. We note that the pipeline developed in \citetalias{paper2} measures the power spectrum only up $k =3~h\,{\rm Mpc}^{-1}$. In this case, both  \forge{} and \emantis{} agree with \mgarepo{} within $1 \%$ up to $k =3~h\,{\rm Mpc}^{-1}$. The fitting formula and \react{} agree with \mgarepo{} within $1 \%$ up to $k =1~h\,{\rm Mpc}^{-1}$. Given this result, the large discrepancy between \forge{} and \mgarepo{} is likely to be caused by emulation errors as well as calibrations using the \halofit{} \lcdm\ power spectra predictions.  

\subsubsection{Comparison in the \Euclid\ reference cosmology}
In this paper, we consider the model called HS6 in \citelist{Euclid:2023tqw}, which has the following parameters: \begin{align}
    \bm\Theta \phantom{_{\rm fid}}&=\{\Omegam,\, \Omegab,\, h,\, n_{\rm s},\, \sigma_8,\, \logfr\}\,,\nonumber\\ 
    \mathrm{HS6}:\bm\Theta_{\rm fid}&=\{ 0.32,\, 0.05,\, 0.67,\, 0.96,\, 0.853,\, -5.301\}\,.
    \label{eq:fiducial-params}
\end{align}
The cosmological parameters are the same ones adopted in \citetalias{Blanchard:2019oqi}. 
\citelist{Euclid:2023tqw} show that this value of $|f_{R0}| = 5 \times 10^{-6}$ can be well constrained by the \Euclid photometric probes. Also the range of $|f_{R0}|$ covered by the four models is wider than the errors predicted in \citelist{Euclid:2023tqw}. Our fiducial cosmology includes massive neutrinos with a total mass of $\sum m_\nu=0.06\,\mathrm{eV}$, but we keep $\sum m_\nu$ fixed in the following analysis. 

\begin{figure*}[ht]
\centering
\includegraphics[width=1\linewidth]{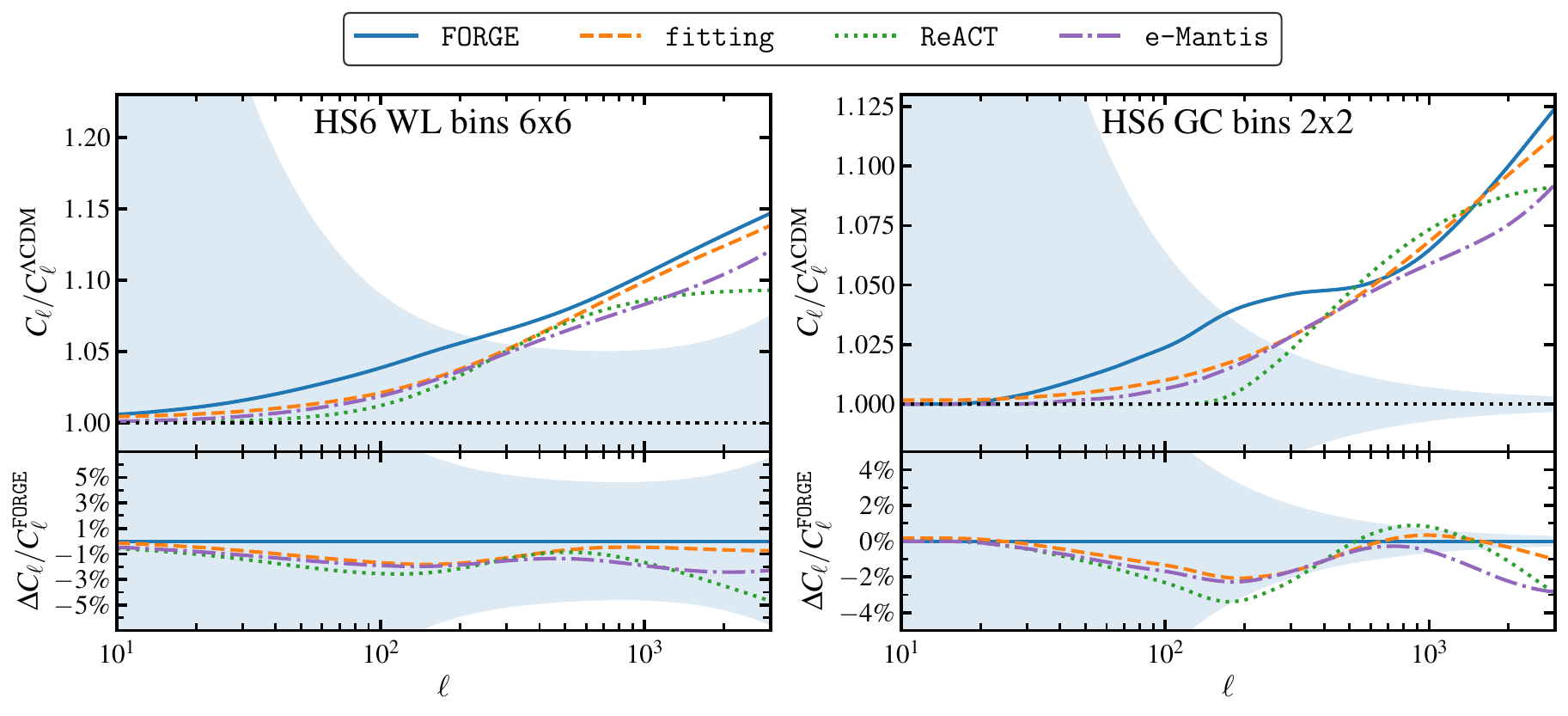}
\includegraphics[width=1\linewidth]{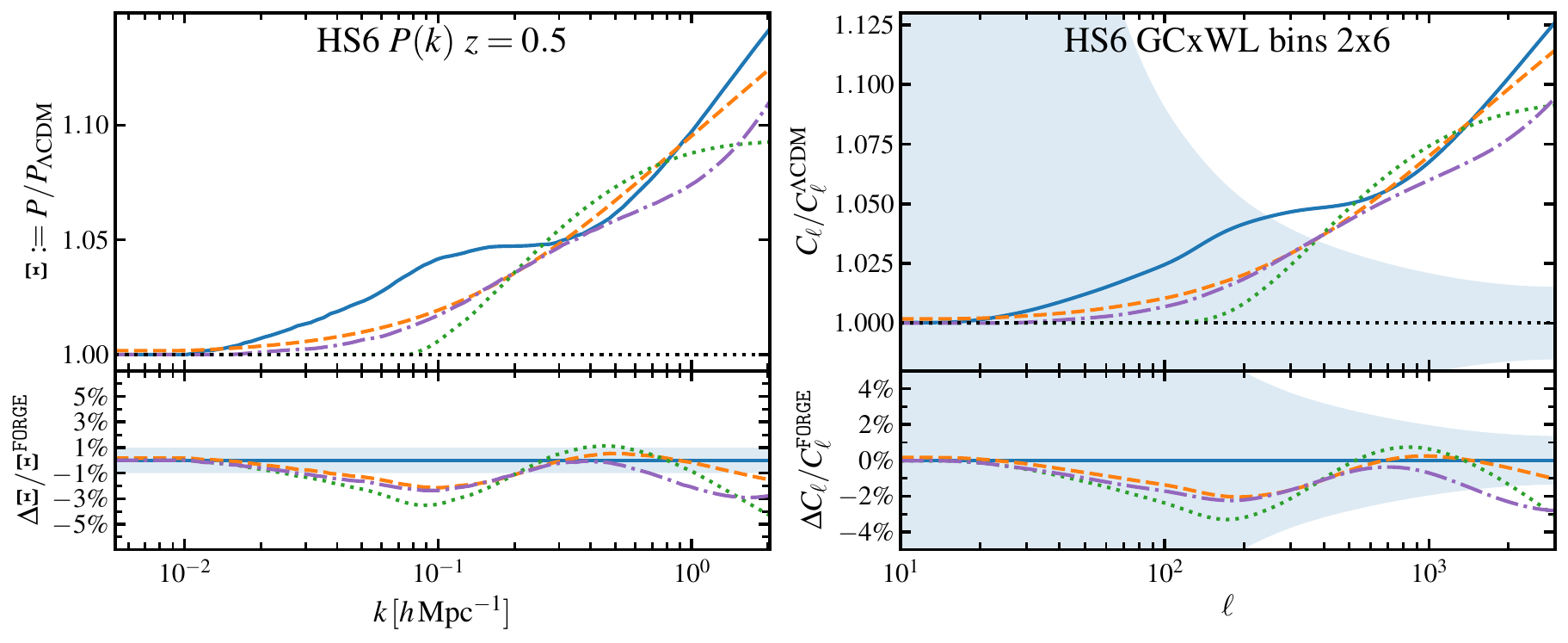}
  \caption{Comparisons between all models considered for the HS6 fiducial cosmology for angular power spectra of WL (\textit{top left}), galaxy clustering (\textit{top right}) and their cross-correlation (\textit{bottom right}) as well as for the matter power spectrum (\textit{bottom left}). We show these for the WL redshift bin 6 and galaxy clustering bin 2, with the matter power spectrum plotted at $z=0.5$, the redshift at which the kernels of both observables peak for the chosen bins. For all angular spectra, the error bands taken from the diagonal of the covariance are also shown.}
   \label{fig:comparison_C_ell}
\end{figure*}

\Cref{fig:comparison_C_ell} shows a comparison of the power spectrum and the angular power spectrum for WL, \GCph\ and their cross-correlation \XCph. In these plots, we show the ratio to the \lcdm\ prediction and error bars from the diagonal part of the covariance matrix. For the power spectrum comparison at $z=0.5$, we see that \emantis{} and \fofrfitting{} agree best. This is not surprising as these two predictions are based on {\it N}-body simulations ran by the same code \ecosmog{}. On the other hand, \forge{} overestimates $\Xi$ at $k= 0.1 h\,{\rm Mpc}^{-1}$. This is similar to the deviation we find in the comparison with the {\it N}-body simulation from \citetalias{paper1} although the deviation is smaller, at the $2 \%$ level. 
This is likely due to the fact that $\Omegam$ in HS6 is closer to $\Omegam$ in the fiducial \lcdm\ \forge\ simulation ($\Omegam=0.31315$). On the other hand, \react{} underestimates $\Xi$ at $k < 0.1 h\,{\rm Mpc}^{-1}$ and even predicts $\Xi <1$. As we discussed in the previous section, we enforce $\Xi \geq  1$.  

\subsection{Forecasting errors and biases} \label{sec:forecast}
\subsubsection{Forecasting errors for WL}
We first compared errors obtained by running MCMC using the synthetic data created by one of the four nonlinear models and fitting it by the same model. In this case, by definition, we recover the input parameters that were used to create the synthetic data. Our interests are constraints on the $|f_{R0}|$ parameter and cosmological parameters. We first considered the WL-only case. In this case, we imposed a tight Gaussian prior on the spectral index, $n_{\rm s}$, taken from the Planck results~\citep{Planck:2018vyg} and on the baryon density parameter, using Big Bang nucleosynthesis constraints \citep{Pisanti:2020efz}: 
\begin{equation}
    n_{\rm s} = 0.96 \pm 0.004, \quad 
    \Omegab h^2 = 0.022445 \pm 0.00036\,,
\end{equation}
as we do not expect to obtain strong constraints on these parameters from WL alone. The parameters that are used in the MCMC runs are summarised in \Cref{tab:MCMCparam}, including their fiducial values and prior ranges. As a convergence criterion, we used a Gelman--Rubin \citep{Gelman:1992zz} value of $R-1 < 0.01$ for each individual sampling parameter using \texttt{MontePython}. For post-processing chains, we used 
\texttt{GetDist} 
\citep[\href{https://github.com/cmbant/getdist}{\faicon{github}}]{Lewis:2019xzd}. 

\begin{table*}[h!]
\centering
\caption{Fiducial values and flat prior ranges for cosmological parameters, $|f_{R0}|$ and IA parameters.}
\begin{tabular}{  c  c  c  c  c  c }
\hline  
 $\Omegac h^2$  & $100h$  & $\ln(10^{10} A_{\rm s})$ & $\mathcal{A}_{\rm IA}$ & $\eta_{\rm IA}$ & $ \logten |f_{R0}|$ \\
\hline
\rowcolor{gray} \multicolumn{6}{l}{Fiducial}\\
0.12056 & 67  & 3.05685 & 1.71 & $-0.41$ & $-5.30103$  \\
\hline
\rowcolor{gray} \multicolumn{6}{l}{Prior}\\
$[0.005, 1]$  & [10, 150] & [2.7, 3.3] & [0, 12.1] & $[-7, 6.17]$  & [$-7$, $-4$]  \\
\end{tabular}
\label{tab:MCMCparam}
\end{table*}

\begin{figure*}[h!]
\centering
	\includegraphics[width=0.49\linewidth]{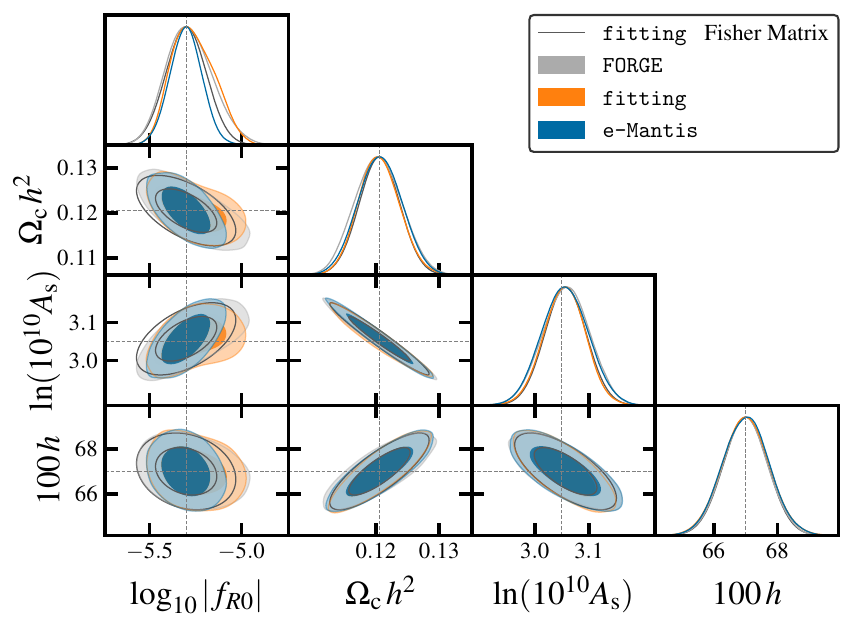}
	\includegraphics[width=0.49\linewidth]{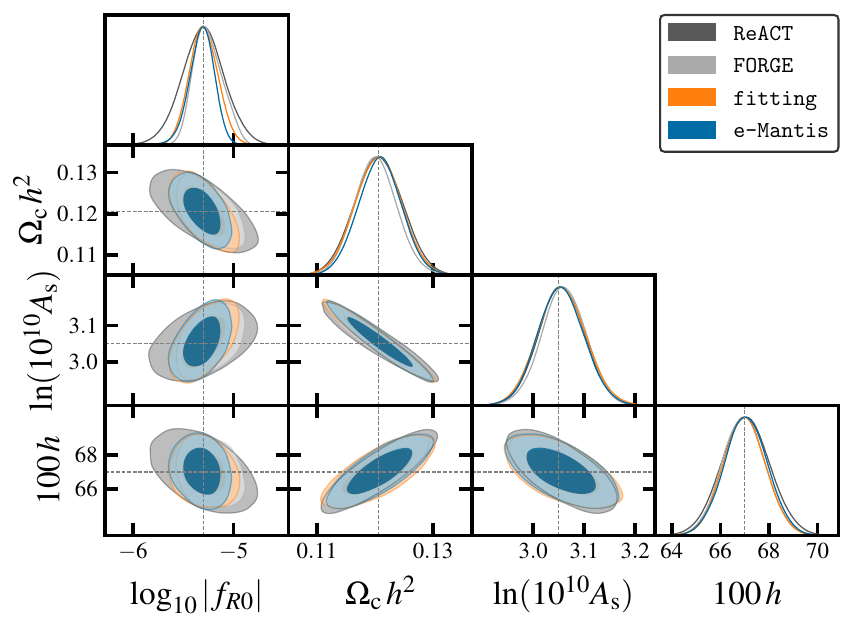}
  \caption{Constraints on parameters where the same model is used to create the synthetic data and perform the fitting such that the input parameters are reproduced. \textit{Left panel}: WL optimistic case. \textit{Right panel}: WL pessimistic case.}
   \label{fig:WLerrors}
\end{figure*}

\Cref{fig:WLerrors} shows the 2D contours of the constraints on parameters,  
and \Cref{tab:WLconstraints1}
and \Cref{tab:WLconstraints2} 
summarise constraints on $\logfr$ for the optimistic and pessimistic settings. The constraints on cosmological parameters are consistent among the four different nonlinear models, while we see some notable differences in the constraints on $|f_{R0}|$. In the case of the optimistic setting, \emantis{} gives the tightest constraints, which is also closer to Gaussian. The fitting formula agrees with \emantis{} for small $\logfr$ but has a weaker constraint for larger $\logfr$. Constraints from \forge{} agree with the fitting formula for large $\logfr$, but give a weaker constraint for small $\logfr$. The degeneracy between $\logfr$, $\Omegac\,h^2$ and $\ln(10^{10} A_{\rm s})$ also presents some notable differences. The degeneracy for larger $\logfr$ is different for the fitting formula when compared with \emantis{} and \forge{}. This could be attributed to the fact that the fitting formula does not include any cosmological parameter dependence in the prediction for $\Xi(k,z)$. We observe similar agreements and disagreements for the pessimistic setting, but the agreement among \emantis{}, \forge{} and the fitting formula is better, particularly for small $\logfr$. \react{} gives a weaker constraint on $\logfr$, but the constraints on cosmological parameters are consistent among the four different models. This is due to the weak cosmology dependence of $\Xi(k,z)$, as discussed in \cite{Winther:2019mus}, and the fact that the constraint on cosmological parameters is coming from the \lcdm \ power spectrum, which is common in all these four models. 

We compared the constraints from the MCMC analysis with the Fisher Matrix forecast and show this in \cref{fig:WLerrors}. The \montepython{} pipeline can be used as a Fisher Matrix forecast tool \citep{Euclid:2023pxu},  which was shown to agree very well with previous Fisher Matrix forecasts given by \citetalias{Blanchard:2019oqi}. We note that in \citelist{Euclid:2023tqw}, no prior was imposed on $n_{\rm s}$ and $\Omegab h^2$, thus no direct comparison is possible. Instead, we used \montepython{} as a Fisher Matrix forecast tool and compared the result with the MCMC analysis to validate the Fisher Matrix forecast. To be consistent with \citelist{Euclid:2023tqw}, we used the fitting formula as the nonlinear model. The $1 \sigma$ error is very consistent: the Fisher Matrix forecast gives $0.111$, while the MCMC analysis gives $0.116$. The constraint from MCMC is non-Gaussian and the 1D posterior has a slightly longer tail for large $\logfr$. The constraints on cosmological parameters agree very well between the Fisher Matrix forecast and the MCMC result. 

\subsubsection{Assessing biases for WL}
Next, we created the synthetic data using \forge{} and we fitted it by different nonlinear models to assess the biases in the recovered parameters due to the difference in the nonlinear modelling. We selected \forge{} to create the data because it has a relatively narrow prior range for $|f_{R0}|$ and we encountered a problem with that range when using \forge{} as the model to fit when including baryonic effects. We should note that \forge{}
has a larger discrepancy with other nonlinear models. As we discussed before, this could be attributed to emulation errors and calibration with \halofit{}. In particular, this choice is disadvantageous to \react{} as the difference of the prediction for $\Xi$ from \forge{} is the largest. 
Thus, the estimation of bias in recovered parameters presented in this section is conservative, particularly for \react{}.

\begin{figure*}[h]
\centering
 \includegraphics[width=0.49\linewidth]{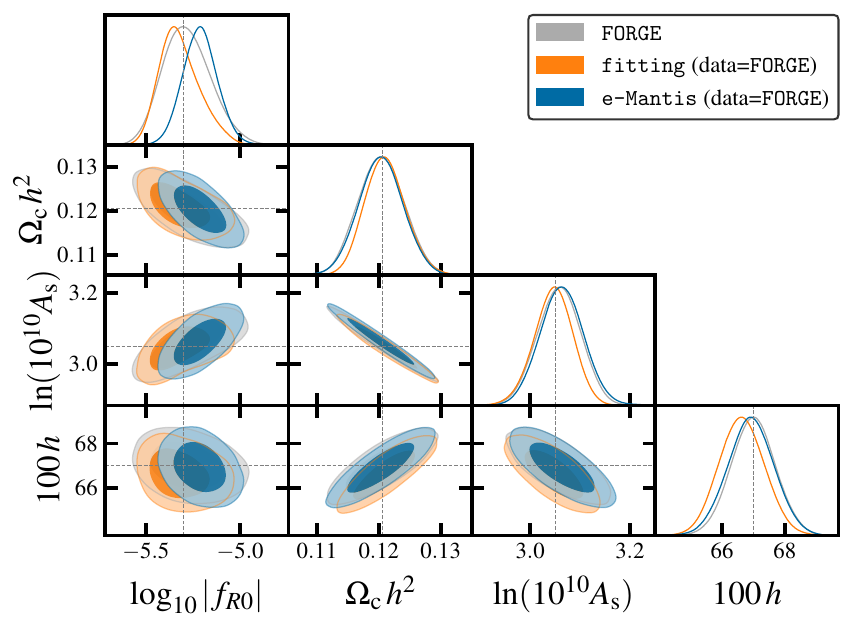}
\includegraphics[width=0.49\linewidth]{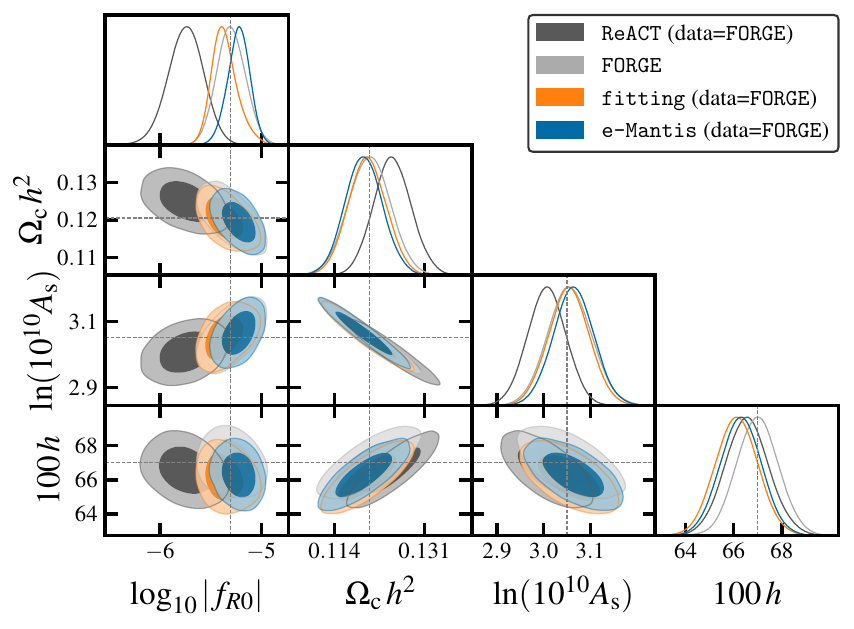}
  \caption{Bias due to different nonlinear modelling. The synthetic data are created by \forge{} and fitted by four different models. \textit{Left panel}: WL optimistic case. \textit{Right panel}: WL pessimistic case.}   \label{fig:errorsforge}
\end{figure*}

\Cref{fig:errorsforge} shows the 2D contours of the constraints on parameters,  
and \Cref{tab:WLconstraints3}
and \Cref{tab:WLconstraints4} 
summarise constraints on $\logfr$ for the optimistic and pessimistic settings. We first start from the optimistic setting. Since \react{} is valid only up to 
$k = 3 h\,{\rm Mpc}^{-1}$, we did not include \react{} in this case. 
The fitting formula and \emantis{} recover the input parameters within $1\sigma$. In the case of \emantis{}, cosmological parameters are well recovered, but there is a slight bias in the recovered $\logfr$. On the other hand, a slight bias appears in $h$ in the case of the fitting formula. This may be attributed to the fact that the fitting formula does not include cosmological parameter dependence in the prediction of $\Xi$. To quantify the bias, we define a 1D bias as 
\begin{equation}
B_{\rm 1D}= \frac{\mu - \mu_{\forge{}}}{\sigma_{\forge{}}}\,,
\end{equation}
where $\mu$ and $\sigma$ are the mean and $1 \sigma$ error computed from the 1D marginalised posteriors. If $\mu > \mu_{\forge{}}$ ($\mu < \mu_{\forge{}}$), we use the upper (lower) $68.3\%$ confidence interval to obtain $\sigma_{\forge{}}$. 
The 1D bias for $\logfr$ is $B_{\rm 1D}=0.273$ and $0.602$ for the fitting formula and \emantis{}, respectively. 

We see a similar result in the pessimistic setting for \emantis{} and the fitting formula. In this case, the input $\logfr$ is well within $1\sigma$ although $h$ is again slightly biased. The 1D bias for $\logfr$ is $B_{\rm 1D}=0.441$ and $0.518$ for the fitting formula and \emantis{}, respectively, while 
the 1D bias for $h$ is 0.988 and 0.812 for the fitting formula and \emantis{}, respectively. 
On the other hand, we find a bias in the recovered $\logfr$ when \react{} is used. As mentioned above, we should bear in mind that the choice of \forge{} as a fiducial is disadvantageous for \react{}. In the case of  \react{}, the difference from \forge{} leads to biases also in the cosmological parameters leading to larger $\Omegac h^2$ and smaller 
$\ln(10^{10} A_{\rm s})$ and $h$. This is due to the different $k$ dependence of $\Xi$ between \react{} and \forge{} and this is compensated by adjusting cosmological parameters as well as $\logfr$ leading to a stronger bias. The 1D bias for $\logfr$ reaches $B_{\rm 1D}=3.11$. We will discuss how to mitigate this bias by correcting the prediction for the fiducial cosmology and including theoretical errors in \cref{sec:theoretical_error}. 

Since the \forge{} prediction deviates from the other three models, we also performed the same analysis using \emantis{} as fiducial data for the pessimistic setting. We show the results in Appendix A. Qualitatively, we obtain consistent results. As we can see from \cref{fig:biasemantis}, the 2D contours overlap in the same way as the case where FORGE is used as fiducial, while the mean of $\logfr$ obtained by \react{} is closer to the input value. However, we still find that the 1D bias is at 3 sigma level due to the smaller errors of \emantis{} compared with \forge{}. 

\begin{table*}[h!]
    \centering
\begin{minipage}[t]{0.45\linewidth}\centering
\caption{Mean, standard deviation, and 68.3\% upper and lower limit of $\logfr$ for the WL optimistic setting }
\begin{tabular}{c c c c c }
\hline
\rowcolor{crisp} \multicolumn{5}{c}{WL optimistic}\\
\rowcolor{crispier} & Mean  & S.d.  & Lower & Upper   \\ 
{\bf \fofrfitting{} }  &  $-5.264$ & 0.116  & $-5.397$   & $-5.160$ \\
{\bf \emantis{}}  & $-5.301$ & 0.088 & $-5.388$ & $-5.214$  \\
{\bf \forge{}}  & $-5.280$ & 0.127  & $-5.419$ & $-5.167$\\
{\bf \fofrfitting{} (Fisher)}  &  $-5.301$ & 0.111   & $-5.412$ & $-5.190$  \\
\end{tabular}
\tablefoot{The same prediction is used to create the data and the model.}
\label{tab:WLconstraints1}
\end{minipage}\hfill%
\begin{minipage}[t]{0.45\linewidth}\centering
\caption{Mean, standard deviation, and 68.3\% upper and lower limit of $\logfr$ for the WL pessimistic setting.}
\label{tab:The parameters 2 }
\begin{tabular}{ c c c c c }
\hline
\rowcolor{crisp} \multicolumn{5}{c}{WL pessimistic}\\
\rowcolor{crispier} & Mean  & S.d.  & Lower & Upper   \\ 
{\bf \fofrfitting{} }  &  $-5.252$ & 0.144 & $-5.419$  & $-5.130$  \\
{\bf \emantis{}}  & $-5.318$ & 0.127 & $-5.430$ & $-5.186$ \\
{\bf \forge{}}  & $-5.296$ & 0.143 & $-5.441$ & $-5.159$\\
{\bf \react{}}  & $-5.305$ & 0.216 & $-5.516$ & $-5.095$
\end{tabular}
\tablefoot{The same prediction is used to create the data and the model.}
\label{tab:WLconstraints2}
\end{minipage}
\end{table*}

\begin{table*}[h!]
\centering
\begin{minipage}[t]{0.46\linewidth}\centering
\caption{Mean, standard deviation, and 68.3\% lower and upper limit of $\logfr$ for the optimistic setting.}
\begin{tabular}{ c c c c c c  }
\hline
\rowcolor{crisp} \multicolumn{6}{c}{WL optimistic}\\
\rowcolor{crispier}  & Mean  & S.d.  & Lower & Upper & $B_{\rm 1D}$ \\ 
\hline 
\rowcolor{gray} \multicolumn{6}{l}{Data = \forge{}}\\
{\bf \fofrfitting{} }  &  $-5.318$ & 0.108 & $-5.443$  & $-5.236$ & 0.273  \\
{\bf \emantis{}}  & $-5.212$ & 0.092 & $-5.302$  & $-5.121$ & 0.602\\
\end{tabular}
\tablefoot{The data are generated by \forge{} and it is fitted by different nonlinear models.}
\label{tab:WLconstraints3}
\end{minipage}\hfill%
\begin{minipage}[t]{0.46\linewidth}\centering
\caption{Mean, standard deviation, and 68.3\% lower and upper limit of $\logfr$ for the pessimistic setting.}
\label{tab:The parameters 3 }
\begin{tabular}{ c c c c c c  }
\hline
\rowcolor{crisp} \multicolumn{6}{c}{WL pessimistic}\\
\rowcolor{crispier}  & Mean  & S.d.  & Lower & Upper & $B_{\rm 1D}$ \\ 
\hline 
\rowcolor{gray} \multicolumn{6}{l}{Data = \forge{}}\\
{\bf \fofrfitting{} }  &  $-5.360$ & 0.129 & $-5.503$ & $-5.258$ & 0.441  \\
{\bf \emantis{}}  & $-5.225$ & 0.111  & $-5.326$ & $-5.109$ & 0.518\\
{\bf \react{}}  & $-5.747$ & 0.178  & $-5.921$ & $-5.572$ & 3.110 \\
\end{tabular}
\tablefoot{The data are generated by \forge{} and it is fitted by different nonlinear models.}
\label{tab:WLconstraints4}
\end{minipage}
\end{table*}

\subsubsection{\texorpdfstring{\threextwo{}}{3x2pt}  analysis}
We considered the constraints from \threextwo{} statistics by adding \GCph{} and its cross-correlation \XCph{} to WL. In this paper, we use a simple model of the scale-independent linear bias as in \lcdm{}. This assumption needs to be reexamined in $f(R)$ gravity as the scale-dependent growth will lead to a scale-dependent bias. The effect of $f(R)$ gravity on the halo bias was studied in simulations \citep{Arnold:2018nmv} and the anlytic model was developed by \cite{Valogiannis:2019xed}. However, the linear bias assumption also needs to be relaxed even in \lcdm{} and it is beyond the scope of this paper to implement a more complete bias description. For this reason, we consider the pessimistic case only and focus our attention on a comparison with the Fisher Matrix forecast and check if the bias when \forge{} is used to create the data becomes worse with the increasing statistical power. We also do not include \react{} as we already observe a significant bias with WL only. 

We have ten scale-invariant bias parameters, one for each redshift bin. For the \threextwo{} analysis, we vary $n_{\rm s}$, but we still impose a tight Gaussian prior on $\Omegab\,h^2$
as it is unlikely to get a tighter constraint than the Big Bang nucleosynthesis constraints.

The parameters used in addition to the WL analysis are listed in \Cref{tab:MCMCparam2}.
We did not change the fiducial bias parameters from \citetalias{Blanchard:2019oqi}  as there is no prediction of the linear bias in  $f(R)$ gravity. The fiducial bias adopted in \citetalias{Blanchard:2019oqi} will need to be improved even in \lcdm. Our prime focus is the study of the effect of nonlinear models and we vary the linear bias in the MCMC analysis resulting in different constraints depending on the nonlinear model used. The observed covariance was built from synthetic data. Thus the effect of $f(R)$ gravity was included in the covariance. We independently checked that the constraints on parameters do not change by using the \lcdm~covariance. Thus we expect that the effect of $f(R)$ on galaxy bias has also negligible effects on the covariance. 

\begin{figure*}
\centering
 \includegraphics[width=0.49\linewidth]{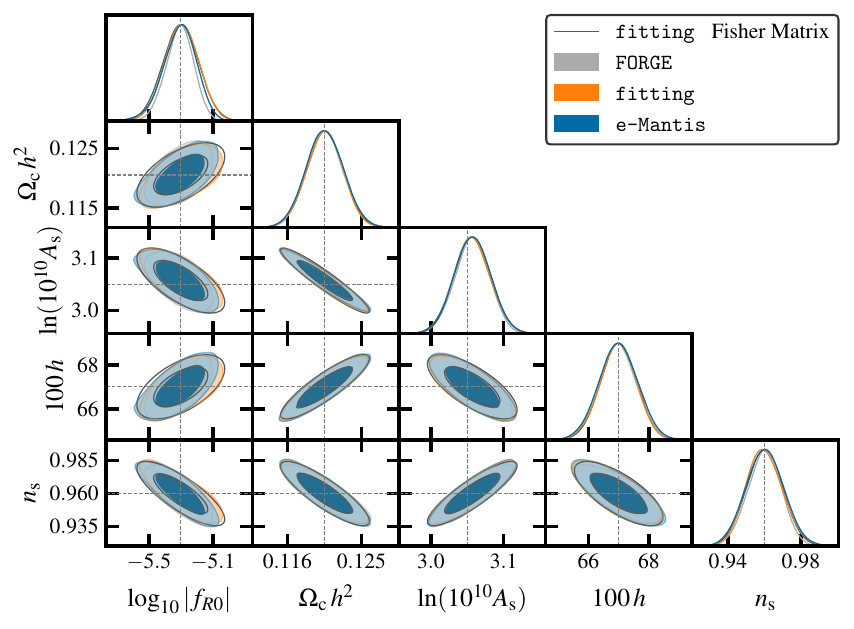}
 \includegraphics[width=0.49\linewidth]{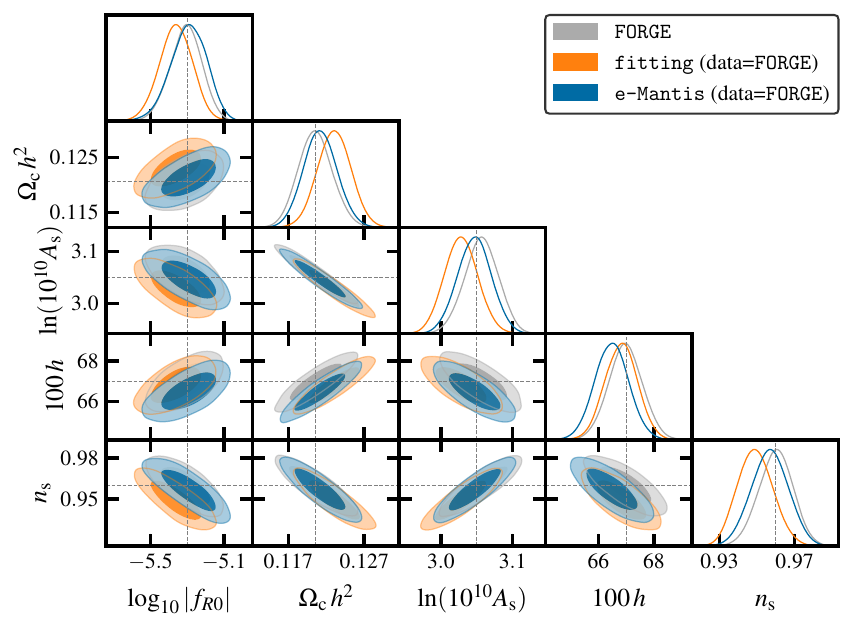}
  \caption{Constraints from \threextwo{} analysis in the pessimistic setting. In the left panel, the same nonlinear model is used for the data and the fitting while in the right panel, the data are generated by \forge{}. }
   \label{fig:errors32}
\end{figure*}

The left-hand side of \cref{fig:errors32} shows the 2D contours of the constraints on parameters, and \Cref{tab:3x2constraints1}
summarises constraints on $\logfr$ where the same nonlinear model is used for the data and the model. 
The right-hand side of \cref{fig:errors32} shows the 2D contours of the constraints on parameters, and \Cref{tab:3x2constraints2}
summarises constraints on $\logfr$ as well as 1D bias for $\logfr$ when \forge{} is used to create the data. 

In the case where the same nonlinear model is used for the data and the model, errors are consistent between the fitting formula and \emantis{}, although we still observe the same longer tail for larger $\logfr$ for the fitting formula as we observe for WL. \forge{} gives slightly tighter constraints on $\logfr$. Constraints on cosmological parameters are very consistent among the three different nonlinear models. The Fisher Matrix forecast 
using the fitting formula is also very consistent with the MCMC results. We note that errors from \threextwo{} analysis in the pessimistic setting are comparable to or better than WL alone in the optimistic setting. This demonstrates the strength of the \threextwo{} analysis although we should bear in mind the limitation of the bias model used in this analysis. 

We find that the increased statistical power does not degrade the agreement between \forge{} and \emantis{}. 
When \forge{} is used to create the data, the input parameters are well recovered by \emantis{}, although $h$ is slightly biased as in the WL-only case. The 1D bias for $\logfr$ is given by 0.150. In the case of the fitting formula, the bias in cosmological parameters becomes worse compared with WL. This could be attributed to the fact that the fitting formula does not take into account the cosmological parameter dependence of $\Xi$ as mentioned before. The 1D bias for $\logfr$ also becomes slightly larger, with a value of $B_{\rm 1D} =0.667$. 

\begin{table*}
\caption{Additional parameters that are varied in the \threextwo{} analysis.}
\centering
\begin{tabular}{ c c c c c c c c c c c }
\hline  
$n_{\rm s}$ & $b_1$ & $b_2$  & $b_3$ & $b_4$ & 
$b_5$ & $b_6$  & $b_7$ & $b_8$ & 
$b_9$ & $b_{10}$ \\ 
\hline 
\rowcolor{gray} \multicolumn{11}{l}{Fiducial}\\
0.96 & 1.0998  & 1.2202 & 1.2724 & 1.3166 & 1.3581 & 1.3998 & 1.4446 &1.4965 &1.5652 &1.7430 \\
\hline
\rowcolor{gray} \multicolumn{11}{l}{Prior}\\
$[0.8, 1.2]$  & N.A.& N.A.& N.A.& N.A.& N.A.& N.A.& N.A.& N.A.& N.A.& N.A.
\end{tabular}
\label{tab:MCMCparam2}
\end{table*}

\begin{table*}
    \centering
\begin{minipage}[t]{0.48\linewidth}\centering
\caption{Mean, standard deviation, and 68.3\% lower and upper limit of $\logfr$ for the \threextwo{} pessimistic setting.}
\begin{tabular}{c c c c c }
\hline
\rowcolor{crisp} \multicolumn{5}{c}{\threextwo{} pessimistic}\\
\rowcolor{crispier} & Mean  & S.d.  & Lower & Upper   \\ 
{\bf \fofrfitting{}}  &  $-5.297$ & 0.107 & $-5.404$  & $-5.191$ \\
{\bf \emantis{}}  & $-5.309$ & 0.107 & $-5.405$ & $-5.198$ \\
{\bf \forge{}}  & $-5.308$ & 0.087 & $-5.389$ & $-5.215$\\
{\bf \fofrfitting{} Fisher}  &  $-5.301$ & 0.112 & $-5.413$ & $-5.189$
\end{tabular}
\tablefoot{The same prediction is used to create the data and the model.}
\label{tab:3x2constraints1}
\end{minipage}\hfill%
\begin{minipage}[t]{0.48\linewidth}\centering
\caption{Mean, standard deviation, and 68.3\% lower and upper limit of $\logfr$ for the  \threextwo{} pessimistic setting}
\label{tab:The parameters 4 }
\begin{tabular}{c c c c c c}
\hline
\rowcolor{crisp} \multicolumn{6}{c}{\threextwo{} pessimistic}\\
\rowcolor{crispier} & Mean  & S.d.  & Lower & Upper    &$B_{\rm 1D}$ \\
 \hline
\rowcolor{gray} \multicolumn{6}{l}{Data = \forge{}}\\
{\bf \fofrfitting{} }  &  $-5.362$ & 0.091  & $-5.451$ & $-5.271$ & 0.667  \\
{\bf \emantis{}}  & $-5.294$ & 0.099 & $-5.382$ & $-5.187$ & 0.151\\
\end{tabular}
\tablefoot{The data are generated by \forge{} and it is fitted by different nonlinear models.}
\label{tab:3x2constraints2}
\end{minipage}
\end{table*}

\section{Baryonic effects} 
\label{sec:baryonic_effects}
\begin{figure*}[h]
\centering
 \includegraphics[width=0.9\linewidth]{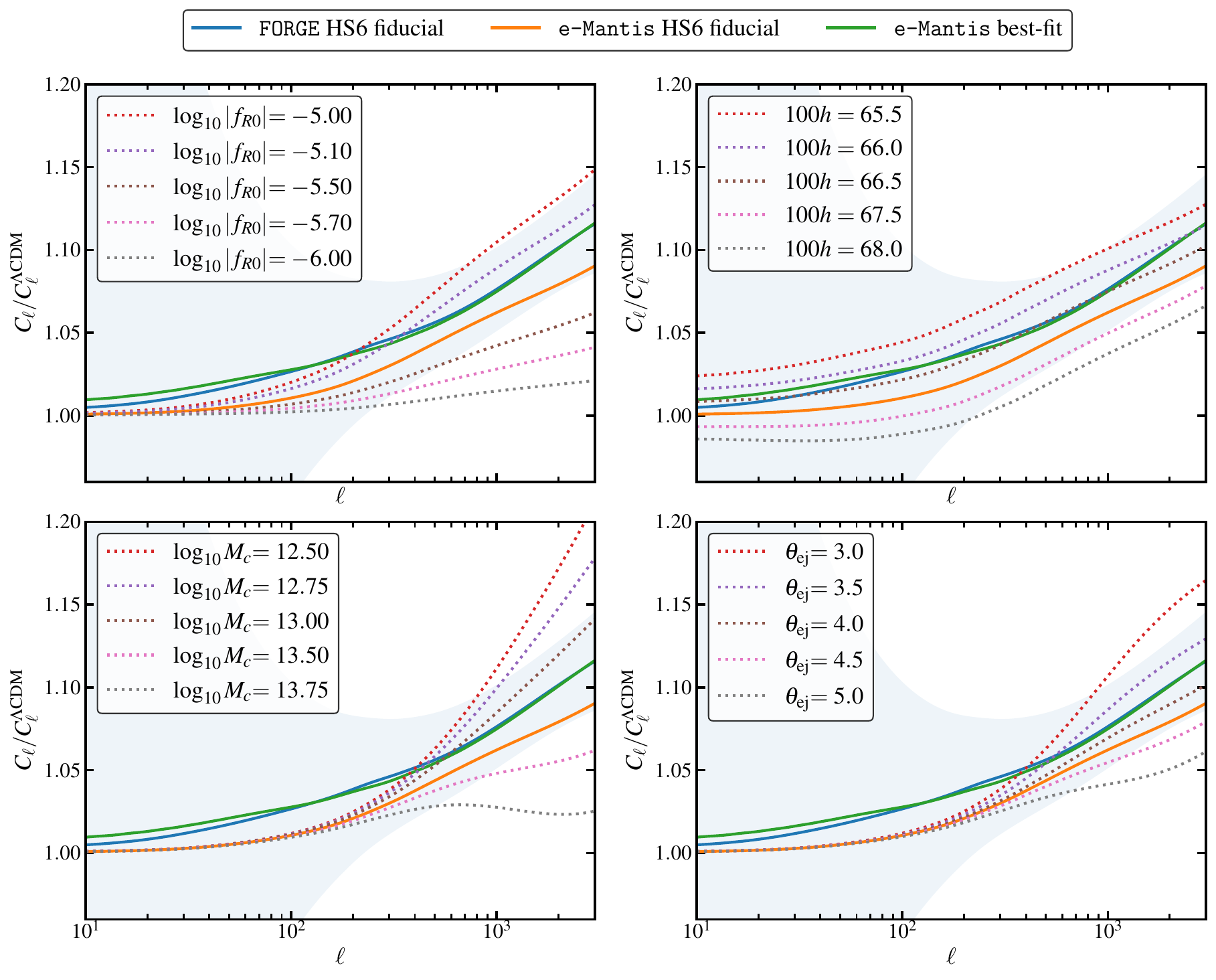}
  \caption{Effect of $\logfr$, two baryonic parameters $\logten M_{\rm c}$ and $\theta_{\rm ej}$, and $h$ on the WL angular power spectrum in bin 10. The dotted lines show predictions of \emantis{} with varying these parameters individually. The plot shows the ratio to the \lcdm \ power spectrum. The plot includes the prediction of \forge{} and \emantis{} with the fiducial parameters and the \emantis{} prediction using the best-fit values 
  ($\logfr=-5.765, 100 h= 65.736,  \logten(M_{\rm c})=12.999$)
with the data created by \forge{}. We can see that the \emantis{} prediction can fit the \forge{} by adjusting these parameters. We note that the cosmological parameters for the \lcdm \ power spectrum, $C_{\ell}^{\lcdm}$ is fixed to be the fiducial ones when we vary $100  h$. }
   \label{fig:baryoneffect}
\end{figure*}

\subsection{Adding baryonic effects using \bcemu{}}
In this section, we study the impact of baryonic effects on the constraints on the $f(R)$ parameter. We use the seven-parameter emulator of baryonic effects called \bcemu{} \citep{Giri_2021} and we assume that the baryonic effects and the modified gravity effects can be treated independently as discussed in \cref{sec:MP} \citep{Arnold:2019zup}. The \bcemu{} parameters govern the gas profiles and stellar abundances in haloes. It is not our intention to study in detail the baryonic effects. We use the default values for the baryonic parameters provided with \bcemu{} and use the full prior range. We choose not to include any redshift dependence. This is because we need to impose a tight prior range for these parameters at $z=0$ to keep them within the prior range and we may miss important degeneracies between baryonic parameters and $\logfr$. 
We found that two baryonic parameters, $M_{\rm c}^{\rm Giri~et. al.}$ and $\theta_{\rm ej}$, have the strongest effects on the matter power spectrum and they are well constrained in the presence of the $f(R)$ parameter. Thus, we only vary these two parameters.
In \bcemu{}, the gas profile is modelled as a cored double power law. $M_{\rm c}^{\rm Giri~et.al.} $ controls the dark matter halo mass dependence of the logarithmic slope of the first-cored power law. It allows the profile to become less steep than the \citelist{Navarro_96} one for $M<M_{\rm c}^{\rm Giri~et.al.}$. The parameter $\theta_{\rm ej}$ determines the scale radius (with respect to the virial radius) of the second-cored power law. \Cref{tab:MCMCparambaryon} summarises the fiducial values and priors for baryonic parameters. We will use the dimensionless parameter $M_{\rm c}$ that is related to the one defined in \cite{Giri_2021} via $M_{\rm c} \coloneqq M_{\rm c}^{\rm Giri~et.al.} / M_{\odot}$ where $M_{\odot}$ is the solar mass.

\begin{table*}
\centering
\caption{Fiducial values and priors for \bcemu{} parameters. }
\begin{tabular}{ c c c c c c c c }
\hline 
 $\logten M_{\rm c}$  & $\theta_{\rm ej}$ & $f_{\rm b}$  & $\mu$ & $\gamma$ & $\delta$ & $\eta$ & $\eta_{\delta}$  \\
\hline
\rowcolor{gray} \multicolumn{8}{l}{Fiducial}\\
 13.32 & 4.235  & 0.186 & 0.93 & 2.25 & 6.40 &0.15 & 0.14  \\
\hline
\rowcolor{gray} \multicolumn{8}{l}{Prior}\\
 $[11, 15]$  & [2 8] &[0.10, 0.25] & N.A. & N.A. & N.A. & N.A. & N.A.
\end{tabular}
\label{tab:MCMCparambaryon}
\end{table*}

\begin{figure*}[h!]
\centering
 \includegraphics[width=0.49\linewidth]{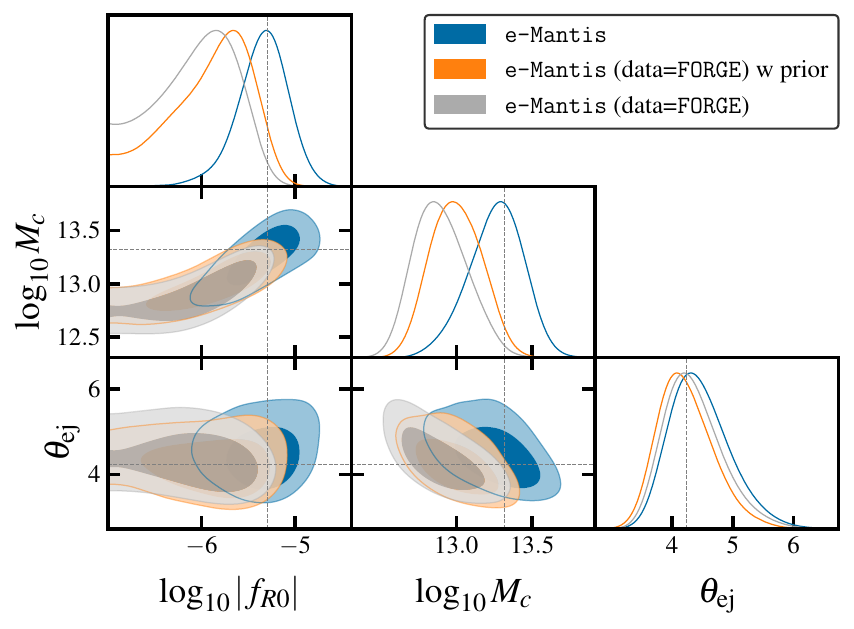}
 \includegraphics[width=0.49\linewidth]{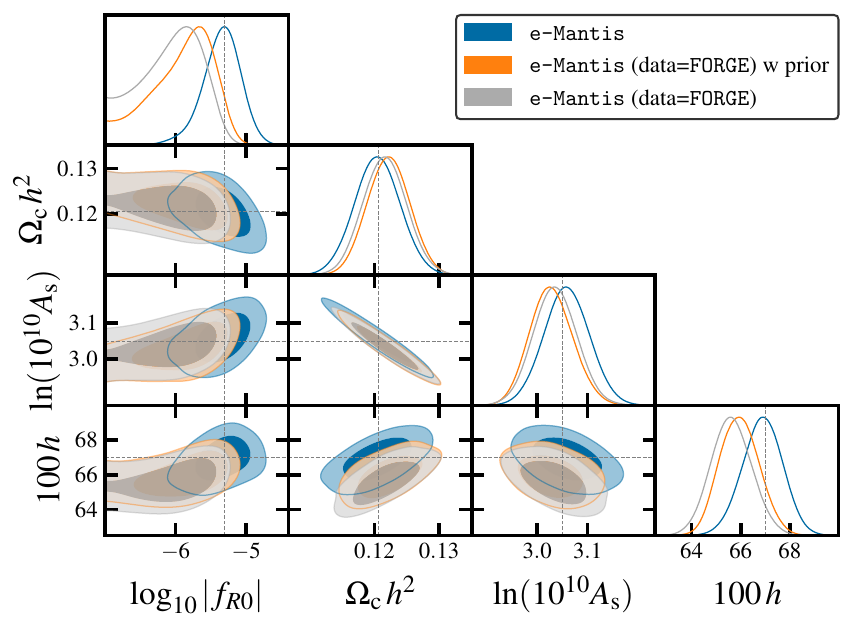}
  \caption{Constraints on parameters from the WL optimistic setting with baryons. We show the case where \emantis{} is used to create both the data and fitting and the case where \forge{} is used to create the data and it is fitted by \emantis{}. We also include the case where we impose a Gaussian prior on $\logten M_{\rm c}$ with the width of 0.2 in the case of \forge{} fiducial data.
  \textit{Left panel}: Constraints on baryonic parameters. \textit{Right panel}: Constraints on cosmological parameters.}
   \label{fig:optbar}
\end{figure*}

\begin{figure}[h!]
\centering \includegraphics[width=1.0\linewidth]{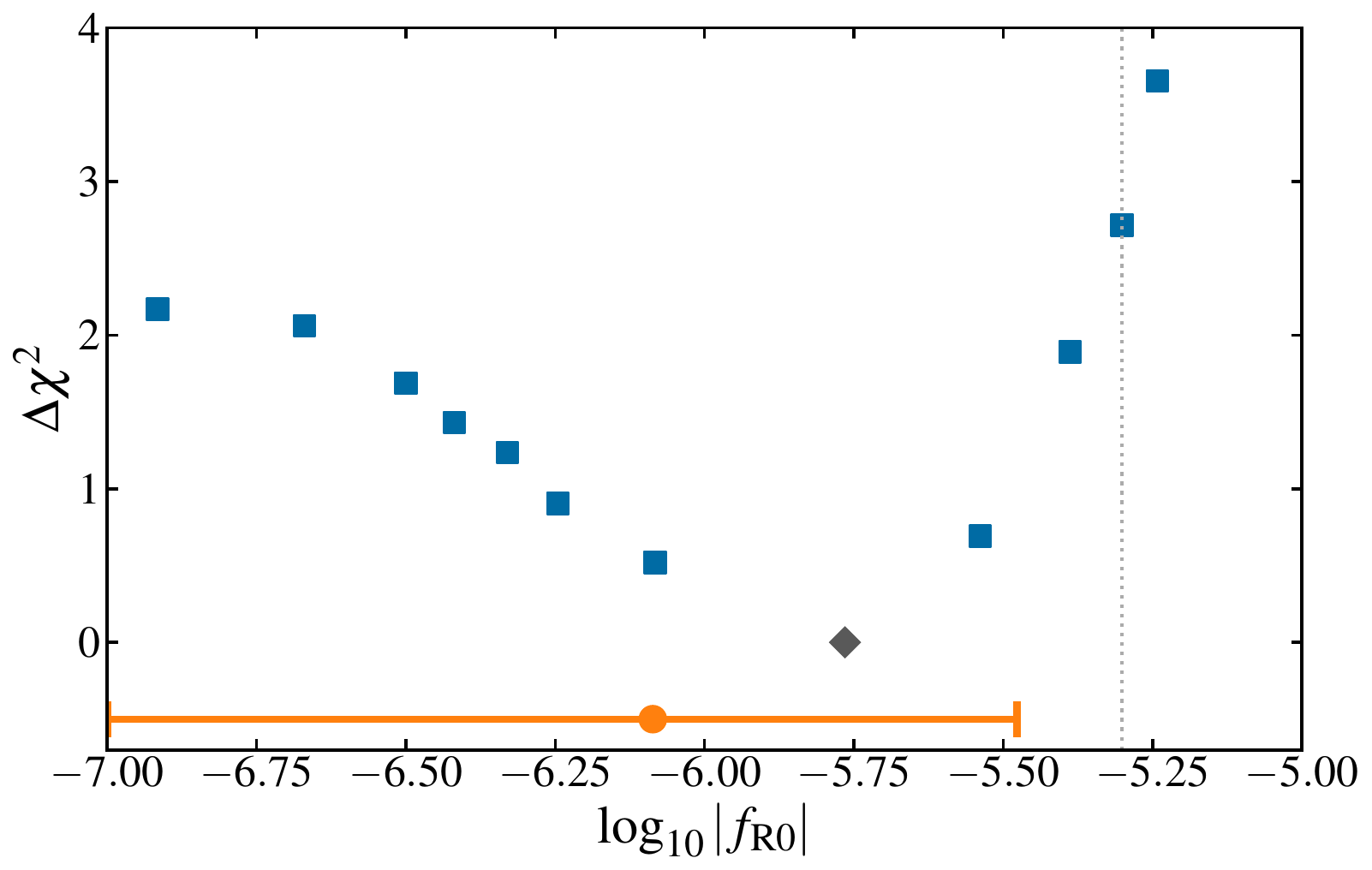}
  \caption{Profile likelihood for $\logfr$ where \forge{} is used to create data fitted by \emantis{} with baryons in the optimistic setting of WL. The square points show $\Delta \chi^2$ from the best-fit indicated by a diamond. The $95.5 \%$ confidence interval from the 1D marginalised posterior is shown as the error bar. The vertical dotted line indicates the input value of $-\logfr = 5.301$. We see that the 1D marginalised posterior is shifted to smaller values of $\logfr$.} 
   \label{fig:profile}
\end{figure}

\begin{figure*}[h!]
\centering
 \includegraphics[width=0.49\linewidth]{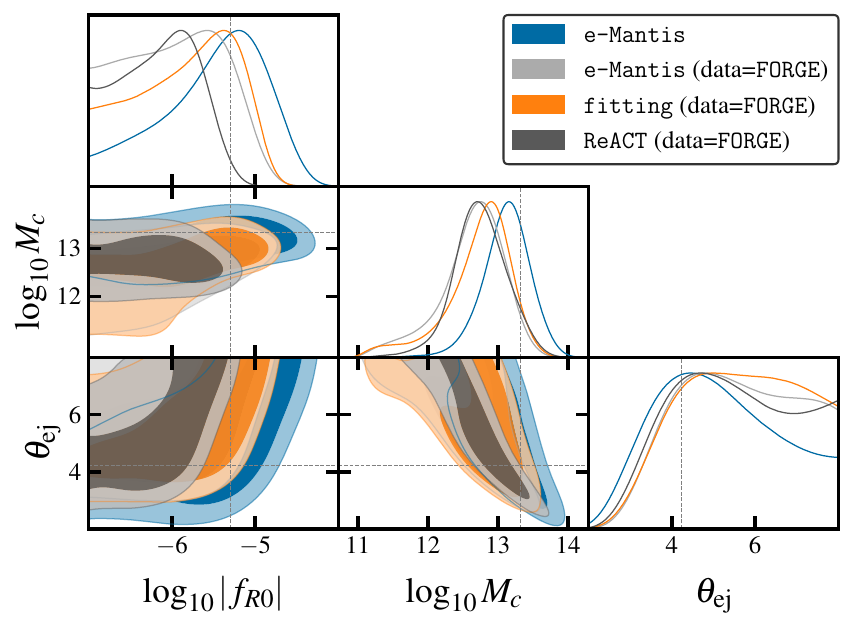}
 \includegraphics[width=0.49\linewidth]{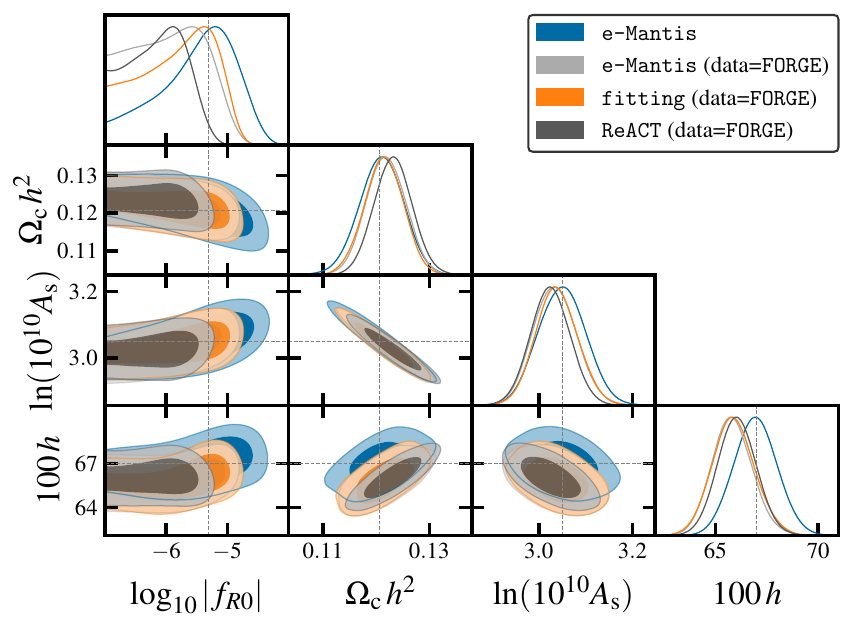}
  \caption{Constraints on parameters from the WL pessimistic setting with baryons. As in \cref{fig:optbar}, we show the case where \emantis{} is used to create both the data and fitting as well as the case where \forge{} is used to create the data and it is fitted by \emantis{}, \react{} and the fitting formula. 
  \textit{Left panel}: Constraints on baryonic parameters. \textit{Right panel}: Constraints on cosmological parameters.}
   \label{fig:pesbar}
\end{figure*}

\begin{table}
\caption{Mean, standard deviation, and 68.3\% lower and upper limit of $\logfr$ for the WL optimistic setting with baryons.}
\begin{tabular}{ c c c c c }
\hline
\rowcolor{crisp} \multicolumn{5}{c}{WL optimistic}\\
\rowcolor{crispier} & Mean  & S.d.  & Lower & Upper   \\

{\bf \emantis{}}  &  $-5.353$ & 0.281 & $-5.586$ & $-5.058$ \\
\hline
\rowcolor{gray} \multicolumn{5}{l}{Data = \forge{}}\\
{\bf \emantis{}}  & $-6.087$ & 0.424  & $-6.417$ & $-5.536$ \\
\end{tabular}
\hfill%
\label{tab:WLbaryonopt}
\end{table}

\begin{table}
\caption{
Mean, standard deviation, and 68.3\% lower and upper limit of $\logfr$ for the WL pessimistic setting with baryons.}
\begin{tabular}{ c c c c c }
\hline  
\rowcolor{crisp} \multicolumn{5}{c}{WL pessimistic}\\
\rowcolor{crispier} & Mean  & S.d.  & Lower & Upper   \\

{\bf \emantis{}}  &  $-5.482$ & 0.615 & $-5.909$ & $-4.699$ \\
\hline
\rowcolor{gray} \multicolumn{5}{l}{Data = \forge{}}\\
{\bf \emantis{}}  & $-5.951$ & 0.575 & $-6.549$ & $-5.262$\\
{\bf fitting}  & $-5.830$ & 0.586 & $-6.305$ & $-5.045$\\
{\bf \react{}}  & $-6.147$ & 0.449  & $-6.581$ & $-5.593$ \\

\end{tabular}
\label{tab:WLbaryonpes}
\end{table}

\subsection{Effects of adding baryons}
\Cref{fig:baryoneffect} shows the effect of changing $\logfr$, $h$, and two baryonic parameters on the ratio of the WL angular power spectrum to the \lcdm\ one. Baryonic effects introduce scale-dependent modifications to the angular power spectrum that are similar to the effect of $f(R)$ at $\ell > 100$. On the other hand, $h$ changes the overall amplitude of the angular power spectrum as it changes $\Omegam$. We will see that the interplay between these parameters leads to interesting degeneracies. 

\Cref{fig:optbar} shows the 2D contours of the WL constraints on parameters for the optimistic settings with baryons and \Cref{tab:WLbaryonopt} summarises constraints on $\logfr$.  In the case of the optimistic setting, when \emantis{} is used as the data and the model, the two baryonic parameters degrade the constraints on $\logfr$: the $1\sigma$ error becomes 0.281 while it was 0.088 without baryons. However, it is important to stress that $\logfr$ and the two baryonic parameters are still constrained well within the prior range of these parameters. It means that the impact of modified gravity on the high-$k$ tail of the matter power spectrum is not washed out by the baryonic feedback, and we can still distinguish between the effect of $f(R)$ and baryons. 

Nevertheless, if \forge{} is used as the data, the difference between \forge{} and \emantis{} in the matter power spectrum can be absorbed by the shift in $\logten M_{\rm c}$ and $h$, and the constraint on $\logfr$ is shifted along the degeneracy direction between $\logfr$ and $\logten M_{\rm c}$ as well as $h$. This can be understood from \cref{fig:baryoneffect}. The difference of the scale dependence between \emantis{} and \forge{} at $\ell >100$ can be adjusted by decreasing $\logfr$ and decreasing $\logten M_{\rm c}$. This leads to a lower amplitude that can be adjusted by decreasing $h$. Due to the combination of these effects, the best-fit $\logfr$ becomes smaller. We note that 
\emantis{} does not include the $h$ dependence in the prediction for $\Xi(k ,z)$ explicitly, but it depends on $h$ implicitly through $\sigma_8^{\lcdm}$. However, this dependence of $h$ on $\Xi(k,z)$ is very weak. To compute the power spectrum in $f(R)$ we use 
$P_{f(R)}
= \Xi(k,z) {P_{\lcdm}(k,z)}$. 
The \emantis{} emulator provides us with the first factor, $\Xi(k,z)$, and a \lcdm{} emulator provides the last factor. The $h$ dependence comes from the \lcdm{} power spectrum not from $\Xi(k,z)$.

As a result, we obtain a 95.5\% upper limit of $\logfr$ as $\logfr < -5.477$, which is incompatible with the input value of $\logfr = -5.301$. This is partly due to prior volume effects shifting the contour to lower values of $\logfr$, caused by the strong degeneracy between $\logfr$ and $\logten M_{\rm c}$, as well as $h$.  

To confirm this, we obtain a profile likelihood for $\logfr$ by fixing $\logfr$ and finding a minimum $\chi^2$ by varying other parameters. The profile likelihood was obtained by \texttt{PROSPECT} 
\citep[\href{https://github.com/AarhusCosmology/prospect_public}{\faicon{github}}]{Holm:2023uwa}. This is shown in \cref{fig:profile}.  
The global best-fit value for $\logfr$ is obtained as $\logfr =-5.765$, which is  larger than the mean of the 1D marginalised posterior $\logfr =-6.087$. \Cref{fig:baryoneffect} also shows the prediction of  \emantis{} with the best-fit values, which agrees well with \forge{}. The $\Delta \chi^2$ for the input value of $\logfr = -5.301$ is found as $\Delta \chi^2 = 2.657$, thus it is still within $2 \sigma$. We also observe that the $\Delta \chi^2$ curve becomes flat for smaller values of $\logfr$. 

We also tested the fitting formula as a model to fit to the data generated by \forge{}. In this case, the posterior is highly non-Gaussian and the chains did not converge well. 

This implies that we can break the degeneracy between $f(R)$ gravity and baryonic effects, but we need to have an accurate nonlinear model for the $f(R)$ gravity model. Otherwise, we could obtain a significantly biased result in terms of the 1D marginalised constraint. We note that if we have a physical prior on $\logten M_{\rm c}$ from baryonic physics, we could break the degeneracy between $\logfr$ and $\logten M_{\rm c}$. For example, in the case where \forge{} was used as the data and they were modelled by \emantis{}, the mean of $\logten M_{\rm c}$ is significantly shifted to a smaller value as we can see in \cref{fig:optbar}. If we had a prior on $\logten M_{\rm c}$ to prevent this, this bias could be avoided. 
To test this idea, we also ran an analysis imposing a prior on $\logten M_{\rm c}$ and showed the result in \cref{fig:optbar}. We used a Gaussian prior with the width of $0.2$, which was estimated from the WL informed gas and stellar mass fraction measurements of massive haloes by \cite{Grandis:2023qwx}. This prior is consistent with the error on $\logten M_{\rm c}$ that we obtained by using \emantis{}  both for data and model. We observe that the bias is relaxed slightly; however, the degeneracy between $\logten M_{\rm c}$ and $\logfr$ is quite strong and this prior is not enough to alleviate the bias in the marginalised constraints. An improved prior from external data is needed to fully break this degeneracy.

We assumed that the effect of $f(R)$ gravity and baryonic effect can be treated independently. 
Recently, the coupling between baryonic feedback and cosmology has been studied showing that the combined effect of baryonic and non-baryonic suppression mechanisms is greater than the sum of its parts for decaying dark matter \citep{Elbers:2024dad}. The effect of this coupling on the degeneracy needs to be studied in the future.

Next, we consider the pessimistic setup. \Cref{fig:pesbar} shows the 2D contours of the WL constraints on parameters for the optimistic settings with baryons and \Cref{tab:WLbaryonpes} summarises constraints on $\logfr$. In this case, due to the weaker constraining power, the $99.7\%$ confidence level upper bound on $\logfr$ reaches the prior boundary of $\logfr=-7$. Also, the $95.5\%$ confidence level upper bound of $\theta_{\rm ej}$ is bounded by the prior. If the data are generated by \forge{}, strong degeneracies appear among $\logfr$ and two baryonic parameters. The input $\logfr$ is consistent within $1\sigma$ for the fitting formula and \emantis{} although the prior bound  $\logfr=-7$ is reached at 95.5\% confidence level lower bound. On the other hand, for \react{}, the input value is outside the 95.5\% confidence level upper bound. As we already observed without baryons, $h$ is biased to a lower value. 

We also checked the case where fiducial data is generated by \emantis{}. As we can see in \cref{fig:biasemantisbaryons}, unlike the case without baryons, the agreements between \emantis{}, \fofrfitting{} and \react{} are improved significantly compared with the case where \forge{} is used for fiducial data. Particularly, for \react{}, the mean of $\logfr$ is now consistent with the input value within $1\sigma$. 

\subsection{Constraints on \texorpdfstring{$|f_{R0}|$}{fR0} with the \lcdm \ data}
\begin{figure*}[h!]
\centering
 \includegraphics[width=0.49\linewidth]{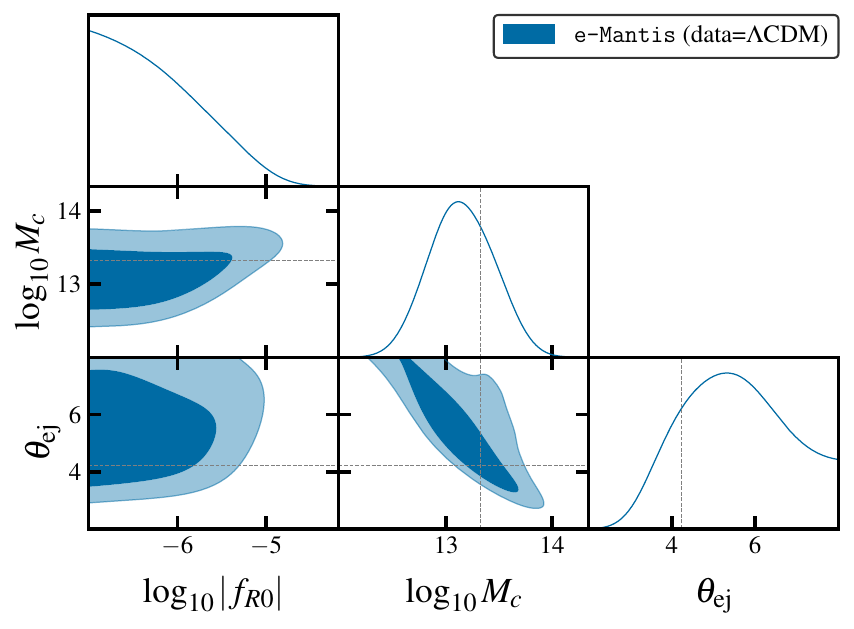}
 \includegraphics[width=0.49\linewidth]{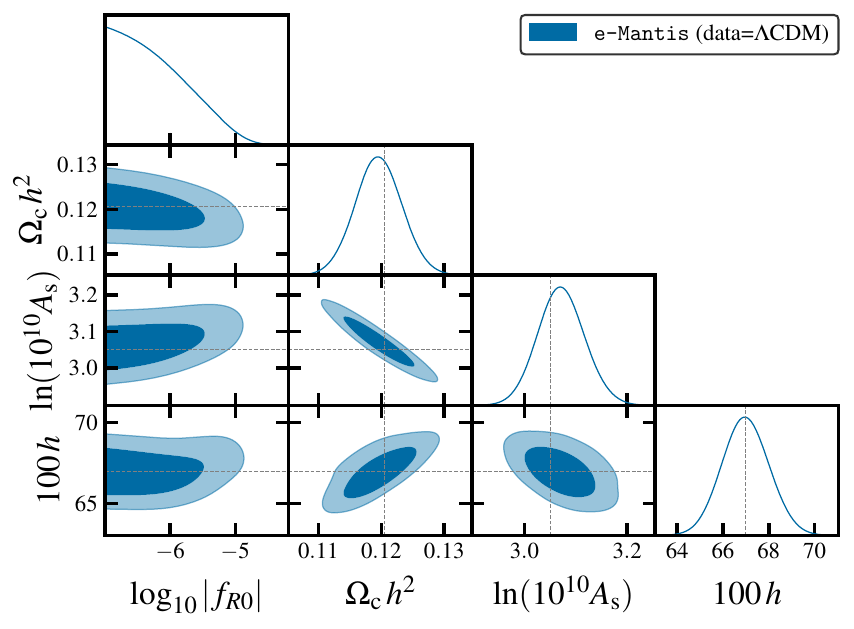}
  \caption{Constraints on parameters from the WL pessimistic setting with baryons where the data are created in \lcdm\ and it is fitted by \emantis{}. \textit{Left panel}: Constraints on baryonic parameters. \textit{Right panel}: Constraints on cosmological parameters. }
   \label{fig:LCDM1}
\end{figure*}

Based on these results, we consider the pessimistic setting and use \emantis{} to obtain an upper limit of $\logfr$ in the presence of baryonic effects to be conservative. \Cref{fig:LCDM1} shows the 2D contours of the WL constraints on parameters for the pessimistic settings where \lcdm\ is used to create the data. 
We find that the recovered cosmological parameters are consistent with the input values. The mean of 1D marginalised constraints of baryonic parameters are slightly biased due to the strong degeneracy between them, but it is still consistent within $1 \sigma$. We obtain the $95.5\%$ confidence level upper bound on $\logfr$ as 
\begin{equation}
\logfr < - 5.21\,.
\end{equation}
We note that this bound depends on the prior $\logfr=-7$. To obtain the prior-independent bound, we follow the approach presented in \citelist{Gordon:2007xm, Piga:2022mge} and \cite{Kou:2023gyc}. 
The ratio of the marginalised posterior and prior is given by 
\begin{equation}
    b(x \mid d, p) = \frac{\mathcal{P}(x \mid d, p)}{p(x)}\,,
\end{equation}
where $x$ is the parameter we are interested in (i.e. $\logfr$), $d$ is the data, $p$ is the prior, and $\mathcal{P}$ is the posterior. The Bayes factor, $B(x_1, x_2)$, which quantifies the support of the models with $x=x_1$ over the models with $x=x_2$, is given by 
\begin{equation}
B(x_1, x_2) = \frac{ b(x_1 \mid d, p) }{b(x_2 \mid d, p)} 
= \frac{ \mathcal{L}(d \mid x_1)}{\mathcal{L}(d\mid x_2)}, 
\end{equation}
where $\mathcal{L}(d\mid x)$ is the marginalised likelihood of the data for parameter $x$. 
For our purpose, we chose $x = \logfr$ and fixed $x_1$ to be the upper bound of the prior, $x_1 = -7$. 
Following \citelist{Gordon:2007xm}, we then used $B(x_1, x_2) =2.5$  to find $x_2$ so that the model with $x=x_1$ is favoured compared to the model with $x=x_2$ at $95.5\%$ confidence level. This gives the $95.5\%$ confidence interval of $\logfr$ that does not depend on the prior. We note that this method applies only to a $\mathcal{L}(d\mid x)$ that is a monotonic function of $x$. Employing this technique, we obtain
\begin{equation}
\logfr < - 5.58\,.
\end{equation}

\begin{figure}
\centering \includegraphics[width=1.0\linewidth]{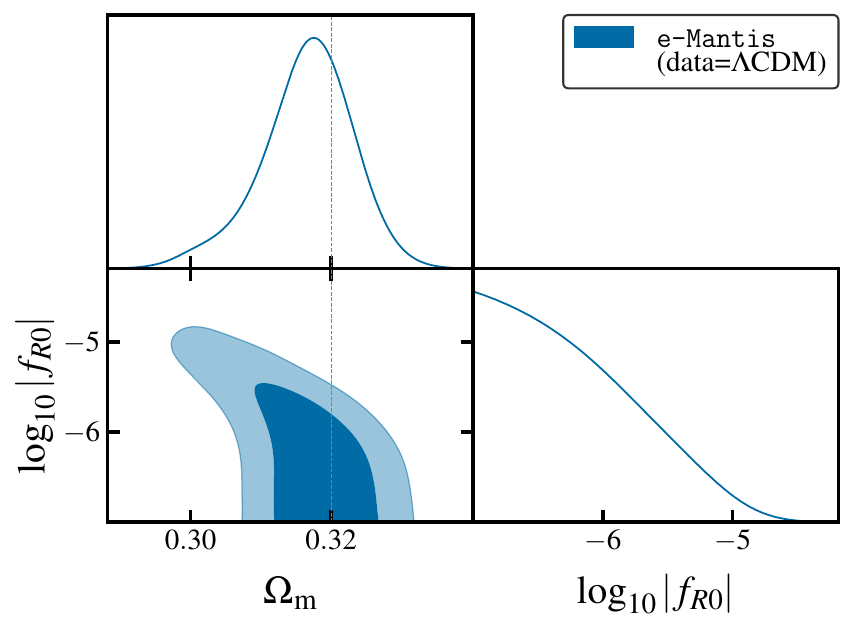}
  \caption{Same as in \cref{fig:LCDM1}, but we show the constraint on the derived quantity $\Omegam$.} 
   \label{fig:LCDM2}
\end{figure}

Finally, in \cref{fig:LCDM2}, we present the constraints on $\Omegam$ and  $\logfr$ so that we can compare these with those in the literature \citep{Schneider:2019xpf, Harnois-Deraps:2022bie, SpurioMancini:2023mpt}. It is not possible to make a direct comparison due to various differences in the setting. Nonetheless, our constraint is comparable to the one presented in \citelist{Schneider:2019xpf} [$\logfr < -5.3$] where a similar analysis was done by combining \bcemu{} and the fitting formula. We cannot also make a comparison with \citelist{Harnois-Deraps:2022bie} as baryonic effects are not included in their analysis, but again the constraint is comparable [$\logfr < -5.24$]. 
On the other hand, our constraint is much weaker than the one found in \citelist{SpurioMancini:2023mpt} that used \react{} with \bcemu{}. We refer the readers to \citelist{SpurioMancini:2023mpt} for possible explanations. In our analysis, \react{} tended to prefer a lower value of $\logfr$ when \forge{} is used as data and this might be one of the reasons for this difference. Finally, a recent study by \citelist{Tsedrik:2024cdi} used similar settings to forecast cosmic shear constraints on model-independent parametrisations of modified gravity theories with scale-independent linear growth. They found that a much better understanding of baryonic feedback is needed in order to detect a screening transition.

\section{Theoretical error}
\label{sec:theoretical_error}
Any theoretical prediction will in general come with an associated error and should be included in the likelihood if it is not subdominant to the statistical error. In this section, we discuss the implementation of theoretical errors in our pipeline. To be conservative, we  estimated the theoretical errors using \react{} and \forge{} as these two give the most discrepant results. We applied this method to 
\react{} for the pessimistic case of WL to check if we could remove the parameter biases that we observe for the data created by \forge{} at the expense of enlarging error bars.

\subsection{Adding uncorrelated theoretical errors to the likelihood}
The implementation presented in this work is based on the work of \citelist{Audren_2013}. We have adjusted the recipe to the full \threextwo{} likelihood, although we only show the result for the WL analysis in this paper. Here, we discuss the idea behind the formulation. We calculated the angular power spectrum from the power spectrum using \cref{eq:ISTrecipe}. If there is some uncertainty in the modelling of the nonlinear power spectrum, this propagates to the angular power spectrum. As \cref{eq:ISTrecipe} is a linear functional of the nonlinear matter power spectrum, $\mathcal{F}[P_{\delta \delta}]$, we can propagate the error on the power spectrum, $\Delta P_{\delta \delta}$, and find \begin{equation}
    \mathcal{F}[P_{\delta \delta}+\Delta P_{\delta \delta}] = \mathcal{F}[P_{\delta \delta}] + \mathcal{F}[\Delta P_{\delta \delta}] \eqqcolon C_{ij}^{XY}(\ell) + E_{ij}^{XY}(\ell)\;,
\end{equation}
where we have defined the angular power spectrum error, $E_{ij}^{XY}(\ell)$. Following \citelist{Audren_2013}, we defined the relative error on the power spectrum:
\begin{equation}
    \Delta P_{\delta \delta}(k,z) \coloneqq \alpha(k,z) \, P_{\delta \delta}(k,z)\;.
\label{eq:alpha}
\end{equation}
We took the most conservative approach and assumed that the error is uncorrelated between different values of $\ell$.
To account for this uncorrelated theoretical error, we added a new nuisance parameter,  $\varepsilon_\ell$, for each multipole and define the shifted covariance as \begin{equation}
    \Tilde{C}^{XY}_{ij}(\ell) \coloneqq \hat{C}^{XY}_{ij}(\ell) + \varepsilon_\ell \, L^{1/2} \, E^{XY}_{ij}(\ell)\;.
\end{equation}
The free parameters, $\varepsilon_\ell$, have the following meaning: they quantify how many standard deviations the shifted covariance matrix is from the theoretically predicted one. We treat them as a random Gaussian variable with zero mean and a standard deviation of one. The normalisation 
factor, $L^{1/2}$, where $L=\ell_\mathrm{max}-\ell_\mathrm{min}+1$, will become clear soon. The log-likelihood function becomes \begin{equation}
    \tilde{\chi}^2 \left(\varepsilon_\ell\right) =  \sum_{\ell =\ell_{\mathrm{min}}}^{\ell_{\mathrm{max}}} \left[ (2\ell+1) f_{\rm sky} \left( \frac{\tilde{d}^{\rm mix}_\ell\left(\varepsilon_\ell\right)}{\tilde{d}^{\rm th}_\ell\left(\varepsilon_\ell\right)} + \ln{ \frac{ \tilde{d}^{\rm th}_\ell\left(\varepsilon_\ell\right)}{d^{\rm obs}_\ell}}-N\right) + \varepsilon_\ell^2\ \right].
\end{equation}
The quantities $\tilde{d}^{\rm th}$ and $\tilde{d}^{\rm mix}$ are constructed in the same way as ${d}^{\rm th}$ and ${d}^{\rm mix}$ 
[see \crefrange{eq:ds1}{eq:ds3}] using the shifted covariance matrix $\Tilde{C}^{XY}_{ij}(\ell)$. To include the theoretical error, we vary $\varepsilon_\ell$ and marginalise over them. To a very good approximation, this is equivalent to minimising the $\tilde{\chi}^2$ with respect to $\varepsilon_\ell$ at the level of the likelihood.  Thus, we define our new $\chi^2$ as the minimum of the $\tilde{\chi}^2$ with respect to $\varepsilon_\ell$:
\begin{equation}
    \chi^2 \coloneqq \min_{\varepsilon_\ell \in \mathbb{R}^L} \tilde{\chi}^2.
\end{equation}
The normalisation factor, $L^{1/2}$, can be explained as follows. If we were to measure $\hat{C}^\mathrm{obs}(\ell) = \hat{C}^\mathrm{th}(\ell) + E(\ell)$, the minimisation would find $\varepsilon_\ell = L^{-1/2}$. The resulting $\chi^2 = \sum_{\ell} \varepsilon_\ell^2 = \sum_{\ell} L^{-1} =1$ would match our expectation that a one-sigma theoretical error for each $\ell$ results in an increase in $\chi^2$ by one. 

The main ingredient of this formulation is the relative error function $\alpha(k,z)$ defined in \cref{eq:alpha}. 
The construction of $\alpha(k,z)$ using \forge{} and \react{} will be discussed in \Cref{app:relerr}. 

\subsection{Numerical implementation}
The numerical computation of the theory error covariance $E^{XY}_{ij}(\ell)$ follows the prescription presented by \citelist{Euclid:2023pxu}. For the minimisation, we can use the fact that all the free $\varepsilon_\ell$ are independent of each other, and we can do the minimisation for each multipole separately. We use Newton's method to find the minimum. For this, we have to compute the first and second derivatives of the likelihood with respect to $\varepsilon_\ell$. For any single multipole, we find \begin{equation}
    \frac{\mathrm{d}\tilde{\chi}}{\mathrm{d} \varepsilon_\ell} = (2\,\ell+1)\,f_{\rm sky}\,\left(\frac{\left(\tilde{d}^\mathrm{mix}_\ell\right)^\prime + \left(\tilde{d}^\mathrm{th}_\ell\right)^\prime}{\tilde{d}^\mathrm{th}_\ell} - \frac{\tilde{d}^\mathrm{mix}_\ell\,\left(\tilde{d}^\mathrm{th}_\ell\right)^\prime}{\left(\tilde{d}^\mathrm{th}_\ell\right)^2} \right) + 2\,\varepsilon_\ell\;.
\end{equation}
The derivatives of the determinants are computed using Jacobi's formula. This gives, for example
\begin{equation}
    \left(\tilde{d}^\mathrm{th}_\ell\right)^\prime = \det\left( \hat{C}_\ell^\mathrm{th}\right)\,L^{1/2}\,\Tr \left[ \left(\hat{C}_\ell^\mathrm{th}\right)^{-1} \, E_\ell\right]\;,
\end{equation}
and a similar expression for $\left(\tilde{d}^\mathrm{mix}_\ell\right)^\prime$. We compute the inverse of the covariance numerically and obtain the second derivatives numerically from the first derivative by doing a double-sided three-point stencil. The minimisation would then need to be done for each multipole. For the pessimistic settings, this would correspond to a minimisation in a 1500-dimensional parameter space. To save time, we only do the minimisation on a logarithmically-spaced grid with 100 discrete values. The other values are obtained from an interpolating function. We checked that the effect of the interpolation does not change the results by more than 1\%. The obtained $\varepsilon_\ell$ tend to vary continuously with $\ell$, as they try to mimic the effects of changing other theory parameters to the observed power spectrum.

\begin{figure*}[ht]
\centering
	\includegraphics[width=0.49\linewidth]{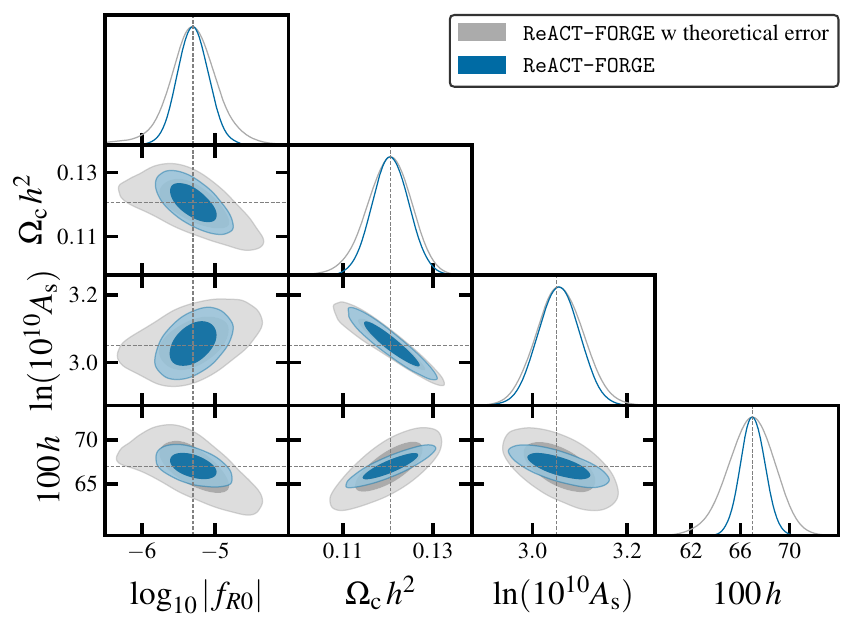}
	\includegraphics[width=0.49\linewidth]{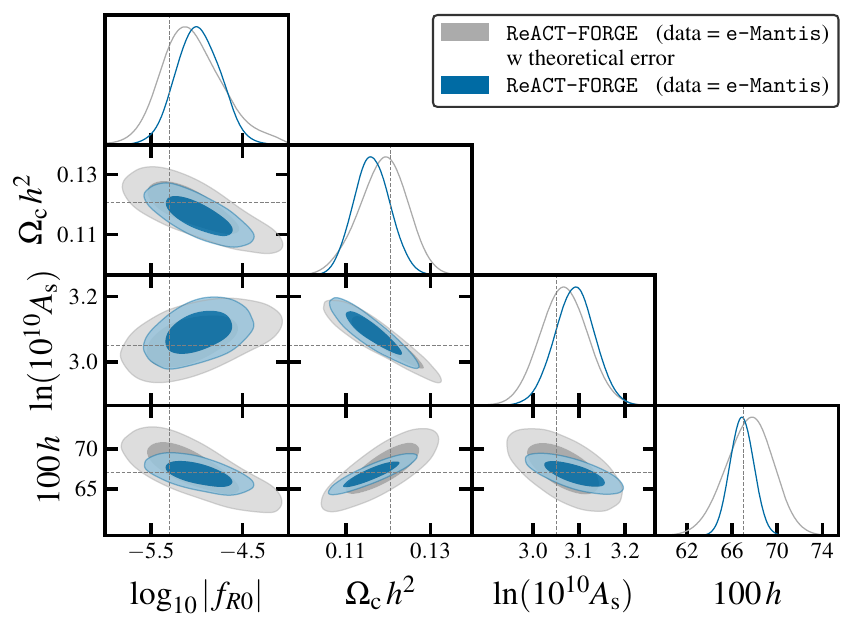}
  \caption{Effect of adding theoretical errors. \textit{Left panel}: Data are created by \forge{} in the WL pessimistic setting and fitted by \react{} corrected by \forge{} for fiducial cosmology, \cref{eq:reactforge}, with and without theoretical errors. \textit{Right panel}: Data are created by \emantis{} in this case. }
   \label{fig:Therrors}
\end{figure*}

\begin{figure}
\centering \includegraphics[width=1\linewidth]{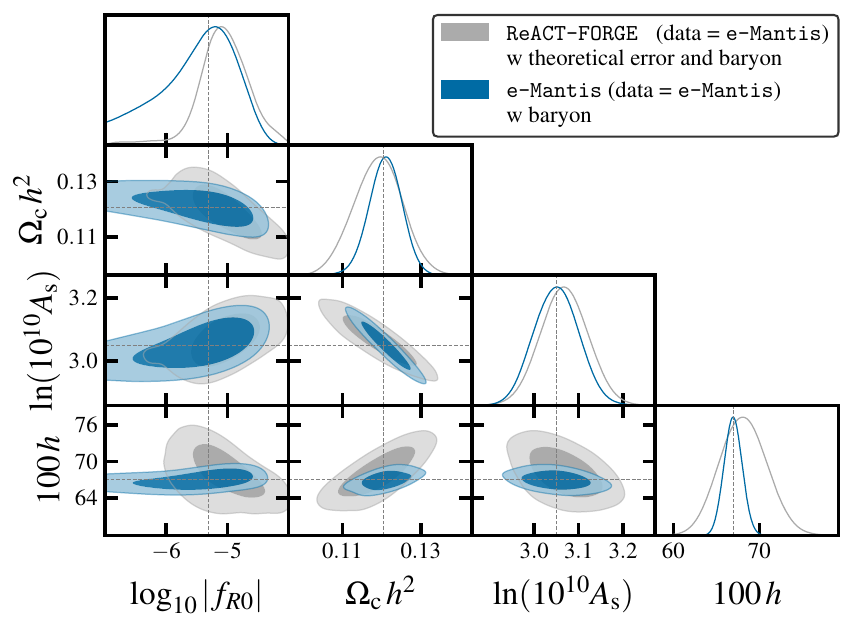}
  \caption{Effect of adding theoretical errors in the presence of baryonic effects. Data are created by \emantis{} and fitted by \react{} corrected by \forge{} for the fiducial cosmology with theoretical errors. For a comparison, we also show the result obtained by using \emantis{} to fit the same data.  
 }
   \label{fig:Therrorsbar}
\end{figure}

\subsection{Application to \react{}}
We first applied this method to the case in which the data are created by \forge{} and the parameter fitting is done by \react{}. Due to the significant bias in the recovered parameters for both $|f_{R0}|$ and cosmological parameters, we find that the inclusion of theoretical errors is not enough to mitigate the bias. Therefore, we additionally correct the prediction of \react{} by \forge{} for the fiducial cosmology as 
\begin{equation}
    \Xi_{\texttt{ReACT-FORGE}}
    = \Xi_{\react{}}
    \left(
\frac{ \Xi_{\forge{}}}{\Xi_{\react{}}} 
    \right)_{\rm fiducial}. 
    \label{eq:reactforge}
\end{equation}
The constraints on cosmological parameters are shown in the left panel of \cref{fig:Therrors}. 
In this case, as expected, we recover the input parameters in an unbiased way when the data are created by \forge{}. We see that theoretical errors affect mainly $\logfr$ and $h$.  In order to check a non-trivial case, next we create the data with \emantis{}. The result is shown in the right panel of \cref{fig:Therrors}. 
Without theoretical errors, the means of 1D marginalised constraints for $\logfr$, $\Omegac h^2$, and $\ln(10^{10} A_{\rm s})$ are slightly biased compared with the input values. The inclusion of the theoretical errors largely resolves these biases not only by enlarging the error bars but also by making the means closer to the input values. We note that the inclusion of theoretical errors in this case is important to justify the rescaling of $\Xi$ in \cref{eq:reactforge}. We also note that the theoretical error included here overestimates the errors significantly now that we corrected the prediction of \react{} by \forge{} for the fiducial cosmology. 

Finally, we study the impact of adding baryonic effects in the presence of theoretical errors. We again use \emantis{} as the data. The result is shown in \cref{fig:Therrorsbar}. Due to the enlarged errors, constraints on parameters are not affected significantly by baryonic effects. Also, the means of the 1D marginalised constraint remain consistent with the input parameters. However, the inclusion of the theoretical errors changes the degeneracies between cosmological parameters, baryonic parameters and $\logfr$. This leads to a tighter lower bound on $\logfr$ compared with the case where the same data are fitted by \emantis{} itself. This reinforces our conclusion that the 1D marginalised constraint on $\logfr$ is sensitive to degeneracies among parameters, and the difference in the theoretical predictions strongly affects the constraint. 

The result shown here is just an illustration of the inclusion of theoretical errors and their effects on the parameter constraints. Our implementation is flexible and it can be applied to any theoretical error described by the relative error function $\alpha(k,z)$ defined in \cref{eq:alpha}. 

\section{Conclusion}
\label{sec:conclusion}
In this paper, we have studied the effect of using different nonlinear predictions for the dark matter power spectrum on the parameter constraints in the Hu--Sawikci $f(R)$ gravity model obtained from \Euclid primary photometric probes. We implemented four different models in the \montepython{} pipeline to predict angular power spectra for WL, photometric galaxy clustering (\GCph{}) and their cross-correlation (\XCph{}). Comparing with the {\it N}-body simulation data obtained in \citetalias{paper1}, we found that \emantis{} agrees very well with \ecosmog{}, which was used to run simulations to construct the emulator, while \forge{} had larger errors compared with the \mgarepo{} simulation. The agreement is better for one of the {\it N}-body simulations used to construct \forge{} obtained in \citetalias{paper2} run by \mgarepo{}. This indicates that the difference between \emantis{} and \forge{} is larger than the one in the baseline {\it N}-body simulations (i.e. \ecosmog{} and \mgarepo{}) mainly due to the way \forge{} was constructed. In the \Euclid\ reference cosmology, \forge{} gives a larger $\Xi$, the ratio between the power spectrum in $f(R)$ and in \lcdm, on all scales compared with \emantis{} and the fitting formula at the $2 \%$ level. \react{} underestimates $\Xi$ compared with \forge{} more than \emantis{} and the fitting formula at intermediate $k$. 

We used the fiducial value of $|f_{R0}| = 5 \times 10^{-6}$ ($\logfr = - 5.301$) and ran MCMC in the \Euclid\ fiducial cosmology defined in \citetalias{Blanchard:2019oqi}. For the fitting formula, the Fisher Matrix forecast and MCMC results generally agree well although the MCMC result is non-Gaussian. This is partly caused by a lack of cosmological parameter dependence of $\Xi$, which affects the degeneracy between $\logfr$ and cosmological parameters for large  $\logfr$. \emantis{} gives more Gaussian constraints with smaller errors. When \forge{} is used to create the data, the 1D mean of $\logfr$ is not strongly biased in the case of \emantis{} and the fitting formula and the 1D bias is at most $0.6 \sigma$. Even for the \threextwo{} analysis including all the probes and their cross-correlations, the 1D bias is $0.15 \sigma$ for $\logfr$ in the case of \emantis{}. 

The impact of baryonic physics on WL was studied by using a baryonification emulator \bcemu{}. For the optimistic setting, the $f(R)$ parameter and two main baryonic parameters are well constrained despite the degeneracies among these parameters. However, the difference in the nonlinear dark matter prediction can be compensated by the adjustment of baryonic parameters as well as cosmological parameters, and the 1D marginalised constraint on $\logfr$ is biased. This bias can be avoided in the pessimistic setting at the expense of weaker constraints. For the pessimistic setting, using the \lcdm\ synthetic data for WL, we obtained the prior-independent bound of $\logfr < -5.6$ using \emantis{}. 

\react{} shows a large bias in $\logfr$ as well as cosmological parameters when \forge{} was used to create the data. 
This is because the prediction of \react{} is furthest away from \forge{}. We implemented a method to include uncorrelated theoretical errors proposed in \citelist{Audren_2013} to address this issue. The method is based on the relative error function for the nonlinear dark matter power spectrum. We estimated this using the difference between \forge{} and \react{}. We found that the inclusion of theoretical errors alone was not enough to mitigate the bias. We then corrected the prediction of \react{} with \forge{} for the fiducial model. We applied this model to the data created by \emantis{}. We found that theoretical errors, in this case, helped reduce the bias not only by enlarging errors but also by making the means of 1D marginalised constraint closer to the input values. When we added baryonic effects with \bcemu{}, errors were not significantly affected and the input values were still recovered. However, the lower bound on $\logfr$ is tighter than the one obtained by applying \emantis{} itself. This reinforces our conclusion that the 1D marginalised constraint on $\logfr$ is sensitive to the degeneracies among parameters. 

Based on the result of this paper, we draw the following conclusions:

\begin{itemize}
\item It is important to check the agreement of different {\it N}-body codes that are used to create theoretical predictions. This means not only the code itself, but also accuracy settings. In \citetalias{paper1}, we found that the accuracy setting such as the refinement criteria in the adaptive mesh refinement method has a large effect on the power spectrum. With the controlled accuracy setting, it is possible to realise $1 \%$ agreements between different {\it N}-body codes in terms of $\Xi(k, z)$ in the Hu--Sawicki $f(R)$ gravity model. 

\item 
We then need to check the accuracy of emulators. We found that \forge{} suffers from larger emulation errors and this leads to a larger difference between \forge{} and \emantis{} compared with the difference in their baseline {\it N}-body code \mgarepo{} and \ecosmog{} in some cosmologies. Improving the emulation technique will make the agreement better. 

\item 
Including baryonic effects can worsen the bias in the 1D marginalised constraint on $\logfr$. This is because the difference in the nonlinear dark matter power spectrum prediction can be compensated by the adjustment of baryonic parameters, and the best-fit values are biased. In addition, the degeneracy between baryonic parameters, cosmological parameters and $\logfr$ leads to a stronger volume effect. This bias can be avoided if there is a prior on baryonic parameters from external datasets for example. 

\item 
To account for the uncertainty in the theoretical prediction for the nonlinear power spectrum, it is safer to include theoretical errors. In this paper, we used a conservative error estimation using \forge{} and \react{}. An improvement in emulators will make the theoretical error smaller, leading to better constraints. However, we still need to check whether or not the inclusion of baryons will change this conclusion. 

\end{itemize}

The pipeline developed in this paper can be used to test the readiness of the nonlinear power spectrum prediction for the application to real data from \Euclid in other extended cosmologies. For the Data Release 1 of \Euclid and future data releases, we plan to improve the 
\forge{} emulator and make the pipeline ready for the data analysis. We also plan to extend the analysis to the models studied in \citelist{Euclid:2023rjj}. 

Finally, we note that our forecasts have ignored observational systematics such as shear and redshift measurement biases. For Stage-IV surveys like \Euclid, it is known that these need to be very well characterised in order to get accurate cosmological parameter estimates. This is a challenge that needs to be addressed for both $\Lambda$CDM and exotic cosmologies. A detailed study of this issue is beyond the scope of our paper, but we refer to \citelist{EuclidSkyOverview} for details. In summary, the mean galaxy redshifts within the bins need to be known with an accuracy better than $\sim 0.002(1+z)$ if errors in cosmological parameters are not to be degraded. On the other hand, \cite{SpurioMancini:2023mpt} included a parameter that shifts the mean of the redshift distribution in each redshift bin and found that these parameters are not strongly degenerate with the $f(R)$ parameter.
\begin{acknowledgements}
\AckEC
K.\ K.\ is supported by STFC grant ST/W001225/1.
B.\ B.\ is supported by a UK Research and Innovation Stephen Hawking Fellowship (EP/W005654/2). 
A.\ P.\ is a UKRI Future Leaders Fellow [grant MR/X005399/1]. P.\ C.\ is supported by grant RF/ERE/221061.
M.\ C.\ acknowledges the financial support provided by the Alexander von Humboldt Foundation through the Humboldt Research Fellowship program, as well as support from the Max Planck Society and the Alexander von Humboldt Foundation in the framework of the Max Planck-Humboldt Research Award endowed by the Federal Ministry of Education and Research. 
F.\ P.\ acknowledges partial support from the INFN grant InDark and the Departments of Excellence grant L.232/2016 of the Italian Ministry of University and Research (MUR). FP also acknowledges the FCT project with ref. number PTDC/FIS-AST/0054/2021.
GR’s research was supported by an appointment to the NASA Postdoctoral Program administered by Oak Ridge Associated Universities under contract with NASA. GR was supported by JPL, which is run under contract by the California Institute of Technology for NASA (80NM0018D0004).
Numerical computations were done on the Sciama High Performance Compute (HPC) cluster which is supported by the ICG, SEPNet, and the University of Portsmouth.
This work has made use of the Infinity Cluster hosted by Institut d'Astrophysique de Paris.
We acknowledge open libraries support \texttt{IPython} \citep{4160251}, \texttt{Matplotlib} \citep{Hunter:2007}, \texttt{Numpy} \citep{Walt:2011:NAS:1957373.1957466}, and \texttt{SciPy} \citep{2019arXiv190710121V}.

For the purpose of open access, we have applied a Creative Commons Attribution (CC BY) licence to any
Author Accepted Manuscript version arising.

Supporting research data are available on reasonable request from the corresponding author.

\end{acknowledgements}


\bibliography{biblio}

\begin{appendix}

\section{\emantis{} fiducial data}
\label{app:emantis}
\begin{figure}
\centering
 \includegraphics[width=0.90\linewidth]{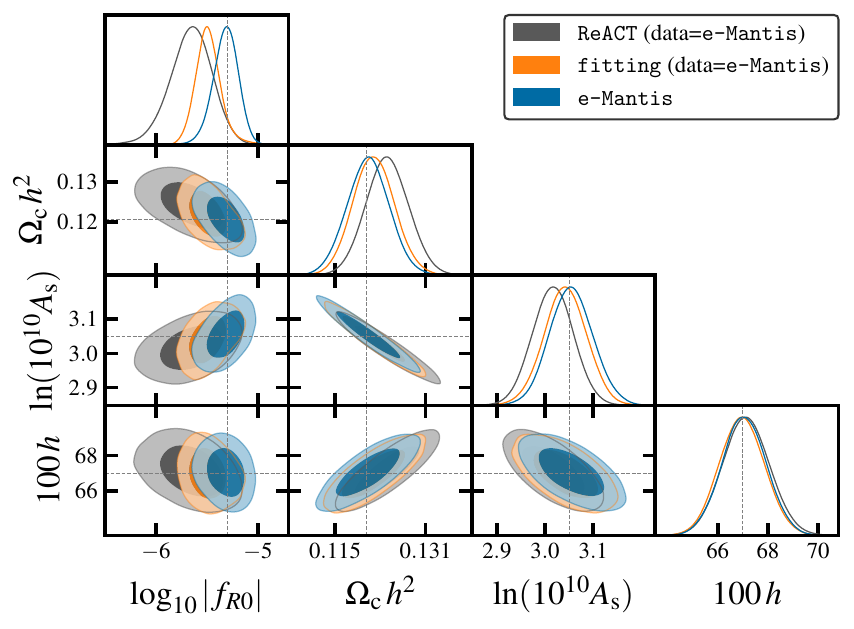}
  \caption{Bias due to different nonlinear modelling. The synthetic data are created by \emantis{} and fitted by two different models in the WL pessimistic case without baryons. }   \label{fig:biasemantis}
\end{figure}

\begin{figure}
\centering
\includegraphics[width=0.90\linewidth]{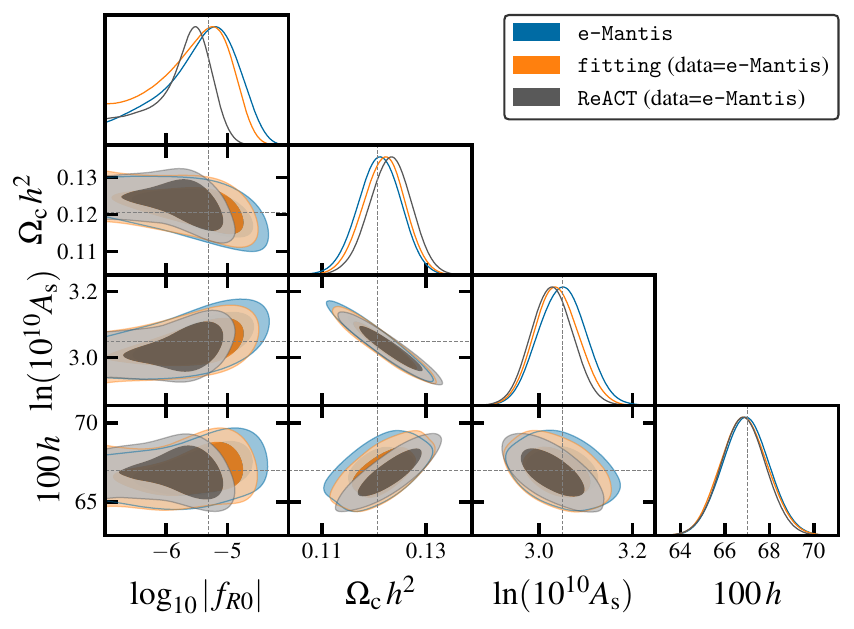}
  \caption{Bias due to different nonlinear modelling. The synthetic data are created by \emantis{} and fitted by two different models in the WL pessimistic case with baryons.}   \label{fig:biasemantisbaryons}
\end{figure}

In this Appendix, we show the result of the cases where we use \emantis{} as fiducial data for the pessimistic setting of WL with and without baryons. \Cref{fig:biasemantis} and \cref{tab:WLpesE} show the results without baryons while the \Cref{fig:biasemantisbaryons} and \cref{tab:WLbaryonpesE} show the results with baryons. We observe that the agreements of the three models are generally better as expected although without baryons, 1D bias is still at the 3$\sigma$ level for \react{} due to smaller errors of \emantis{}. On the other hand, in the case with baryons, 1D bias is reduced to less than 1$\sigma$. 

\begin{table}
\caption{
Mean, standard deviation, and 68.3\% lower and upper limit of $\logfr$ for the WL pessimistic setting.}
\begin{tabular}{ c c c c c c}
\hline  
\rowcolor{crisp} \multicolumn{6}{c}{WL pessimistic}\\
\rowcolor{crispier} & Mean  & S.d.  & Lower & Upper &  $B_{\rm 1D}$  \\
\rowcolor{gray} \multicolumn{6}{l}{Data = \emantis{}}\\
{\bf fitting}  & $-5.482$ & 0.135 & $-5.615$ & $-5.376$ & 1.464   \\
{\bf \react{}}  & $-5.663$ & 0.216 & $-5.851$ & $-5.440$ & 3.080 \\
\end{tabular}
\label{tab:WLpesE}
\end{table}

\begin{table}
\caption{
Mean, standard deviation, and 68.3\% lower and upper limit of $\logfr$ for the WL pessimistic setting with baryons.}
\begin{tabular}{ c c c c c c}
\hline  
\rowcolor{crisp} \multicolumn{6}{c}{WL pessimistic with baryons}\\
\rowcolor{crispier} & Mean  & S.d.  & Lower & Upper &  $B_{\rm 1D}$  \\
\rowcolor{gray} \multicolumn{6}{l}{Data = \emantis{}}\\
{\bf fitting} & $-5.626$ & $0.637$ & $-6.085$ & $-4.785$ &  0.337  \\
{\bf \react{}}  & $-5.781$ & 0.495 & $-6.053$ & $-5.147$ & 0.700  \\
\end{tabular}
\label{tab:WLbaryonpesE}
\end{table}

\section{Construction of the relative error} \label{app:relerr}

\begin{figure}[h]
	\centering
	\includegraphics[width=\linewidth]{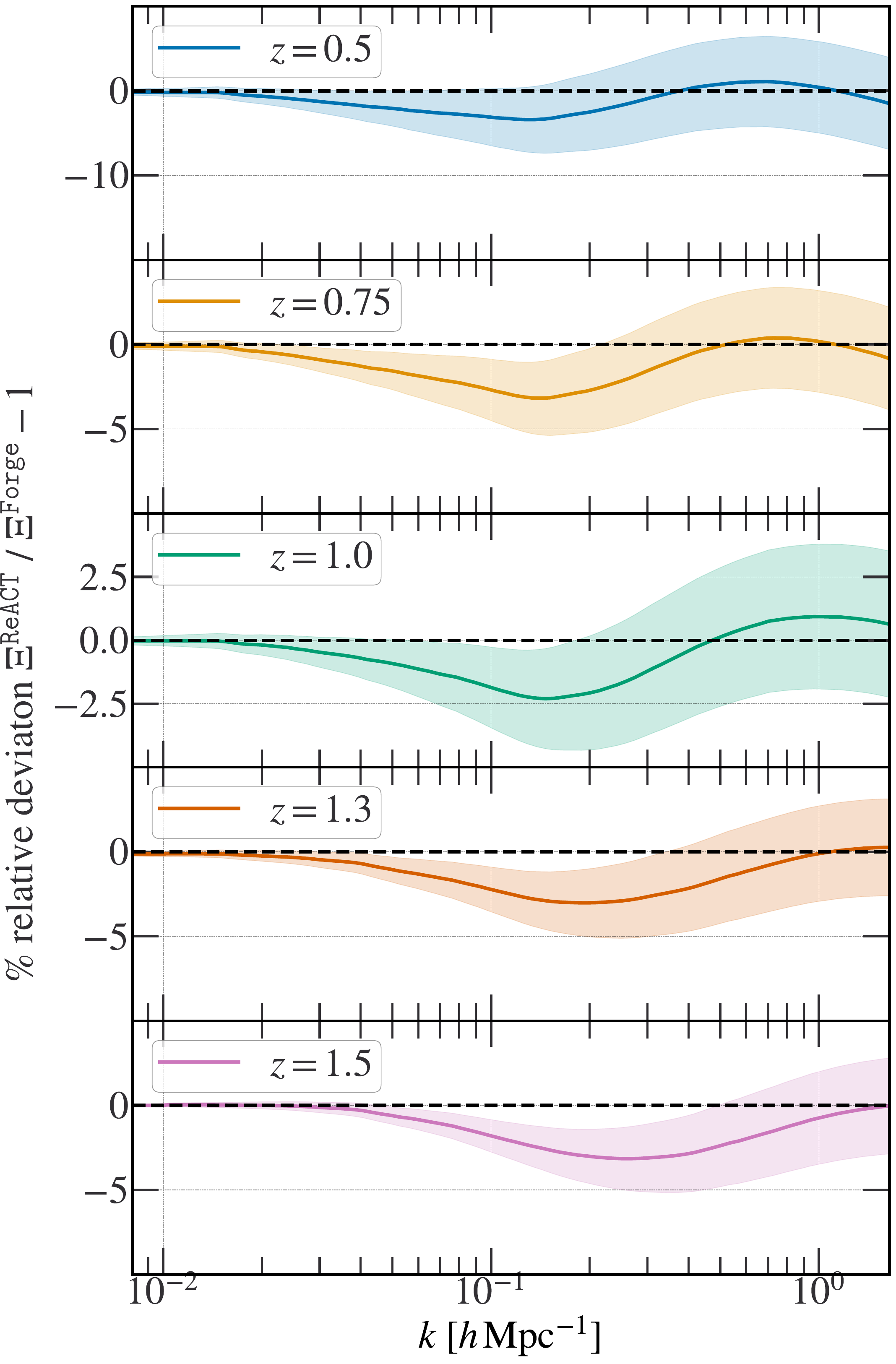}
	\caption{Relative difference between the power spectra from \forge{} and \react{}. We show the relative deviation as the lines and the 68.3\% confidence level as the shaded area. The boosts were all computed at the fiducial cosmology.}
	\label{fig:App_theo_error}
\end{figure}

To construct the relative error function, $\alpha(k,z)$, we can use the fact that $\Xi(k, z)$ is constructed as the ratio to the same \lcdm \ nonlinear power spectrum calculated with the \halofit{} `Takahashi' prescription in all models. This means that any deviations of the \lcdm \ power spectrum from $N$-body simulations is also absorbed in $\Xi$. We can thus write the relative error as
\begin{equation}
    \alpha(k,z) = \frac{\Delta P_{\delta \delta}(k,z)}{ P_{\delta \delta}(k,z)} = \frac{\Delta \Xi(k,z)}{ \Xi(k,z)}\,.
\end{equation}
To construct $\Delta P_{\delta \delta}$ we follow the basic idea that all predictions for $\Xi$ are equally accurate in calculating the modified gravity power spectrum. The true power spectrum is thus only known up to the spread of the predictions. To be conservative with our forecast, we choose to calculate the error from the difference between the \forge{} predictions and the \react{} predictions, which is the largest among these predictions. This is shown in \cref{fig:App_theo_error}. The \react{} prescription underpredicts the power spectrum at intermediate scales and overpredicts slightly at smaller scales. This is most likely due to \react{} handling the one- and two-halo power spectra separately. \texttt{HMcode2020} predictions for the pseudo-cosmology nonlinear power spectrum also contribute to these inaccuracies as we can see by comparing the left and right panels of Figure 4 in \citelist{Cataneo:2018cic}. 
To simplify the exact difference between these two codes, we argue that the theoretical error should plateau at the maximum deviation. This is because at nonlinear scales, the modes of different wave numbers are no longer independent of each other. By plateauing, we essentially say that we can no longer be more precise in the computation of the power spectrum on smaller scales. 

Finally, we construct $\alpha$ using only the difference of the predictions at a fiducial of $|f_{R0}| = 5 \times 10^{-6}$ in order not to bias our results. Our final fit has the form
\begin{align}
    \alpha(k,z) &= A(z)\,\frac{x^2+x}{x^2+x+1}\,,\\
    A(z) &= \frac{A_1}{\exp\left(\frac{z-A_2}{A_3}\right)+1}+A_4\,, \\
    k_{\rm p}(z) & = B_1 \, \exp\left[\mathrm{tanh}\left(\frac{z-B_2}{B_3}\right)\right]\,,
\end{align} 
where we use $x \coloneqq k / k_{\rm p}$. We separately fit the amplitude function $A(z)$ to the maximum deviation and the plateau wavenumber $k_{\rm p}(z)$ to the wavenumber of maximum deviation. The best-fit values can be found in \Cref{tab:app_best_fit_theo_err} and are shown as the shaded region in \cref{fig:App_theo_error}. We see that the zero line is within the 68.3\% confidence bounds most of the time. This is in accordance with our construction that the two emulators should differ by the theoretical error.

\begin{table}[t]
    \centering
    \caption{Best-fit values of the relative error function.}
    \begin{tabular}{cl}
        \hline
         \rowcolor{gray} \multicolumn{1}{c}{Parameter}& \multicolumn{1}{c}{Best-fit} \\

         $A_1$&$3.56\phantom{0} \times 10^{-2}$\\
         $A_2$&$0.562\phantom{\times 10^{-2}}$\\
         $A_3$&$6.1\phantom{00} \times 10^{-2}$\\
         $A_4$&$2.90\phantom{0} \times 10^{-2}$\\
         \hline
         \noalign{\vskip 1pt}
         $B_1$& $0.202$\\
         $B_2$& $1.56  $\\
         $B_3$& $0.55$\\
    \end{tabular}
    \label{tab:app_best_fit_theo_err}
\end{table}

\end{appendix}

\label{LastPage}

\end{document}